\documentclass[traditabstract]{aa} 

\usepackage{graphicx, times, txfonts, float, rotating, color, url, lscape, bm, mathtools}
\usepackage{caption, subcaption, mwe, hyperref, gensymb, booktabs, dirtytalk, lmodern, fancyhdr, mhchem, tabularx}
\usepackage{wrapfig, floatflt, natbib}
\usepackage{placeins}
\usepackage{longtable}

\newcommand{\msol}{\mbox{M$_{\odot}$}}

\newcommand{\lsol}{\mbox{L$_{\odot}$}}

\newcommand{\Ks}{K_{\rm s}}

\newcommand{\muas}{$\mu$as} 

\newcommand{\G}{{\it Gaia}}

\begin{document} 
 
\title{Calibration of the JAGB method for the Magellanic Clouds and Milky Way from {\it Gaia} DR3, considering the role of oxygen-rich AGB stars
}  
 
\author{ 
  E.~Magnus\inst{1,2} 
  \and
  M.~A.~T.~Groenewegen\inst{2}
  \and
  L.~Girardi\inst{3}
  \and
  G.~Pastorelli\inst{4,3}
  \and
  P.~Marigo\inst{4}
  \and
  M.~L.~Boyer\inst{5}
}

\institute{ 
  Vrije Universiteit Brussel, Dienst ELEM, Pleinlaan 2, B--1050 Brussels, Belgium
  \and
  Koninklijke Sterrenwacht van Belgi\"e, Ringlaan 3, B--1180 Brussels, Belgium  \email{martin.groenewegen@oma.be}
  \and
  INAF - Osservatorio Astronomico di Padova, Vicolo dell'Osservatorio 5, I--35122 Padova, Italy
  \and
  Dipartimento di Fisica e Astronomia "Galileo Galilei", Universit\`a di Padova, Vicolo dell'Osservatorio 3, I--35122 Padova, Italy
  \and
  Space Telescope Science Institute, 3700 San Martin Drive, Baltimore, MD 21218, USA
  \\
} 
 
\date{received: ** 2024, accepted: * 2024} 
 
\offprints{Martin Groenewegen} 
 
\titlerunning{JAGB method for MCs and MW} 
 
\abstract
{
The JAGB method is a new way of measuring distances in the Universe with use of AGB stars that are
situated in a selected region in a $J$ versus $J-\Ks$ colour-magnitude diagram (CMD), using the fact that the
absolute $J$ magnitude is (nearly) constant.
It is implicitly assumed in the method that the selected stars are carbon-rich AGB stars (carbon stars).
However, as the sample selected to determine $M_{\rm J}$ is purely colour based there can also be contamination by
oxygen-rich AGB stars in principle. As the ratio of carbon to oxygen-rich stars is known to depend on metallicity and initial mass,
the star formation history and age-metallicity relation in a galaxy should influence the value of $M_{\rm J}$.
The aim of this paper is to look at mixed samples of oxygen-rich and carbon-rich stars for the
Large Magellanic Cloud (LMC), Small Magellanic Cloud (SMC) and Milky way (MW), using the {\em Gaia} catalogue of
long period variables (LPVs) as basis. The advantage of this catalogue is that it contains a classification of O- and C-stars
based on an analysis of {\em Gaia} Rp spectra.
The LPV catalogue is correlated with data from the Two Micron All Sky Survey (2MASS) and samples in the LMC, SMC, and the MW
are retrieved.
Following methods proposed in the literature we report the mean and median magnitudes of the selected sample using
different colour and magnitudes cuts and the results of fitting Gaussian and Lorentzian profiles to the luminosity function (LF). 
For the SMC and LMC we confirm the previous results in the literature.
The LFs of the SMC and LMC JAGB stars are clearly different, yet the mean magnitude inside a selection box
can be argued to agree at the 0.021~mag level.
The analysis of the MW sample is less straightforward. The contamination by O-rich stars is substantial for a classical
lower limit of  $(J-\Ks)_0= 1.3$, and becomes less than 10\% only for $(J-\Ks)_0= 1.5$.
The sample of AGB stars is smaller than for the MCs for two reasons. Nearby AGB stars (with potentially the best determined parallax)
tend to be absent as they saturate in the 2MASS catalogue, and the parallax errors of AGB stars tend to be larger compared to
non-AGB stars.
Several approaches have been taken to improve the situation but in the end the JAGB LF for the MW contains about 130 stars, and
the fit of Gaussian and Lorentzian profiles is essentially meaningless. The mean and median magnitudes are fainter than
for the MC samples by about 0.4~mag which is not predicted by theory.
We do not confirm the claim in the literature that the absolute calibration of the JAGB method is independent of metallicity up to solar metallicity.
A reliable calibration of the JAGB method at (near) solar metallicity should await further \G\ data releases,
or should be carried out in another environment.
}


\maketitle

\section{Introduction}
\label{sec:introduction}

\cite{Nikolaev_2000} presented an infrared $\Ks$ vs. $J-\Ks$ colour-magnitude diagram (CMD) of stars
in the Large Magellanic Cloud (LMC) based on Two Micron All Sky Survey (2MASS, \citealt{Skrutskie98, Skrutskie06}) data
and reported on a region that (almost solely) consists of carbon-rich asymptotic giant branch (AGB) stars, that they
denoted as the J region. This, and the fact that the infrared $J$-band is used in this method [see later], gives the name JAGB stars
in the context of these stars as a distance indicator that has appeared in the literature over the past few years.
The name is actually confusing as the class of J-type carbon-stars is long known as a separate spectroscopic group
(together with R, N and CH-type C stars; \citealt{Keenan93, Bouigue54}) that are characterised by strong $^{13}$C-lines.
The origin of this class is not well known and possibly related to binary evolution.
In the LMC J-type carbon-stars partly occupy the same region as the JAGB stars in the ($\Ks$, $J-\Ks$) CMD \citep{Morgan03}
which may add to the confusion.

JAGB stars are characterised by their high infrared luminosities and red colours and take only certain values of
those variables. As can be seen in the original CMD of \cite{Nikolaev_2000} (their Figure 3), the J region is easily
separated from the other groups by considering a range in colour.
Blue wards of  $J-\Ks \sim 1.3-1.4$~mag (regions F and G), the distribution of oxygen-rich AGB stars and
RGB stars stops abruptly at this boundary, illustrating indeed the difference in colour between O-rich and
C-rich AGB stars in the general population. Red wards, the limit of the JAGB stars is put at $J-\Ks = 2.0$~mag to minimise the
contamination of extreme carbon stars (region K) whose dusty envelopes cause the extreme red colours.
The location of predominantly C stars in this region is understood theoretically \citep{Marigo2003}.

\cite{Nikolaev_2000} and \cite{Weinberg_2001} discuss the use of the stars in region J as standard candles, but not in terms of their
$J$-band magnitude, but in connection to the fact that many AGB stars are long period variables (LPVs), in particular large amplitude Mira variables.
They take 79 C- and O-rich LPVs from \citet{Glass90} and when restricting the analysis to the 14 Miras in the range $1.4 < (J-\Ks) <1.9$~mag
they derive the relation
$K_{\rm s} = -(0.99 \pm 0.80) \; (J-\Ks)+ (12.36 \pm 1.33)$. 
From this it actually follows that
$\Ks + 0.99 \; (J-\Ks) \approx J = 12.36$~mag and with the distance modulus (DM) of 18.5 ($\pm$ 0.1) (used by \citealt{Nikolaev_2000}) one finds that
$M_{\rm J} \approx -6.15$~mag.
This result was not explicitly remarked in \cite{Weinberg_2001} and went seemingly unnoticed until \cite{madore2020calibration} pointed out this property
and introduced the JAGB method for distance determination.

It is of historical interest to point out that \citet{Richer1984} used the mean $I$-band magnitude of seven carbon stars in NGC 205
to derive the distance to this galaxy. This was possible due to the use of narrow-band filters 
(centred on a TiO and a CN band, thus distinguishing directly between C and late-type M stars; see \citealt{PW82,Aaronson1984,Richer1984,Cook86})
developed at that time and the $V, R$ and/or $I$-bands as continuum filters.
Two decades later \cite{Battinelli05,Battinelli2005} presented the final results of their homogeneous survey of carbon stars in nearby galaxies using
these filters. They also focussed on the  $I$ band (although they did point out the potential of using NIR photometry), and
found a weak dependence of $M_{\rm I}$ on metallicity, the metal-poorer systems being brighter\footnote{\cite{Battinelli2005} provide a fit but without error bars.
  Refitting their data on 16 galaxies gives $M_{\rm I} = (-4.32 \pm 0.07) + (0.27 \pm 0.05)$ [Fe/H] with an rms of 0.09~mag.}.

\cite{madore2020calibration} selected stars in the box  $1.30 < (J - K) < 2.00$, corrected for extinction, and determined mean magnitudes
of $-6.22 \pm 0.01 \pm 0.03$~mag for the LMC and $-6.18 \pm 0.01 \pm 0.05$~mag for the SMC (see Table~\ref{tab:comparison}).
\cite{freedman2020application} provide some additional details on the calibration of the LMC, adopt $M_{\rm J} = -6.20 \pm 0.037$~mag based on the SMC and LMC,
and then apply the method to 14 nearby galaxies.

Independently, the $J$ mag luminosity function (LF) of C stars in the SMC, LMC and Milky Way (MW) was investigated by \cite{ripoche2020carbon} using
very similar techniques and 2MASS data. They used a slightly different colour box, namely $1.4 < (J-\Ks)_{\rm 0} < 2.0$ and determined the absolute magnitude
using the median. They also used catalogues of spectroscopically confirmed C stars to provide the absolute magnitude of both all stars in the colour box and
the confirmed C stars. They find $M_{\rm J}$ values for the SMC and LMC that are significantly different from each other, and that are different from
the value in \cite{madore2020calibration} and \citet{freedman2020application}. Their value for the MW ($-5.601 \pm 0.026$)~mag is significantly fainter
than that for the MCs.
The width of the LF in the MW (0.67~mag) is also significantly wider than they find for LMC (0.35~mag) and SMC (0.37~mag). 
They attribute this to differences in metallicity and star formation history.

\citet{Parada21} follow \cite{ripoche2020carbon} in using the median magnitude but additionally also fit a Lorentzian profile to determine
skewness and kurtosis of the LF. Depending on the skewness of the LF in the target galaxy, either the LMC or the SMC is used as calibrator galaxy.
This method is used to derive the distance to IC 1613 and NGC 6822.
\citet{Parada23} expand this work to a larger sample of galaxies and introduced an un-binned maximum likelihood estimator to determine the parameters of
the Lorentzian profile. The method is applied to NGC 6822, IC 1613, NGC 3109, and WLM. The distance to these galaxies is also estimated via
the tip of the red giant branch (TRGB) method, and good agreement is found.

\citet{Zgirski21} fit the LF with a function consisting of a Gaussian plus a second degree polynomial.
The calibration in the LMC and the application to the SMC
  WLM, NGC 6822 and NGC 3109
used a colour box of $1.30 < (J - K)_0 < 2.0$ mag, while
for M33, 
  NGC 55, NGC 247, NGC 300 and NGC 7793,
which show visible contamination on the blue side, the box is
$1.45 < (J - K)_0 < 2.0$.
In all cases they adopt a selection box of 2.5~mag height in $J$.

\cite{lee2021preliminary} present a calibration of the JAGB method for the MW using {\it Gaia} \citep{GC2016a}
data release 3 (GDR3; \citealt{GaiaDR3Vallenari22}). They used two catalogues of
confirmed C stars\footnote{239 stars from \citet{Whitelock06} and 972 from \citet{Chen12}.}
and correlated them with 2MASS and  {\it Gaia} based distances from \cite{bailerjones2021}. A selection on parallax error
(stars with $\sigma_{\pi}/\pi >0.2$ were eliminated) and on quality of the astrometric solution ({\tt RUWE} $< 2.0$) was made.
In the box $1.40 < (J-\Ks) < 2.00$ 153 stars remained and based on the median their
final result was $M_{\rm J} = -6.14 \pm 0.05$ (stat.) $\pm 0.11$ (sys.)~mag. Note that this value is the apparent absolute magnitude,
i.e., not corrected for extinction (see the last sentence in their Sect.~3)\footnote{We retrieved the list of C-stars from the publishers
  website that contained the \G\ source\_id, the $J$ and $(J-K)$ colour (transformed to the 2MASS system when required), and the distance they used.
  We then independently made the correlation with the \G\ main catalogue to get the parallax and the coordinates, with \cite{bailerjones2021} to
  also obtain the error on the distance, and with the 2MASS catalogue to obtain the errors on the photometry and the 2MASS quality flag.
  In fact, only 29 of the 106 sources for which the original 2MASS photometry was kept have a quality flag "AAA", which is the criterion we will use below.
}.

\citet{Madore22OC} consider the calibration in the MW by using AGB stars in MW open clusters (OCs), as compiled by \citet{Marigo22}.
Based on 17 JAGB stars they derive $M_{\rm J} = -6.40$~mag with a scatter of 0.40 and an error on the mean of 0.10~mag.
Considering the MW calibration by \cite{lee2021preliminary} mentioned above, \citet{Madore22OC} obtain a weighted average of
$M_{\rm J} = -6.20 \pm 0.01$ (stat) $\pm 0.04$ (sys)~mag to claim consistency between the SMC, LMC, and MW with no evidence for a dependence on metallicity.
In trying to reproduce figure~1 in \citet{Madore22OC} (see Fig.~\ref{Fig:Marigo}) it was discovered that they must have used a colour selection of 
$1.20 \le (J-\Ks)_0 < 2.0$, which was not explicitly mentioned in their paper, and is different from what was used in \cite{lee2021preliminary}.
In addition, Table~2 in  \citet{Marigo22} contains the spectral types
of a subset of AGB stars in OCs that shows that 4 of the 17 stars are of spectral type M, MS, or S, and only 5 are confirmed C stars.
Repeating the analysis using only the C stars, excluding the known non-C stars, or using  redder lower limit results in
brighter magnitudes ($\approx -6.5$ to $-6.7$~mag) and larger errors in the mean.

In this paper we investigate the calibration of the JAGB method using the latest results from {\it Gaia} in particular the
second catalogue of LPVs \citep{lebzelter2022lpvcatalog} that contains almost 1.7 million LPVs with variability amplitudes
in the $G$ band larger than 0.1~mag. It is therefore a very reliable catalogue for AGB stars in general.
This catalogue also contains a flag indicating whether an object is a C star (see below), and thus makes it possible to better
investigate the contamination of O stars in any colour box that is chosen to contain C stars.

It is well know that the ratio of C to O-rich AGB stars depends on metallicity \citep{Cook86, Gr2002, Mouhcine03, Battinelli05, Boyer19}
with a C to late-M number ratio of about 5, 1, and 0.2 in the SMC, LMC, and solar neighbourhood, respectively \citep{Gr2002},
and therefore the number of O-rich contaminants is potentially larger at higher metallicities.
A clear example of contamination of the JAGB region by M-type stars is provided by the M31 disk: While the adaptive-optics $JHK$ photometry
by \citet{Davidge05} revealed the presence of hundreds of candidate C stars with $J - K >$ 1.3~mag, the medium-band HST photometry
by \cite{Boyer13} reclassified most of those candidate C stars as being late M-type stars.
The conclusion by Boyer et al. was that at higher metallicities the M-type stars of later types became more frequent,
more easily contaminating the CMD region of $J-K > 1.3$. Although this effect is striking at the near-solar metallicities of M31,
it could also affect observations in galaxies of sub solar metallicities such as the LMC.

Secondly, O-rich AGB stars also lose mass and can attain red colours, in its extreme form the so-called OH/IR stars (see \citealt{Hyland74}
for an early review). These stars are typically more massive than the C stars in a given galaxy and therefore less numerous but as they
evolve in time, mass loss increases, and colours become redder (e.g. \citealt{Jones82}) and they potentially enter (and possibly cross) the J region as well.
As \citet{Madore22OC} showed that there is a trend of $M_{\rm J}$ with turn-off mass in the OCs containing AGB stars and as increasing metallicity
is making the mass range for AGB stars to turn into C stars narrower (leading to the smaller C/M ratio observed in galaxies), one
might actually expect a dependence of $M_{\rm J}$ on metallicity.

In Section~\ref{sec:data} the sample and the input data is introduced, and the models to fit the data are described in Section~\ref{sec:model}.
Subsequently, our analysis is performed in Section~\ref{S:Res} and the conclusions are reported in Section~\ref{sec:conclusions}.

\section{Data} 
\label{sec:data}

\cite{lebzelter2022lpvcatalog} provide the second catalogue of LPVs (hereafter LPV2 catalogue) based on the 34 months of data from the \G\ third
data release \citep{GaiaDR3Vallenari22}. Compared to the 22 months of data that was at the basis of GDR2 and the first
LPV catalogue \citep{mowlavi2018gaia} it contains more LPVs (1.7 million versus 150~000, of which 390~000 vs. 89~000 with periods), probes
to lower variability amplitudes (0.1 vs. 0.2~mag), and unlike the first catalogue it classifies about 545~000 objects as C stars.
The identification is based on the shape of \G\ low-resolution Rp spectra, in particular on the difference in pseudo-wavelength between the two highest
peaks in the spectrum ({\tt median\_delta\_wl\_rp}) which are very different for O- and C-rich AGB stars (see section~2.4 in \citealt{lebzelter2022lpvcatalog}).
When {\tt median\_delta\_wl\_rp} $> 7$ the star is assumed to be a C star and the flag $\textit{isCstar}$ is set to $\mathrm{TRUE}$.

In a first step, all data fields from the $\textit{vari\_long\_period\_variable}$ table were extracted as well as selected fields from the 
$\textit{vari\_summary}$ and the main $\textit{gaia\_source}$ tables\footnote{The ASDL query was
SELECT gs.source\_id, gs.ra, gs.dec, gs.parallax, gs.parallax\_error, gs.pmra, gs.pmdec, gs.ruwe, gs.astrometric\_gof\_al, gs.astrometric\_params\_solved,
  gs.nu\_eff\_used\_in\_astrometry, gs.pseudocolour, gs.phot\_g\_mean\_mag, gs.phot\_bp\_mean\_mag, gs.phot\_rp\_mean\_mag, gs.non\_single\_star,
  gs.in\_andromeda\_survey, gs.radial\_velocity, gs.radial\_velocity\_error, gs.l, gs.b, gs.ecl\_lat, vs.trimmed\_range\_mag\_g\_fov, lpv.*
FROM            gaiadr3.vari\_summary                AS vs
LEFT JOIN       gaiadr3.gaia\_source                 AS gs  ON  gs.source\_id = vs.source\_id
LEFT OUTER JOIN gaiadr3.vari\_long\_period\_variable AS lpv ON lpv.source\_id = vs.source\_id
WHERE lpv.source\_id = vs.source\_id
}
for the 1.7 million LPVs.
In a second step, the sources were correlated with the 2MASS catalogue, using the cross-correlation table provided by the \G\ team, and retaining
only the 1.5 million objects with a photometric quality of `A' in all three bands.
In a third step the SMC, LMC and MW samples were constructed.

The SMC and LMC samples are selected according to the criteria on position, proper motion, and parallax as outlined in \citet{Mowlavi2019} resulting
in 4973 and 39014 sources, respectively.
As a check, the radial velocity (RV) data available in GDR3 allowed us to verify that the distribution of the available RVs in these sample is
consistent with that expected for the SMC and LMC, see Appendix~\ref{App:RV}.
The distances to the SMC and LMC are based on the works on EBs \citep{Pietrzynski19,graczyk20}.
Reddening is based on the nearest match in the  reddening maps of stars in the Magellanic Clouds  \citep{skowron2021ogle}, and we adopt
a selected reddening of 3.1 and $E(B-V) = E(V-I) / 1.318$.

The selection of the MW sample is done in two steps.
First an all-sky selection is made on parallax and parallax error. It is well known that there is an parallax zero-point offset (PZPO).
For (faint) QSO this is about $-17$~\muas\ \citep{GEDR3_LindegrenZP}, but it is more negative at brighter magnitudes
(e.g. \citealt{GEDR3_LindegrenZP, Gr21, MaizA22}).
A generous cut of $(\pi +0.1$ (mas)) $> 0$ and $\sigma_{\pi} / (\pi +0.1) < 0.2$ (or equivalently, $R_{\rm plx}$ $\equiv (\pi +0.1)/\sigma_{\pi}  >= 5$)
is made resulting in 258~000 sources. 
This choice has no influence on the final results, as the adopted distances to the MW sample will not be based on the observed parallax but on the
Bayesian inference of the distance from \cite{bailerjones2021}.
In a second step objects in the MCs and in the
Sgr dSph, M31, and M33 galaxies are eliminated according to the selection rules in  \citet{Mowlavi2019}
and \cite{lebzelter2022lpvcatalog}\footnote{Other galaxies containing LPVs are listed in \cite{lebzelter2022lpvcatalog} but their numbers
are insignificant.}.

The MW sample is correlated with \cite{bailerjones2021} to obtain the geometric distance to the source.
The error is estimated from half the difference between the 84 and 16 percentiles.
Reddening is based on the map described in \cite{Lallement18}\footnote{\url{https://stilism.obspm.fr/} (version 4.1).
} (hereafter STILISM) and is based on {\it Gaia}, 2MASS and APOGEE-DR14 data.
For a given galactic longitude, latitude and distance, the tool returns the value of $E(B-V)$ and an error, as well as the distance to which these
values refer. If this distance is smaller than the input distance the returned value for the reddening is a lower limit.
Selected reddenings of $A_{\rm J} = 0.243 \cdot A_{\rm V}$ and $A_{\rm K} = 0.078 \cdot A_{\rm V}$ were adopted \citep{wang2019optical}.

As mentioned earlier, the second LPV catalogue classifies stars as C stars based on properties in the Rp spectra.
However, the combination of \G\ and 2MASS data also allows an independent classification scheme as introduced by \citet{lebzelter2018new}
and slightly refined by \citet{Mowlavi2019}. It is based on a diagram (hereafter the \G-2M diagram) where the ordinate is the $\Ks$-band magnitude
and the abscissa is the difference between two Wesenheit indices,
\begin{equation}
  \Delta W_{\rm G2M} = W_{\rm Bp,Rp} - W_{\rm K,J-\Ks},
  \label{Eq:WG2M}
\end{equation}
where 
\begin{equation}
  W_{\rm Bp,Rp} =  Rp - 1.3 \cdot (Bp - Rp)
  \label{Eq:WG}
\end{equation}
and
\begin{equation}
  W_{\rm K,J-\Ks} =  \Ks - 0.686 \cdot (J-\Ks)
  \label{Eq:W2M}
\end{equation}
are two Wesenheit functions using \G\ and 2MASS colours.
Based on the position in the \G-2M diagram the star is classified into C-rich and O-rich, and subdivided into extreme and standard C-rich,
respectively, low mass, intermediate mass, and massive O-rich AGB plus red supergiants \citep{lebzelter2018new,Mowlavi2019}.
The original separation between the different classes is based on the observed magnitudes in the LMC.
Here we will use dereddened absolute magnitudes which might shift these boundaries a little.
However, the typical value of $A_{\rm K}$ is only 0.02~mag for the LMC sample and this is smaller than the uncertainty in
determining these boundaries. For completeness the adopted boundaries are given in Table~\ref{App:Tab:Boun}.

\section{Model} 
\label{sec:model}

The references on the JAGB method mentioned in the introduction highlight that no yet standard method has been developed to determine the absolute
$J$ magnitude. Authors have used the mean, the median, the mode, have fitted Lorentzian and Gaussian functions, applied different ranges in $J-\Ks$, and sometimes
in the $J$ magnitude as well. All methods are explored below.

One fitted model is a Gaussian plus a  quadratic function \citep{Zgirski21}
\begin{eqnarray}
  G & = & \frac{N}{\sqrt{2\, \pi} \; \sigma} \exp{\left(-\frac{1}{2} \left( \frac{x - \mu}{\sigma}\right)^2 \right)} + \nonumber \\
    &   &  \hspace{25mm} + b + c \; (x - x_0) + d \; (x - x_0)^2,
\label{Eq:G}
\end{eqnarray}
where $x$ refers to the $J$ magnitude, $N$ is a scaling number, $\mu$ is the sought after absolute magnitude, $\sigma$ is the width,
$b, c$, and $d$ represent the background terms  (taking into account any contaminants of the sample of carbon stars),
and $x_{\rm o}$ is a constant chosen to be $-6.2$~mag.
The second fitted model is a modified Lorentzian profile introduced by \citet{Parada21}, and adding the background terms our implementation reads
\begin{eqnarray}
  L & = & \frac{N}{1 +  \left(\frac{x - \mu}{w}\right)^2 + s \; \left(\frac{x - \mu}{w}\right)^3 + k \; \left(\frac{x - \mu}{w}\right)^4} + \nonumber \\
    &   &  \hspace{25mm}  + b + c \; (x - x_0) + d \; (x - x_0)^2, 
\label{Eq:L}
\end{eqnarray}
where $\mu$ is the sought after absolute magnitude, $w$ is the width, $s$ is the skewness, and $k$ is the kurtosis of the distribution.
The fitting is done with the Levenberg-Marquardt algorithm (the routine {\it mrqmin} as implemented in Fortran by \citealt{Press1992}).

\section{Results}
\label{S:Res}

\subsection{Magellanic Clouds}
\label{subsec:LMC}

\begin{figure*}
    \centering
    \begin{minipage}{0.45\textwidth}
       \resizebox{\hsize}{!}{\includegraphics{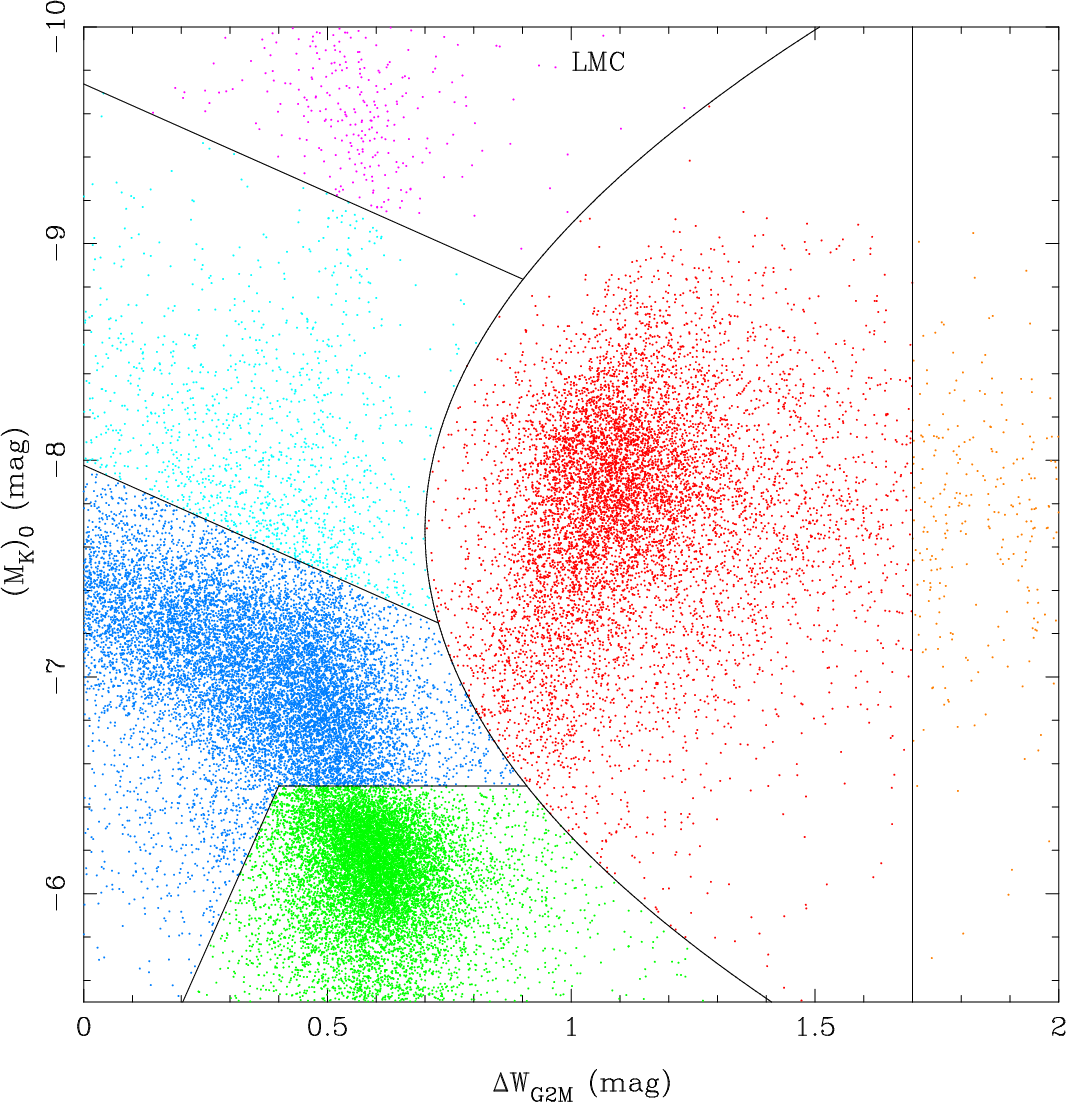}}
    \end{minipage}
    \begin{minipage}{0.45\textwidth}
       \resizebox{\hsize}{!}{\includegraphics{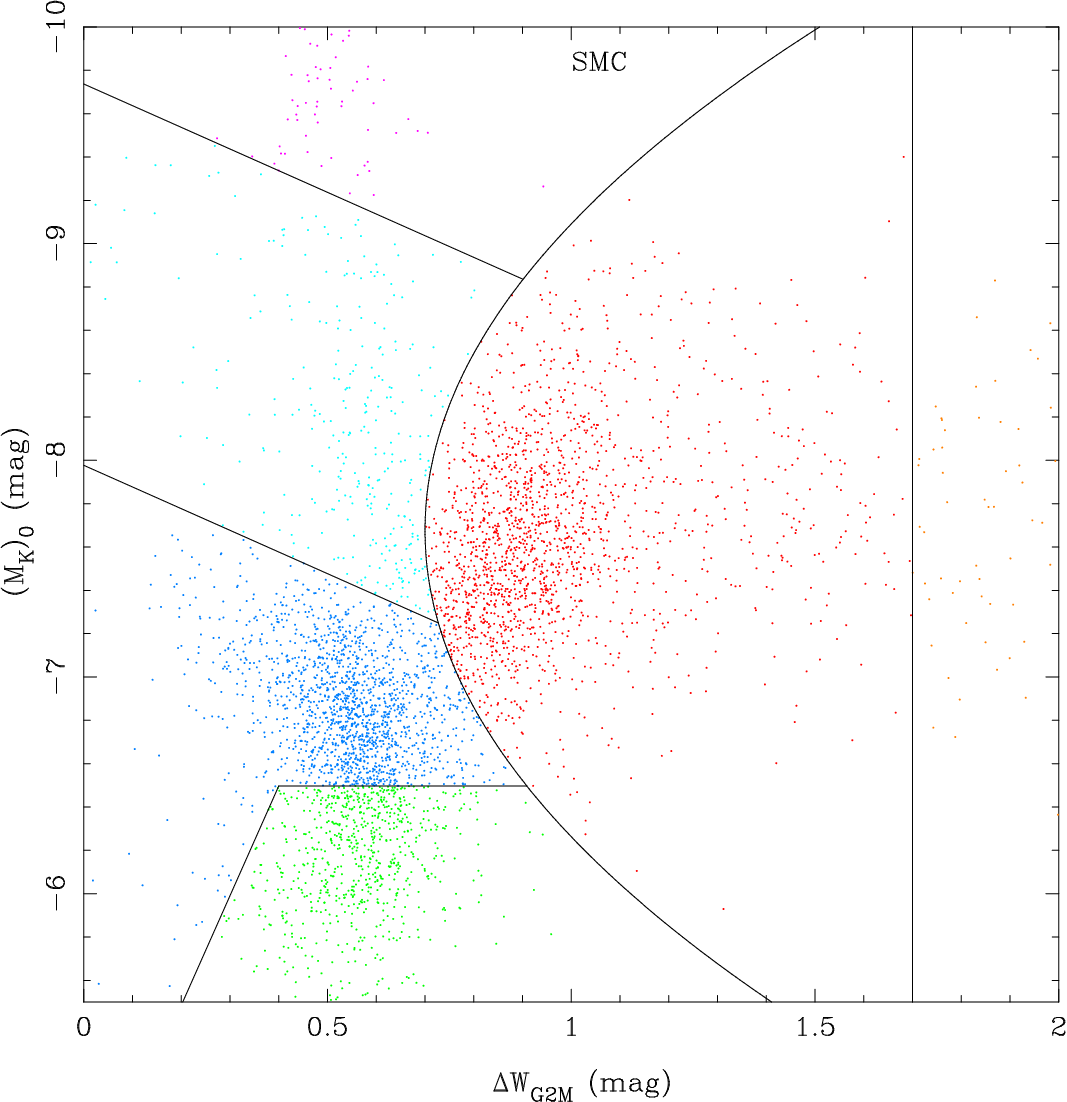}}
    \end{minipage}
        
    \caption{\G-2M diagram for the LMC and SMC (models 1 and 2). Boxes and colours indicate the various classes according
      to \citet{lebzelter2018new} and \citet{Mowlavi2019}.
    }
      
        \label{Fig:LebMC}
    \end{figure*}

\begin{figure*}
    \centering
    \begin{minipage}{0.45\textwidth}
       \resizebox{\hsize}{!}{\includegraphics{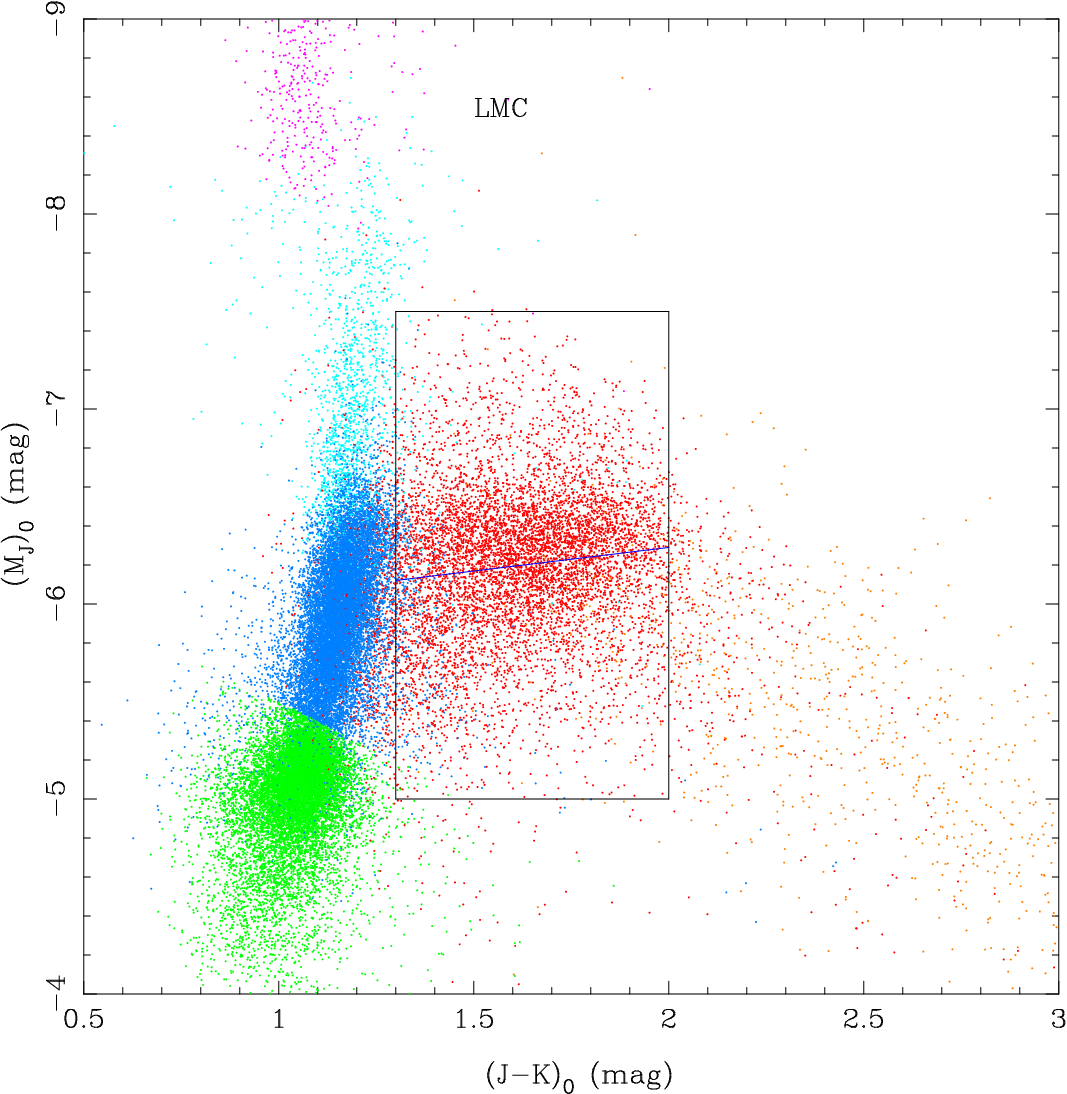}}
    \end{minipage}
    \begin{minipage}{0.45\textwidth}
       \resizebox{\hsize}{!}{\includegraphics{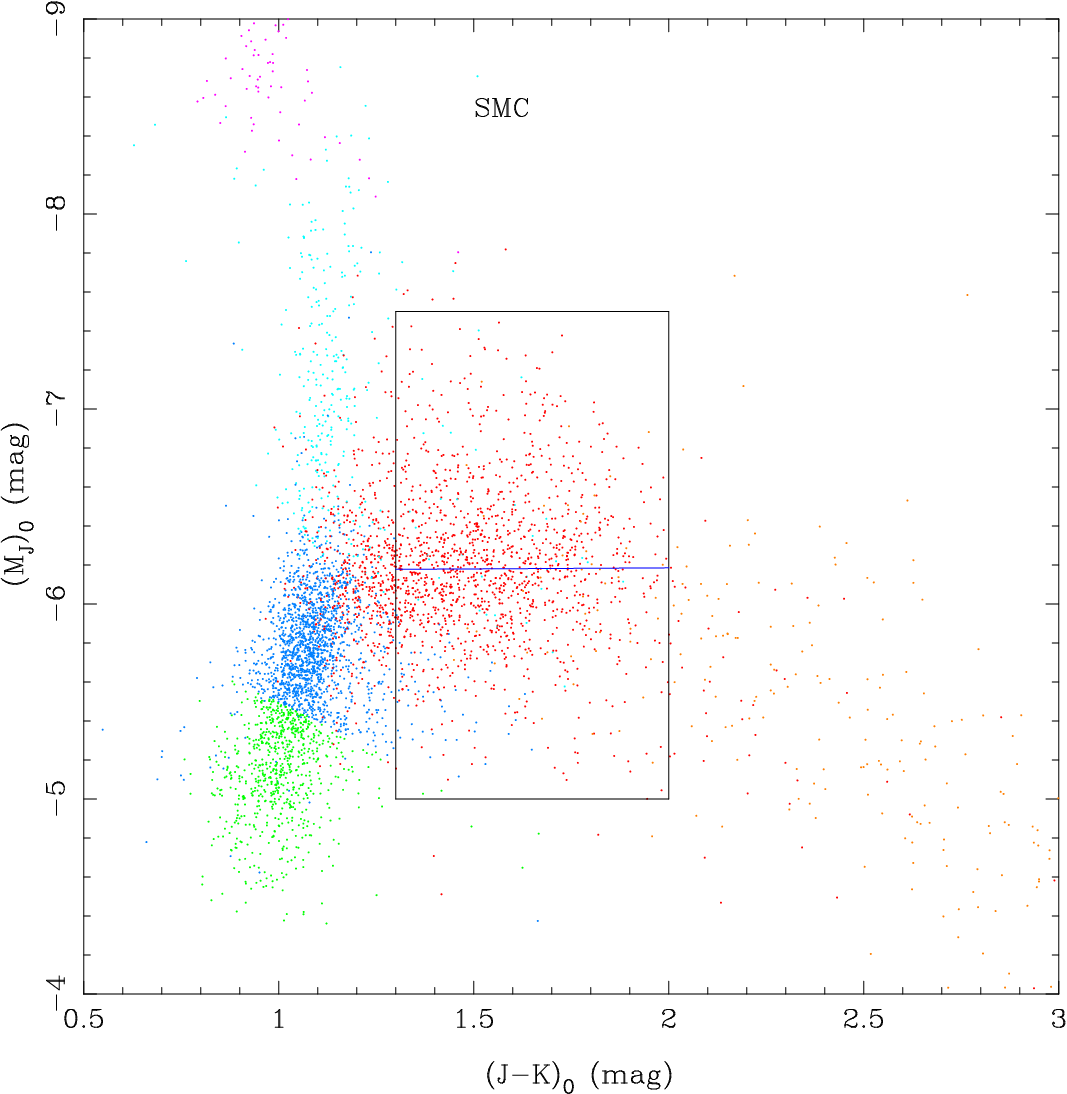}}
    \end{minipage}
        
    \caption{CMD for the LMC and SMC (models 1 and 2). Colours correspond to the classes in the  \G-2M diagram. The black box
      indicates which stars are selected to construct the $J$-band LF. The blue line indicates a linear fit to all stars inside this box.
    }
        \label{Fig:JJKMC}
    \end{figure*}

\begin{figure*}
    \centering
    \begin{minipage}{0.49\textwidth}
       \resizebox{\hsize}{!}{\includegraphics{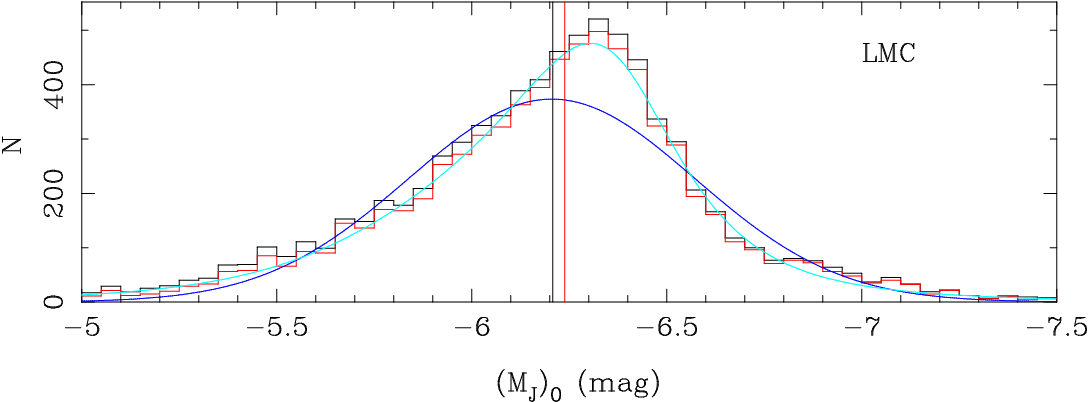}}
    \end{minipage}
    \begin{minipage}{0.49\textwidth}
       \resizebox{\hsize}{!}{\includegraphics{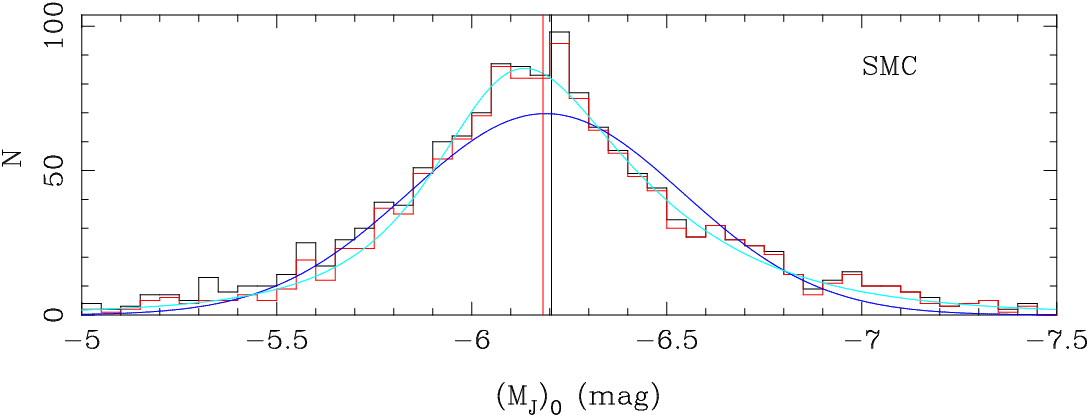}}
    \end{minipage}
        
    \caption{Fit to the $J$-band LF (models 1 and 2).
      The red and black histograms refer to the C stars and all stars, respectively.
      The bin width is 0.05~mag.
      The dark and light-blue line refer to the best fit Gaussian and Lorentzian profile to the C star LF, respectively.
      The vertical black and red line refer to the weighted mean and median $\bm{ (M_{\rm J})_0}$ magnitudes of
      the C stars in the selection box, respectively.
      }
        \label{Fig:FitMC}
    \end{figure*}

In a first step the results of a standard model are presented where stars are selected in a box $1.3 < (J-\Ks)_0 < 2.0$~mag and $-5.0 < (M_{\rm J})_0 < -7.5$~mag.
C stars are selected as those that are C stars according to the  \G-2M diagram AND based on the classification in the LPV2 catalogue.
No background terms are included in the fitting (Equations~\ref{Eq:G} and \ref{Eq:L}).
Figure~\ref{Fig:LebMC} shows the  \G-2M diagram, Figure~\ref{Fig:JJKMC} the $K_0- (J-\Ks)_0$ CMD, and Figure~\ref{Fig:FitMC} the fit to the
$J$-band LF of stars in the box for the LMC and SMC (models 1 and 2), respectively.
Results of the fitting are listed in Tables~\ref{Tab:Fit}-\ref{Tab:FitL}, which give general results, the results from
fitting the Gaussian model, and the results from fitting the Lorentzian model, respectively.
The latter two tables list the reduced $\chi^2$ ($\chi^2_{\rm r}$) and also the value of the 
Bayesian information criteria (BIC, \citealt{Schwarz1978}). This is a useful parameter to decide whether the lower $\chi^2$ expected for models with
more parameters is actually significant. 

Tables~\ref{Tab:Fit}-\ref{Tab:FitL} include several fitted models, starting with the standard model
(models 1 and 2 for for LMC and SMC, respectively)\footnote{Additional details on the parameters that were varied between the different models
    can be found in the footnotes to the tables which, for latex technical and clearer presentation, are given in the appendix,
    Tables~\ref{Tab:FitNotes}-\ref{Tab:FitLNotes}.}.
The first parameter to vary is the lower limit of the selection box (models 3-8). 
Visually, a smaller lower-limit would encompass the SMC C stars better, while a higher one is more suitable for the LMC.
With a lower limit of 1.2 in $(J-\Ks)_0$ the contamination of O-rich stars is 13\% for the SMC and 34\% for the LMC.
For  a lower limit of 1.5 in $(J-\Ks)_0$ this is 4\% and 1\%, respectively.
Another interesting parameter is the slope of $(M_{\rm J})_0$ versus $(J-\Ks)_0$ of the stars inside the selection box.
This slope increases (from negative to zero to positive) when increasing the lower limit of the selection box.
The CMD indicates why this is the case as most contaminants inside the selection box occur at fainter and bluer magnitudes than
the typical C star.
A non-zero slope indicates that the distribution of stars is not symmetric around $(M_{\rm J})_0$, as assumed in fitting
a Gaussian distribution.
Indeed, the fitted models indicate that a Gaussian distribution is a poor fit to the LMC C stars LF
(comparing models 1 with 13 for the LMC, and models 2 and 14 for the SMC).
The Lorentzian distribution clearly gives the better fit compared to a Gaussian distribution.
Adding the background terms marginally improves the $\chi^2_{\rm r}$ in the case of the Lorentzian distribution, but the
BIC values are larger indicating that the data does not require the addition of these terms.
In the case of the Gaussian distribution the BIC values are smaller when including the background terms.
  Adding the background terms could also influence the fitted values for the width of the Gaussian profile and the width,
  skewness and kurtosis of the Lorentizan profile. Models 23 and 24 for the LMC are similar to models 3 and 7 (with the lower limits of
  $(J-\Ks)_0$ of 1.2 and 1.5~mag, respectively), but including the background terms.
  The effect on the fitted parameters are largest for the bluest lower-limit on $(J-\Ks)_0$, where the contamination of M-stars is largest.
  The BIC values are lower when including the background terms, but the effect is marginal for the model fitting the Lorentzian profile
  which results in the lowest BIC value.

Purely in terms of fitting the LF the Lorentzian distribution is best, followed by
the Lorentzian plus background terms, the Gaussian distribution plus background terms, and the Gaussian distribution.

The consequence of adopting the Lorentzian distribution is the conclusion that the LMC and SMC LF are
different, as already concluded by \citet{Parada21,Parada23}. They use a lower limit of $(J-\Ks)_0 > 1.4$~mag (our models 5 and 6).
The parameter values are similar and the trends identical, namely that the mode and the skewness of the LMC and SMC distributions
are significantly different. In terms of calibration \citet{Parada21,Parada23} propose to take the LMC as calibrator if the skewness
of the distribution in the target galaxy is $< -0.2$, and otherwise the SMC.
Based on models 1-8 the results in the present paper put this value for the skewness at $-0.10$.
Alternatively, one could assume and adopt a linear relation between the skewness and the mode, and estimate the mode based on the skewness
of the LF in the target galaxy.

The conclusion in the papers by Madore, Freedman and coworkers is that the mean magnitude of the selected SMC and LMC stars is very similar, and can be averaged
to a single value that can be applied to any target galaxy.
The current results confirm this. Considering models 1-8, that vary the most important parameter, namely the lower limit to the $(J-\Ks)$ colour, and
considering all stars in the selection box, 
the weighted difference in the weighted mean magnitudes between LMC and SMC is $-0.025 \pm 0.007$~mag, and this is smaller (in absolute sense) than
the weighted difference in the median        magnitudes                     of $-0.066 \pm 0.005$~mag, or  
the weighted difference in the peaks of the Gaussian distribution           of $-0.031 \pm 0.007$~mag.
In this paper we introduced also another approach, namely to fit a linear relation between the dereddened absolute $J$ magnitude and the $((J-\Ks)_0 -1.6)$ colour
to all stars in the selection box.
The average difference in the ZPs between LMC and SMC is even smaller at $-0.016 \pm 0.006$~mag.

Averaging therefore the weighted mean magnitudes or the ZPs of the LMC and SMC (models 1-8) gives a weighted mean of $-6.212$~mag with a rms of 0.021~mag,
and a total error that is driven by the systematic error in the adopted DM to the LMC and SMC (cf. \citealt{freedman2020application}).

The parameters were fitted based on all stars in the selection box, with the underlying assumption that these are in majority C stars.
In typical applications the selection is purely a photometric one with no a-priori knowledge of the spectral type.
The average difference over models 1-8 for the weighted mean of all stars minus the weighted mean of the C stars is $+0.010$~mag with an error in the mean of
0.002~mag and a  rms of 0.012~mag, that is, C stars are on average slightly brighter than the average star in the selection box, as expected,
as the contaminants come mostly from fainter (and bluer) stars.
The way the C stars are selected (models 1-2, 17-22) plays a minor role. The rms in the weighted average is 0.010~mag.

\begin{table*}
	\centering
  \setlength{\tabcolsep}{1.4mm}
  \tiny
\caption{\label{Tab:Fit} General results}
\begin{tabular}{rcrrccccccccccccccllllllllllllllllll}
\hline\hline
M & Galaxy & $N_{\rm all}$ & $N_{\rm C}$ & slope & offset & mean$_{\rm all}$ & mean$_{\rm C}$ & median$_{\rm all}$ & median$_{\rm C}$   \\
  &        &             &            &        &  (mag.)  & (mag.) & (mag.) &  (mag.)    &  (mag.)  \\
\hline\hline
1 & LMC & 7827  & 7306  & -0.24 $\pm$ 0.06  & -6.1932 $\pm$ 0.0018  & -6.2154 $\pm$ 0.0003  & -6.2263 $\pm$ 0.0003  & -6.2310   & -6.2390 \\  
2 & SMC & 1412  & 1316  & -0.01 $\pm$ 0.18  & -6.1808 $\pm$ 0.0049  & -6.1995 $\pm$ 0.0008  & -6.2215 $\pm$ 0.0008  & -6.1690   & -6.1827 \\  
3 & LMC & 11878 & 7860  & -0.18 $\pm$ 0.03  & -6.2003 $\pm$ 0.0017  & -6.2019 $\pm$ 0.0002  & -6.2123 $\pm$ 0.0003  & -6.2112   & -6.2292 \\  
4 & SMC & 1815  & 1588  & -0.18 $\pm$ 0.12  & -6.1735 $\pm$ 0.0055  & -6.1735 $\pm$ 0.0007  & -6.2167 $\pm$ 0.0007  & -6.1410   & -6.1731 \\  
5 & LMC & 6532  & 6391  & -0.17 $\pm$ 0.09  & -6.2031 $\pm$ 0.0023  & -6.2314 $\pm$ 0.0003  & -6.2384 $\pm$ 0.0003  & -6.2423   & -6.2476 \\  
6 & SMC & 1110  & 1052  & +0.11 $\pm$ 0.27  & -6.1913 $\pm$ 0.0050  & -6.2078 $\pm$ 0.0009  & -6.2211 $\pm$ 0.0009  & -6.1823   & -6.1957 \\  
7 & LMC & 5244  & 5192  & -0.01 $\pm$ 0.13  & -6.2297 $\pm$ 0.0036  & -6.2469 $\pm$ 0.0003  & -6.2496 $\pm$ 0.0003  & -6.2554   & -6.2566 \\  
8 & SMC &  808  & \phantom{1}773  & +0.31 $\pm$ 0.46  & -6.2173 $\pm$ 0.0082  & -6.2113 $\pm$ 0.0010  & -6.2203 $\pm$ 0.0010  & -6.1827   & -6.1899 \\  
9 & LMC & 7182  & 6783  & -0.19 $\pm$ 0.04  & -6.2187 $\pm$ 0.0012  & -6.2328 $\pm$ 0.0003  & -6.2375 $\pm$ 0.0003  & -6.2420   & -6.2470 \\  
10 & SMC & 1291 & 1226  & -0.13 $\pm$ 0.11  & -6.1930 $\pm$ 0.0031  & -6.1984 $\pm$ 0.0008  & -6.2086 $\pm$ 0.0008  & -6.1731   & -6.1823 \\  
11 & LMC & 7948  & 7337 & -0.26 $\pm$ 0.07  & -6.1836 $\pm$ 0.0021  & -6.2107 $\pm$ 0.0003  & -6.2233 $\pm$ 0.0003  & -6.2287   & -6.2370 \\  
12 & SMC & 1429  & 1325 & +0.06 $\pm$ 0.20  & -6.1811 $\pm$ 0.0054  & -6.2052 $\pm$ 0.0008  & -6.2237 $\pm$ 0.0008  & -6.1690   & -6.1827 \\  
13 & LMC & 7827  & 7306 & -0.24 $\pm$ 0.06  & -6.1932 $\pm$ 0.0018  & -6.2154 $\pm$ 0.0003  & -6.2263 $\pm$ 0.0003  & -6.2310   & -6.2390 \\  
14 & SMC & 1412  & 1316 & -0.01 $\pm$ 0.18  & -6.1808 $\pm$ 0.0049  & -6.1995 $\pm$ 0.0008  & -6.2215 $\pm$ 0.0008  & -6.1690   & -6.1827 \\  
15 & LMC & 7948  & 7337 & -0.26 $\pm$ 0.07  & -6.1836 $\pm$ 0.0021  & -6.2107 $\pm$ 0.0003  & -6.2233 $\pm$ 0.0003  & -6.2287   & -6.2370 \\  
16 & SMC & 1429  & 1325 & +0.06 $\pm$ 0.20  & -6.1811 $\pm$ 0.0054  & -6.2052 $\pm$ 0.0008  & -6.2237 $\pm$ 0.0008  & -6.1690   & -6.1827 \\  
17 & LMC & 7827  & 7346 & -0.24 $\pm$ 0.06  & -6.1932 $\pm$ 0.0018  & -6.2154 $\pm$ 0.0003  & -6.2239 $\pm$ 0.0003  & -6.2310   & -6.2366 \\  
18 & SMC & 1412  & 1326 & -0.01 $\pm$ 0.18  & -6.1808 $\pm$ 0.0049  & -6.1995 $\pm$ 0.0008  & -6.2192 $\pm$ 0.0008  & -6.1690   & -6.1823 \\  
19 & LMC & 7827  & 7371 & -0.24 $\pm$ 0.06  & -6.1932 $\pm$ 0.0018  & -6.2154 $\pm$ 0.0003  & -6.2222 $\pm$ 0.0003  & -6.2310   & -6.2360 \\  
20 & SMC & 1412  & 1401 & -0.01 $\pm$ 0.18  & -6.1808 $\pm$ 0.0049  & -6.1995 $\pm$ 0.0008  & -6.2010 $\pm$ 0.0008  & -6.1690   & -6.1693 \\  
21 & LMC & 7827  & 7411 & -0.24 $\pm$ 0.06  & -6.1932 $\pm$ 0.0018  & -6.2154 $\pm$ 0.0003  & -6.2199 $\pm$ 0.0003  & -6.2310   & -6.2346 \\  
22 & SMC & 1412  & 1411 & -0.01 $\pm$ 0.18  & -6.1808 $\pm$ 0.0049  & -6.1995 $\pm$ 0.0008  & -6.1989 $\pm$ 0.0008  & -6.1690   & -6.1674 \\

23 & LMC & 11878  & 7860  & -0.18 $\pm$ 0.03  & -6.2003 $\pm$ 0.0017  & -6.2019 $\pm$ 0.0002  & -6.2123 $\pm$ 0.0003  & -6.2112   & -6.2292 \\  
24 & SMC &  5244  & 5192  & -0.01 $\pm$ 0.13  & -6.2297 $\pm$ 0.0036  & -6.2469 $\pm$ 0.0003  & -6.2496 $\pm$ 0.0003  & -6.2554   & -6.2566 \\  

& \\
101 & MW & 915  &  483  & -0.72 $\pm$ 0.15  & -5.6200 $\pm$ 0.0084  & -5.5392 $\pm$ 0.0051  & -5.6913 $\pm$ 0.0067  & -5.4606   & -5.7020 \\  
102 & MW & 461  &  404  & -0.24 $\pm$ 0.32  & -5.7300 $\pm$ 0.0094  & -5.7100 $\pm$ 0.0070  & -5.7443 $\pm$ 0.0074  & -5.6914   & -5.7385 \\  
&  \\
103 & MW & 536  &  442  & -0.21 $\pm$ 0.40  & -5.8078 $\pm$ 0.0108  & -5.8912 $\pm$ 0.0059  & -5.8731 $\pm$ 0.0066  & -5.7570   & -5.7747 \\  
104 & MW & 541  &  444  & -0.23 $\pm$ 0.40  & -5.8100 $\pm$ 0.0107  & -5.8964 $\pm$ 0.0058  & -5.8800 $\pm$ 0.0065  & -5.7628   & -5.7751 \\  
105 & MW & 548  &  445  & -0.20 $\pm$ 0.40  & -5.8071 $\pm$ 0.0107  & -5.8834 $\pm$ 0.0057  & -5.8790 $\pm$ 0.0065  & -5.7570   & -5.7749 \\  
106 & MW & 404  &  366  & +0.02 $\pm$ 0.73  & -5.8420 $\pm$ 0.0213  & -5.9139 $\pm$ 0.0067  & -5.9273 $\pm$ 0.0071  & -5.8103   & -5.8271 \\  
107 & MW & 353  &  321  & -0.03 $\pm$ 0.81  & -5.8058 $\pm$ 0.0224  & -5.8676 $\pm$ 0.0072  & -5.8907 $\pm$ 0.0077  & -5.7726   & -5.7922 \\  
108 & MW & 168  &  144  & -0.24 $\pm$ 0.97  & -5.7720 $\pm$ 0.0336  & -5.9017 $\pm$ 0.0094  & -5.9343 $\pm$ 0.0100  & -5.7751   & -5.8271 \\  
109 & MW & 307  &  282  & -0.12 $\pm$ 0.84  & -5.8240 $\pm$ 0.0226  & -5.8915 $\pm$ 0.0076  & -5.9103 $\pm$ 0.0080  & -5.8220   & -5.8422 \\ 
110 & MW & 159  &  146  & -0.00 $\pm$ 0.94  & -5.8475 $\pm$ 0.0322  & -5.9671 $\pm$ 0.0097  & -5.9788 $\pm$ 0.0102  & -5.8006   & -5.8120 \\  
111 & MW & 154  &  140  & -0.18 $\pm$ 1.06  & -5.8882 $\pm$ 0.0304  & -6.0245 $\pm$ 0.0098  & -6.0109 $\pm$ 0.0103  & -5.8913   & -5.9103 \\ 
112 & MW &  30  &   26  & +0.29 $\pm$ 4.02  & -6.2273 $\pm$ 0.1650  & -6.2242 $\pm$ 0.0143  & -6.2032 $\pm$ 0.0150  & -6.0862   & -6.1139 \\  
113 & MW & 132  &  122  & -0.16 $\pm$ 0.58  & -5.8225 $\pm$ 0.0145  & -5.9016 $\pm$ 0.0108  & -5.9104 $\pm$ 0.0112  & -5.8369   & -5.8765 \\
114 & MW & 132  &  122  & -0.16 $\pm$ 0.58  & -5.8225 $\pm$ 0.0145  & -5.9016 $\pm$ 0.0108  & -5.9104 $\pm$ 0.0112  & -5.8369   & -5.8765 \\

&  \\
25 & LMC & 4652  & 4640  & -0.08 $\pm$ 0.07  & -6.2353 $\pm$ 0.0018  & -6.2514 $\pm$ 0.0004  & -6.2520 $\pm$ 0.0004  & -6.2606   & -6.2617 \\
   &     &       &       & -0.05 $\pm$ 0.02  & -6.2386 $\pm$ 0.0040  & -6.2518 $\pm$ 0.0035  & -6.2525 $\pm$ 0.0037  & -6.2609 $\pm$ 0.0039 & -6.2616 $\pm$ 0.0040 \\
26 & SMC &  688  &  688  & -0.03 $\pm$ 0.22  & -6.1904 $\pm$ 0.0038  & -6.1995 $\pm$ 0.0011  & -6.1995 $\pm$ 0.0011  & -6.1830   & -6.1830  \\
   &     &       &       & +0.02 $\pm$ 0.04  & -6.1941 $\pm$ 0.0133  & -6.1992 $\pm$ 0.0132  & -6.1992 $\pm$ 0.0132  & -6.1863 $\pm$ 0.0147 & -6.1862  $\pm$ 0.0147 \\
115 & MW &  126  &  117  & -0.09 $\pm$ 0.50  & -5.8291 $\pm$ 0.0133  & -5.8875 $\pm$ 0.0111  & -5.8936 $\pm$ 0.0115  & -5.8369   & -5.8429 \\
    &    &       &       & -0.01 $\pm$ 0.19  & -5.847 $\pm$  0.022   & -5.897  $\pm$ 0.023   & -5.908  $\pm$ 0.021   & -5.853 $\pm$ 0.030 & -5.859 $\pm$ 0.035 \\
        \hline
\end{tabular}
\tablefoot{The notes to this table are given in Table~\ref{Tab:FitNotes}}
\end{table*}








\begin{sidewaystable*}
	\centering
\caption{\label{Tab:FitG} Results of the Gaussian fit.}
\small

\begin{tabular}{rccrccccccrcllllllllllllllll}
\hline
M  &  $\mu_{\rm all}$ & $\sigma_{\rm all}$ &  $\chi^2_{\rm r}$ & BIC & $\mu_{\rm C}$ & $\sigma_{\rm C}$  & $b_{\rm C}$ & $c_{\rm C}$ & $d_{\rm C}$ & $\chi^2_{\rm r}$   & BIC  \\
   &  (mag.) & (mag.) &                       &    &   (mag.) & (mag.) &               &  (mag.$^{-1}$) & (mag.$^{-2}$) &   \\
 \hline\hline

1 & -6.1952 $\pm$ 0.0045  &  0.3810 $\pm$ 0.0038  & 10.11 & 801.0  & -6.2053 $\pm$ 0.0045  &  0.3684 $\pm$ 0.0037  & 0.0 F & 0.0 F & 0.0 F & 8.84  & 734.2   \\ 
2 & -6.1692 $\pm$ 0.0107  &  0.3851 $\pm$ 0.0092  & 1.78  & 340.6  & -6.1898 $\pm$ 0.0101  &  0.3516 $\pm$ 0.0084  & 0.0 F & 0.0 F & 0.0 F & 1.86  & 335.7   \\ 
3 & -6.1719 $\pm$ 0.0038  &  0.3889 $\pm$ 0.0033  & 17.50 & 1173.8 & -6.1898 $\pm$ 0.0044  &  0.3788 $\pm$ 0.0037  & 0.0 F & 0.0 F & 0.0 F & 9.23  & 757.6   \\ 
4 & -6.1407 $\pm$ 0.0095  &  0.3867 $\pm$ 0.0083  & 2.68  & 382.4  & -6.1799 $\pm$ 0.0088  &  0.3343 $\pm$ 0.0073  & 0.0 F & 0.0 F & 0.0 F & 2.54  & 374.4   \\ 
5 & -6.2148 $\pm$ 0.0046  &  0.3554 $\pm$ 0.0039  & 9.04  & 738.0  & -6.2192 $\pm$ 0.0046  &  0.3505 $\pm$ 0.0039  & 0.0 F & 0.0 F & 0.0 F & 8.48  & 708.8   \\ 
6 & -6.1857 $\pm$ 0.0119  &  0.3786 $\pm$ 0.0101  & 1.52  & 316.1  & -6.1971 $\pm$ 0.0115  &  0.3580 $\pm$ 0.0095  & 0.0 F & 0.0 F & 0.0 F & 1.49  & 307.6   \\ 
7 & -6.2346 $\pm$ 0.0048  &  0.3319 $\pm$ 0.0041  & 7.24  & 639.4  & -6.2346 $\pm$ 0.0047  &  0.3302 $\pm$ 0.0040  & 0.0 F & 0.0 F & 0.0 F & 6.97  & 624.5   \\ 
8 & -6.1900 $\pm$ 0.0139  &  0.3765 $\pm$ 0.0119  & 1.30  & 290.5  & -6.2021 $\pm$ 0.0140  &  0.3666 $\pm$ 0.0119  & 0.0 F & 0.0 F & 0.0 F & 1.26  & 297.6   \\ 
9 & -6.2142 $\pm$ 0.0044  &  0.3313 $\pm$ 0.0041  & 5.21  & 585.8  & -6.2200 $\pm$ 0.0044  &  0.3251 $\pm$ 0.0040  & 0.0 F & 0.0 F & 0.0 F & 4.89  & 567.1   \\ 
10 & -6.1617 $\pm$ 0.0105 &  0.3343 $\pm$ 0.0100  & 1.45  & 323.4  & -6.1764 $\pm$ 0.0100  &  0.3179 $\pm$ 0.0092  & 0.0 F & 0.0 F & 0.0 F & 1.43  & 318.6   \\ 
11 & -6.1932 $\pm$ 0.0046 &  0.3874 $\pm$ 0.0038  & 12.02 & 866.7  & -6.2045 $\pm$ 0.0045  &  0.3725 $\pm$ 0.0037  & 0.0 F & 0.0 F & 0.0 F & 9.05  & 785.6   \\ 
12 & -6.1692 $\pm$ 0.0108 &  0.3932 $\pm$ 0.0092  & 1.94  & 380.0  & -6.1897 $\pm$ 0.0102  &  0.3560 $\pm$ 0.0084  & 0.0 F & 0.0 F & 0.0 F & 1.87  & 408.2   \\ 
13 & -6.2288 $\pm$ 0.0048 &  0.2629 $\pm$ 0.0058  & 4.88  & 552.4  & -6.2318 $\pm$ 0.0048  &  0.2637 $\pm$ 0.0057  & 61.3 $\pm$  4.1  & +1.3 $\pm$  0.9  & -37.5 $\pm$ 3.4  & 4.77  & 540.3   \\ 
14 & -6.1529 $\pm$ 0.0119 &  0.2515 $\pm$ 0.0148  & 0.68  & 298.7  & -6.1657 $\pm$ 0.0111  &  0.2488 $\pm$ 0.0134  & 12.3 $\pm$  1.8  & -0.8 $\pm$  0.4  & -8.0 $\pm$  1.5  & 0.87  & 297.8   \\ 
15 & -6.2077 $\pm$ 0.0046 &  0.3326 $\pm$ 0.0047  & 8.18  & 673.5  & -6.2187 $\pm$ 0.0046  &  0.3163 $\pm$ 0.0046  & 32.3 $\pm$  2.5  & +2.4 $\pm$  0.4  & -14.0 $\pm$ 1.2  & 6.53  & 658.8   \\ 
16 & -6.1610 $\pm$ 0.0110 &  0.3227 $\pm$ 0.0120  & 1.18  & 352.2  & -6.1783 $\pm$ 0.0104  &  0.3064 $\pm$ 0.0109  &  5.3 $\pm$  1.2  & -0.2 $\pm$  0.3  & -1.8 $\pm$  0.6  & 1.26  & 386.9   \\ 
17 & -6.1952 $\pm$ 0.0045 &  0.3810 $\pm$ 0.0038  & 10.11 & 801.0  & -6.2038 $\pm$ 0.0045  &  0.3703 $\pm$ 0.0038  & 0.0 F & 0.0 F & 0.0 F & 9.09  & 746.9   \\ 
18 & -6.1692 $\pm$ 0.0107 &  0.3851 $\pm$ 0.0092  & 1.78  & 340.5  & -6.1871 $\pm$ 0.0102  &  0.3561 $\pm$ 0.0086  & 0.0 F & 0.0 F & 0.0 F & 1.89  & 339.0   \\ 
19 & -6.1952 $\pm$ 0.0045 &  0.3809 $\pm$ 0.0038  & 10.11 & 801.0  & -6.2028 $\pm$ 0.0045  &  0.3730 $\pm$ 0.0038  & 0.0 F & 0.0 F & 0.0 F & 9.32  & 758.5   \\ 
20 & -6.1692 $\pm$ 0.0107 &  0.3851 $\pm$ 0.0092  & 1.78  & 340.5  & -6.1719 $\pm$ 0.0106  &  0.3811 $\pm$ 0.0090  & 0.0 F & 0.0 F & 0.0 F & 1.75  & 337.1   \\ 
21 & -6.1952 $\pm$ 0.0045 &  0.3809 $\pm$ 0.0038  & 10.11 & 801.0  & -6.2013 $\pm$ 0.0045  &  0.3748 $\pm$ 0.0038  & 0.0 F & 0.0 F & 0.0 F & 9.57  & 771.2   \\ 
22 & -6.1692 $\pm$ 0.0107 &  0.3851 $\pm$ 0.0092  & 1.78  & 340.5  & -6.1691 $\pm$ 0.0107  &  0.3857 $\pm$ 0.0092  & 0.0 F & 0.0 F & 0.0 F & 1.77  & 340.2   \\

23 & -6.2094 $\pm$ 0.0040 &  0.2591 $\pm$ 0.0049  & 6.00  & 627.2  & -6.2197 $\pm$ 0.0048  &  0.2738 $\pm$ 0.0059  & 63.6 $\pm$ 4.6 & 3.1 $\pm$ 1.0 & -37.2 $\pm$ 3.7  & 5.46  & 575.6 \\ 
24 & -6.2509 $\pm$ 0.0050 &  0.2448 $\pm$ 0.0056  & 2.80  & 434.4  & -6.2505 $\pm$ 0.0050  &  0.2432 $\pm$ 0.0055  & 41.1 $\pm$ 2.9 & -0.9 $\pm$ 0.7 & -26.3 $\pm$ 2.3 & 2.82  & 432.7  \\ 

& \\
101 & -5.9212 $\pm$ 0.2394  &  0.35 F  & 0.53  & 347.7  & -5.8777 $\pm$ 0.0575  &  0.35 F  &  3.4 $\pm$  1.1  &  6.3 $\pm$  0.9  &  4.1 $\pm$  1.6  & 0.44  & 341.3   \\ 
102 & -5.8804 $\pm$ 0.0868  &  0.35 F  & 0.66  & 335.4  & -5.8682 $\pm$ 0.0686  &  0.35 F  &  2.8 $\pm$  1.0  &  4.7 $\pm$  1.0  &  2.9 $\pm$  1.5  & 0.65  & 331.1   \\ 
103 & -5.9229 $\pm$ 0.0855  &  0.35 F  & 0.78  & 284.5  & -5.8895 $\pm$ 0.0637  &  0.35 F  &  3.4 $\pm$  1.0  &  4.8 $\pm$  1.0  &  2.6 $\pm$  1.4  & 0.71  & 285.6   \\ 
104 & -5.9594 $\pm$ 0.0820  &  0.35 F  & 0.72  & 282.0  & -5.9106 $\pm$ 0.0626  &  0.35 F  &  3.3 $\pm$  1.0  &  5.0 $\pm$  1.0  &  2.8 $\pm$  1.4  & 0.70  & 285.2   \\ 
105 & -5.9726 $\pm$ 0.0850  &  0.35 F  & 0.69  & 281.2  & -5.9109 $\pm$ 0.0625  &  0.35 F  &  3.3 $\pm$  1.0  &  5.0 $\pm$  1.0  &  2.8 $\pm$  1.4  & 0.71  & 285.4   \\ 
106 & -5.9671 $\pm$ 0.0712  &  0.35 F  & 0.57  & 262.4  & -5.9126 $\pm$ 0.0651  &  0.35 F  &  2.6 $\pm$  0.9  &  3.8 $\pm$  0.8  &  2.3 $\pm$  1.3  & 0.54  & 270.7   \\ 
107 & -6.0500 $\pm$ 0.3656  &  0.35 F  & 0.95  & 330.1  & -6.0500 $\pm$ 0.3529  &  0.35 F  &  6.0 $\pm$  1.3  &  5.6 $\pm$  0.7  &  0.5 $\pm$  1.7  & 1.05  & 342.1   \\ 
108 & -5.7897 $\pm$ 0.1101  &  0.35 F  & 0.50  & 344.7  & -5.7305 $\pm$ 0.0734  &  0.35 F  &  0.6 $\pm$  0.7  & -0.5 $\pm$  0.6  & -0.0 $\pm$  1.0  & 0.36  & 346.7   \\ 
109 & -5.9653 $\pm$ 0.0911  &  0.35 F  & 0.80  & 303.1  & -5.9865 $\pm$ 0.0747  &  0.35 F  &  1.7 $\pm$  1.1  &  2.8 $\pm$  0.7  &  2.4 $\pm$  1.5  & 0.86  & 301.6   \\ 
110 & -5.6715 $\pm$ 0.0687  &  0.35 F  & 0.55  & 329.8  & -5.7084 $\pm$ 0.0722  &  0.35 F  &  0.2 $\pm$  0.8  & -1.2 $\pm$  0.8  &  0.5 $\pm$  1.3  & 0.46  & 365.2   \\ 
111 & -5.7274 $\pm$ 0.0863  &  0.35 F  & 0.47  & 287.1  & -5.8741 $\pm$ 0.0909  &  0.35 F  &  0.3 $\pm$  1.0  & -0.4 $\pm$  0.7  &  1.3 $\pm$  1.6  & 0.64  & 332.2   \\
112 & -5.9500 $\pm$ 0.3330  &  0.35 F  & 0.10  & 531.7  & -5.9708 $\pm$ 0.2358  &  0.35 F  &  0.3 $\pm$  1.2  & -0.6 $\pm$  0.7  &  1.1 $\pm$  1.7  & 0.05  & 556.1   \\ 
113 & -5.8459 $\pm$ 0.0486  &  0.35 F  & 0.91  & 223.5  & -5.8863 $\pm$ 0.0490  &  0.35 F  &  0.0 F           &  0.0 F           &  0.0 F           & 0.85  & 244.7   \\ 
114 & -5.8283 $\pm$ 0.0998  &  0.60 F  & 0.82  & 218.9  & -5.8722 $\pm$ 0.1053  &  0.60 F  &  0.0 F           &  0.0 F           &  0.0 F           & 0.79  & 241.5  \\

 & \\
25 & -6.2468 $\pm$ 0.0046  &  0.2794 $\pm$ 0.0044  & 2.76  & 450.6  & -6.2476 $\pm$ 0.0046  & 0.2782 $\pm$ 0.0044  &  0.0 F &  0.0 F &  0.0 F & 2.77  & 451.0   \\
   & -6.2454 $\pm$ 0.0045  &  0.2847 $\pm$ 0.0026  &       &        & -6.2462 $\pm$ 0.0048  & 0.2837 $\pm$ 0.0026  &            &                 &        &    &    & \\
26 & -6.1788 $\pm$ 0.0142  &  0.3128 $\pm$ 0.0149  & 0.69  & 260.3  & -6.1788 $\pm$ 0.0142  & 0.3128 $\pm$ 0.0149  &  0.0 F &  0.0 F &  0.0 F & 0.69  & 260.3   \\
   & -6.1761 $\pm$ 0.0195  &  0.3134 $\pm$ 0.0149  &       &        & -6.1761 $\pm$ 0.0191  & 0.3135 $\pm$ 0.0149  &            &                 &        &    &    & \\
115 & -5.8116 $\pm$ 0.0520  &  0.35 F & 0.70  & 228.1  & -5.8793 $\pm$ 0.0565  &  0.35 F    &  0.0 F          &  0.0 F  &  0.0 F   & 0.78  & 227.3   \\
    & -5.838  $\pm$ 0.041 &           &       &        & -5.854  $\pm$ 0.038   &            &                 &        &    &    & \\
\hline
\end{tabular}
\tablefoot{The notes to this table can be found in Table~\ref{Tab:FitGNotes}}
\end{sidewaystable*}

\begin{sidewaystable*}
	\centering
\caption{\label{Tab:FitL} Results of the Lorentzian fit.}

\small
\setlength{\tabcolsep}{1.6mm}
\begin{tabular}{cccccccccccllllllllllllllllll}
\hline \hline
M &  $\mu_{\rm all}$ & $w_{\rm all}$ &  $s_{\rm all}$ & $k_{\rm all}$ &  $\chi^2_{\rm r}$ & BIC & $\mu_{\rm C}$ & $w_{\rm C}$ & $s_{\rm C}$ & $k_{\rm C}$ & $\chi^2_{\rm r}$ & BIC  \\
   &   (mag.) &  (mag.) &              &             &                 &     &   (mag.) &  (mag.) &        &            &             &   \\
\hline
1 & -6.2987 $\pm$ 0.0070  &  0.308 $\pm$ 0.009  & -0.41 $\pm$ 0.04  &  0.12 $\pm$ 0.02  & 1.80  & 414.6  & -6.3046 $\pm$ 0.0075  &  0.309 $\pm$ 0.010  & -0.46 $\pm$ 0.05  &  0.17 $\pm$ 0.02  & 1.99  & 416.4  \\ 
2 & -6.1452 $\pm$ 0.0163  &  0.331 $\pm$ 0.025  & +0.10 $\pm$ 0.07  &  0.08 $\pm$ 0.04  & 0.59  & 291.4  & -6.1357 $\pm$ 0.0173  &  0.316 $\pm$ 0.024  & +0.28 $\pm$ 0.10  &  0.14 $\pm$ 0.06  & 0.73  & 288.9  \\ 
3 & -6.2680 $\pm$ 0.0058  &  0.316 $\pm$ 0.007  & -0.32 $\pm$ 0.03  &  0.09 $\pm$ 0.01  & 2.93  & 491.4  & -6.3034 $\pm$ 0.0074  &  0.313 $\pm$ 0.010  & -0.50 $\pm$ 0.04  &  0.17 $\pm$ 0.02  & 2.00  & 421.6  \\ 
4 & -6.1369 $\pm$ 0.0139  &  0.326 $\pm$ 0.021  & +0.01 $\pm$ 0.05  &  0.07 $\pm$ 0.04  & 0.99  & 308.9  & -6.1204 $\pm$ 0.0145  &  0.291 $\pm$ 0.019  & +0.30 $\pm$ 0.08  &  0.12 $\pm$ 0.04  & 0.97  & 306.7  \\ 
5 & -6.3019 $\pm$ 0.0074  &  0.297 $\pm$ 0.010  & -0.39 $\pm$ 0.04  &  0.13 $\pm$ 0.02  & 1.93  & 408.1  & -6.3043 $\pm$ 0.0078  &  0.301 $\pm$ 0.011  & -0.43 $\pm$ 0.05  &  0.17 $\pm$ 0.03  & 2.10  & 412.3  \\ 
6 & -6.1651 $\pm$ 0.0189  &  0.336 $\pm$ 0.030  & +0.08 $\pm$ 0.08  &  0.09 $\pm$ 0.06  & 0.64  & 281.3  & -6.1569 $\pm$ 0.0210  &  0.340 $\pm$ 0.033  & +0.22 $\pm$ 0.13  &  0.17 $\pm$ 0.10  & 0.77  & 280.1  \\ 
7 & -6.2992 $\pm$ 0.0080  &  0.290 $\pm$ 0.011  & -0.33 $\pm$ 0.05  &  0.14 $\pm$ 0.03  & 1.58  & 377.9  & -6.3018 $\pm$ 0.0083  &  0.293 $\pm$ 0.012  & -0.37 $\pm$ 0.06  &  0.17 $\pm$ 0.03  & 1.65  & 378.8  \\ 
8 & -6.1697 $\pm$ 0.0222  &  0.334 $\pm$ 0.035  & +0.07 $\pm$ 0.10  &  0.10 $\pm$ 0.07  & 0.68  & 267.4  & -6.1639 $\pm$ 0.0227  &  0.328 $\pm$ 0.034  & +0.17 $\pm$ 0.11  &  0.11 $\pm$ 0.07  & 0.62  & 273.8  \\ 
9 & -6.3009 $\pm$ 0.0077  &  0.299 $\pm$ 0.010  & -0.41 $\pm$ 0.06  &  0.11 $\pm$ 0.03  & 1.29  & 406.7  & -6.3048 $\pm$ 0.0079  &  0.296 $\pm$ 0.011  & -0.42 $\pm$ 0.06  &  0.13 $\pm$ 0.03  & 1.22  & 400.2  \\ 
10 & -6.1536 $\pm$ 0.0173 &  0.312 $\pm$ 0.030  & -0.00 $\pm$ 0.08  &  0.05 $\pm$ 0.07  & 1.08  & 311.7  & -6.1440 $\pm$ 0.0189  &  0.312 $\pm$ 0.031  & +0.16 $\pm$ 0.13  &  0.12 $\pm$ 0.10  & 1.07  & 307.1  \\ 
11 & -6.3060 $\pm$ 0.0067 &  0.303 $\pm$ 0.009  & -0.46 $\pm$ 0.03  &  0.14 $\pm$ 0.02  & 2.03  & 401.1  & -6.3231 $\pm$ 0.0074  &  0.303 $\pm$ 0.011  & -0.64 $\pm$ 0.04  &  0.22 $\pm$ 0.03  & 3.46  & 523.5  \\ 
12 & -6.1401 $\pm$ 0.0180 &  0.354 $\pm$ 0.027  & +0.14 $\pm$ 0.09  &  0.15 $\pm$ 0.07  & 0.66  & 326.4  & -6.1260 $\pm$ 0.0180  &  0.323 $\pm$ 0.025  & +0.36 $\pm$ 0.11  &  0.18 $\pm$ 0.07  & 0.70  & 359.2  \\ 
13 & -6.2851 $\pm$ 0.0088 &  0.304 $\pm$ 0.013  & -0.30 $\pm$ 0.05  &  0.07 $\pm$ 0.03  & 2.17  & 520.6  & -6.2948 $\pm$ 0.0087  &  0.291 $\pm$ 0.012  & -0.34 $\pm$ 0.06  &  0.09 $\pm$ 0.03  & 1.85  & 417.3  \\ 
14 & -6.1497 $\pm$ 0.0646 &  0.316 $\pm$ 0.650  & +0.05 $\pm$ 1.19  &  0.00 $\pm$ 4.11  & 0.59  & 303.6  & -6.1457 $\pm$ 0.0236  &  0.313 $\pm$ 0.061  & +0.17 $\pm$ 0.09  &  0.02 $\pm$ 0.21  & 0.74  & 325.6  \\ 
15 & -6.2958 $\pm$ 0.0079 &  0.312 $\pm$ 0.010  & -0.39 $\pm$ 0.05  &  0.11 $\pm$ 0.02  & 2.05  & 407.8  & -6.3003 $\pm$ 0.0089  &  0.315 $\pm$ 0.011  & -0.43 $\pm$ 0.07  &  0.16 $\pm$ 0.04  & 2.08  & 466.4  \\ 
16 & -6.1527 $\pm$ 0.0170 &  0.347 $\pm$ 0.024  & +0.04 $\pm$ 0.08  &  0.04 $\pm$ 0.17  & 1.47  & 931.2  & -6.1398 $\pm$ 0.0182  &  0.325 $\pm$ 0.022  & +0.19 $\pm$ 0.10  &  0.05 $\pm$ 0.06  & 0.64  & 365.7  \\ 
17 & -6.2987 $\pm$ 0.0070 &  0.308 $\pm$ 0.009  & -0.41 $\pm$ 0.04  &  0.12 $\pm$ 0.02  & 1.80  & 414.6  & -6.3038 $\pm$ 0.0074  &  0.307 $\pm$ 0.010  & -0.45 $\pm$ 0.04  &  0.15 $\pm$ 0.02  & 1.93  & 414.3  \\ 
18 & -6.1452 $\pm$ 0.0163 &  0.331 $\pm$ 0.025  & +0.10 $\pm$ 0.07  &  0.08 $\pm$ 0.04  & 0.59  & 291.2  & -6.1390 $\pm$ 0.0166  &  0.312 $\pm$ 0.023  & +0.22 $\pm$ 0.09  &  0.10 $\pm$ 0.05  & 0.65  & 287.1  \\ 
19 & -6.2987 $\pm$ 0.0070 &  0.308 $\pm$ 0.009  & -0.41 $\pm$ 0.04  &  0.12 $\pm$ 0.02  & 1.80  & 414.6  & -6.3038 $\pm$ 0.0072  &  0.305 $\pm$ 0.009  & -0.43 $\pm$ 0.04  &  0.14 $\pm$ 0.02  & 1.83  & 410.8  \\ 
20 & -6.1452 $\pm$ 0.0163 &  0.331 $\pm$ 0.025  & +0.10 $\pm$ 0.07  &  0.08 $\pm$ 0.04  & 0.59  & 291.2  & -6.1425 $\pm$ 0.0174  &  0.340 $\pm$ 0.027  & +0.14 $\pm$ 0.09  &  0.13 $\pm$ 0.06  & 0.75  & 296.5  \\ 
21 & -6.2987 $\pm$ 0.0070 &  0.308 $\pm$ 0.009  & -0.41 $\pm$ 0.04  &  0.12 $\pm$ 0.02  & 1.80  & 414.6  & -6.3031 $\pm$ 0.0071  &  0.302 $\pm$ 0.009  & -0.42 $\pm$ 0.04  &  0.13 $\pm$ 0.02  & 1.79  & 409.5  \\ 
22 & -6.1452 $\pm$ 0.0163 &  0.331 $\pm$ 0.025  & +0.10 $\pm$ 0.07  &  0.08 $\pm$ 0.04  & 0.59  & 291.2  & -6.1455 $\pm$ 0.0163  &  0.331 $\pm$ 0.025  & +0.10 $\pm$ 0.07  &  0.08 $\pm$ 0.05  & 0.59  & 291.1  \\

23 & -6.2477 $\pm$ 0.0064 &  0.331 $\pm$ 0.023  & -0.23 $\pm$ 0.14  &  0.02 $\pm$ 0.03 & 13.01  & 576288  & -6.2660 $\pm$ 0.0063  &  0.325 $\pm$ 0.015  & -0.30 $\pm$ 0.11  &  0.03 $\pm$ 0.01  & 2.65  & 561.7  \\ 
24 & -6.2866 $\pm$ 0.0098 &  0.287 $\pm$ 0.013  & -0.20 $\pm$ 0.09  &  0.10 $\pm$ 0.07  & 2.54  & 433.4 & -6.3200 $\pm$ 0.0107  &  0.263 $\pm$ 0.015  & -0.58 $\pm$ 0.12  &  0.24 $\pm$ 0.07  & 1.32  &  371.7  \\

& \\
101 & -5.8231 $\pm$ 0.1816  &  0.32 F  & +0.0 F  &  0.0 F  & 0.52  & 347.7  & -5.8405 $\pm$ 0.0460  &  0.32 F   & +0.0 F  &  0.0 F   & 0.48  & 343.1  \\ 
102 & -5.8333 $\pm$ 0.0673  &  0.32 F  & +0.0 F  &  0.0 F  & 0.70  & 337.4  & -5.8278 $\pm$ 0.0524  &  0.32 F   & +0.0 F  &  0.0 F   & 0.70  & 333.3  \\ 
103 & -5.8556 $\pm$ 0.0685  &  0.32 F  & +0.0 F  &  0.0 F  & 0.82  & 286.3  & -5.8489 $\pm$ 0.0501  &  0.32 F   & +0.0 F  &  0.0 F   & 0.78  & 288.4  \\ 
104 & -5.8859 $\pm$ 0.0686  &  0.32 F  & +0.0 F  &  0.0 F  & 0.79  & 285.0  & -5.8647 $\pm$ 0.0503  &  0.32 F   & +0.0 F  &  0.0 F   & 0.79  & 289.2  \\ 
105 & -5.8898 $\pm$ 0.0717  &  0.32 F  & +0.0 F  &  0.0 F  & 0.75  & 284.0  & -5.8665 $\pm$ 0.0500  &  0.32 F   & +0.0 F  &  0.0 F   & 0.79  & 289.2  \\ 
106 & -5.9036 $\pm$ 0.0575  &  0.32 F  & +0.0 F  &  0.0 F  & 0.63  & 264.9  & -5.8799 $\pm$ 0.0515  &  0.32 F   & +0.0 F  &  0.0 F   & 0.60  & 273.5  \\ 
107 & -5.9302 $\pm$ 0.3993  &  0.32 F  & +0.0 F  &  0.0 F  & 0.91  & 340.4  & -5.9302 $\pm$ 0.3892  &  0.32 F   & +0.0 F  &  0.0 F   & 1.08  & 351.6  \\ 
108 & -5.8116 $\pm$ 0.0686  &  0.32 F  & +0.0 F  &  0.0 F  & 0.44  & 342.1  & -5.7635 $\pm$ 0.0513  &  0.32 F   & +0.0 F  &  0.0 F   & 0.34  & 345.6  \\ 
109 & -5.9500 $\pm$ 0.0654  &  0.32 F  & +0.0 F  &  0.0 F  & 0.74  & 300.6  & -5.9666 $\pm$ 0.0583  &  0.32 F   & +0.0 F  &  0.0 F   & 0.83  & 299.9  \\
110 & -5.6921 $\pm$ 0.0529  &  0.32 F  & +0.0 F  &  0.0 F  & 0.61  & 332.4  & -5.7268 $\pm$ 0.0549  &  0.32 F   & +0.0 F  &  0.0 F   & 0.50  & 366.7  \\ 
111 & -5.7371 $\pm$ 0.0658  &  0.32 F  & +0.0 F  &  0.0 F  & 0.50  & 288.8  & -5.8621 $\pm$ 0.0712  &  0.32 F   & +0.0 F  &  0.0 F   & 0.65  & 332.6  \\ 
112 & -6.0308 $\pm$ 0.6983  &  0.32 F  & +0.0 F  &  0.0 F  & 0.11  & 532.8  & -6.0308 $\pm$ 0.6983  &  0.32 F   & +0.0 F  &  0.0 F   & 0.07  & 557.1  \\ 
113 & -5.8577 $\pm$ 0.0471  &  0.32 F  & +0.0 F  &  0.0 F  & 0.97  & 226.5  & -5.8861 $\pm$ 0.0460  &  0.32 F   & +0.0 F  &  0.0 F   & 0.88  & 245.9  \\ 
114 & -5.8322 $\pm$ 0.0850  &  0.60 F  & +0.0 F  &  0.0 F  & 0.81  & 218.7  & -5.8817 $\pm$ 0.0872  &  0.60 F   & +0.0 F  &  0.0 F   & 0.78  & 240.9  \\

& \\
 25 & -6.3128 $\pm$ 0.0089  &  0.270 $\pm$ 0.013  & -0.41 $\pm$ 0.08  &  0.13 $\pm$ 0.05  & 1.08  & 377.3  & -6.3132 $\pm$ 0.0089  &  0.269 $\pm$ 0.013  & -0.41 $\pm$ 0.08  &  0.13 $\pm$ 0.05  & 1.08  & 377.3  \\
    & -6.3104 $\pm$ 0.0065  &  0.281 $\pm$ 0.008  & -0.42 $\pm$ 0.06  &  0.14 $\pm$ 0.04  &       &        & -6.3107 $\pm$ 0.0064  &  0.281 $\pm$ 0.007  & -0.42 $\pm$ 0.06  &  0.15 $\pm$ 0.04  & &  \\
 26 & -6.1613 $\pm$ 0.0243  &  0.309 $\pm$ 0.046  & +0.10 $\pm$ 0.14  &  0.02 $\pm$ 0.12  & 0.61  & 263.0  & -6.1613 $\pm$ 0.0243  &  0.309 $\pm$ 0.046  & +0.10 $\pm$ 0.14  &  0.02 $\pm$ 0.12  & 0.61  & 263.0  \\
    & -6.1732 $\pm$ 0.0191  &  0.311 $\pm$ 0.034  & +0.04 $\pm$ 0.11  &  0.03 $\pm$ 0.10  &       &        & -6.1733 $\pm$ 0.0194  &  0.311 $\pm$ 0.034  & +0.04 $\pm$ 0.11  &  0.03 $\pm$ 0.10  & & \\
 115 & -5.7722 $\pm$ 0.0460  &  0.32 F  & 0.0 F   &  0.0 F  & 0.75  & 230.7  & -5.8859 $\pm$ 0.0509  &  0.32 F  & +0.0 F   &  0.0 F   & 0.85  & 231.0  \\
     &  -5.830 $\pm$ 0.086  &    & & &    &                                 & -5.839  $\pm$ 0.085  & & & &  \\

\hline
\end{tabular}
\tablefoot{The notes to this table can be found in Table~\ref{Tab:FitLNotes}}
\end{sidewaystable*}

\subsection{MW}

\begin{figure}
    \centering
    \begin{minipage}{0.45\textwidth}
       \resizebox{\hsize}{!}{\includegraphics{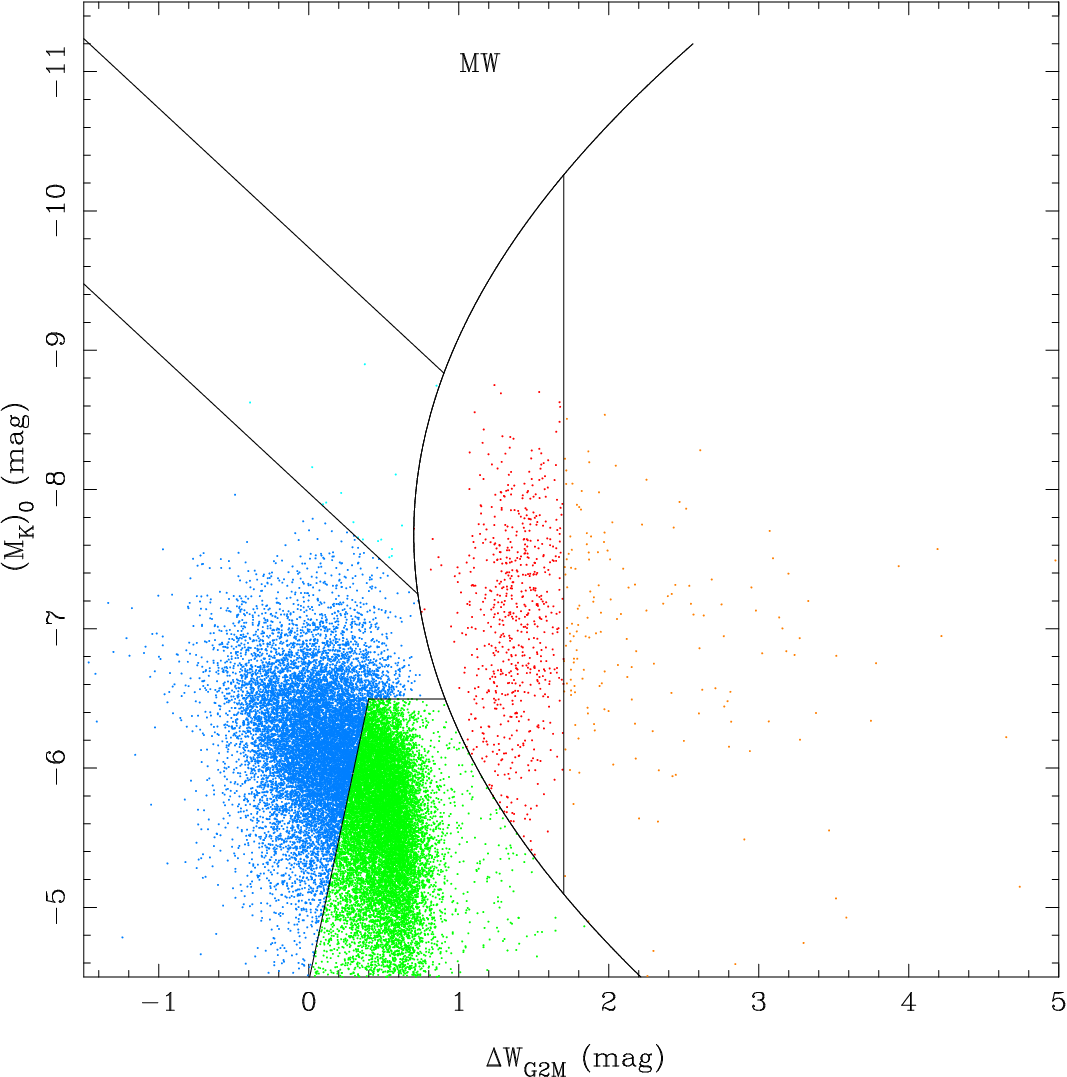}}
    \end{minipage}
    
    \begin{minipage}{0.45\textwidth}
       \resizebox{\hsize}{!}{\includegraphics{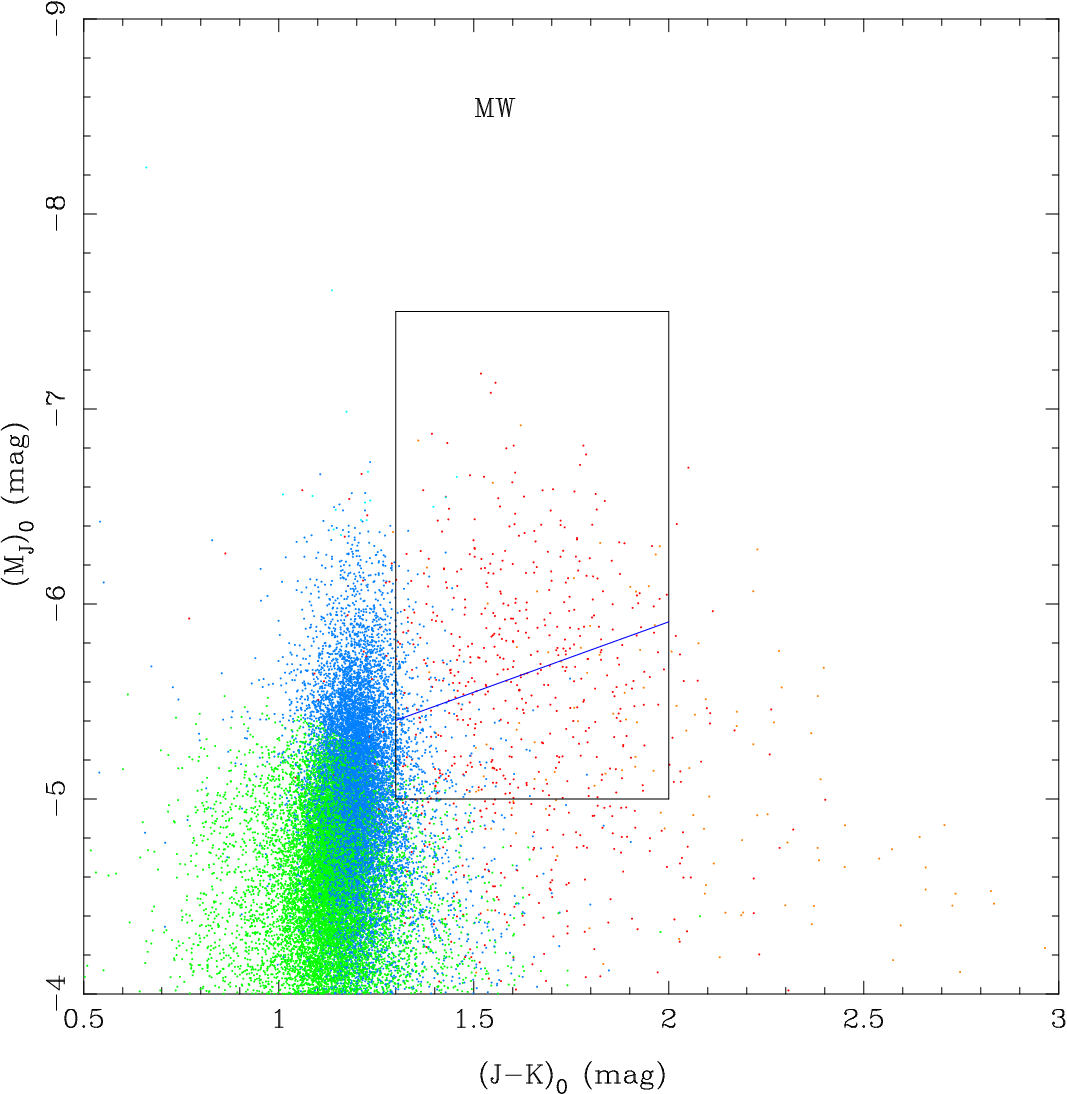}}
    \end{minipage}
        
    \begin{minipage}{0.46\textwidth}
       \resizebox{\hsize}{!}{\includegraphics{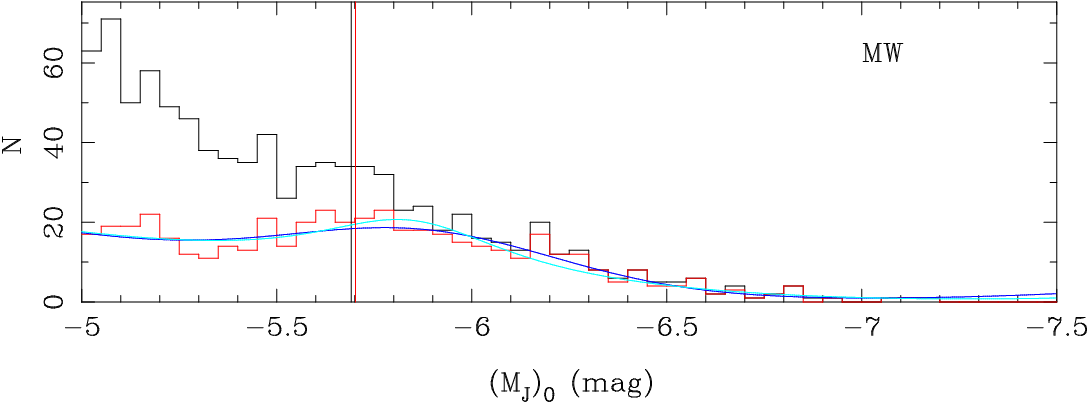}}
    \end{minipage}
    
    \begin{minipage}{0.46\textwidth}
       \resizebox{\hsize}{!}{\includegraphics{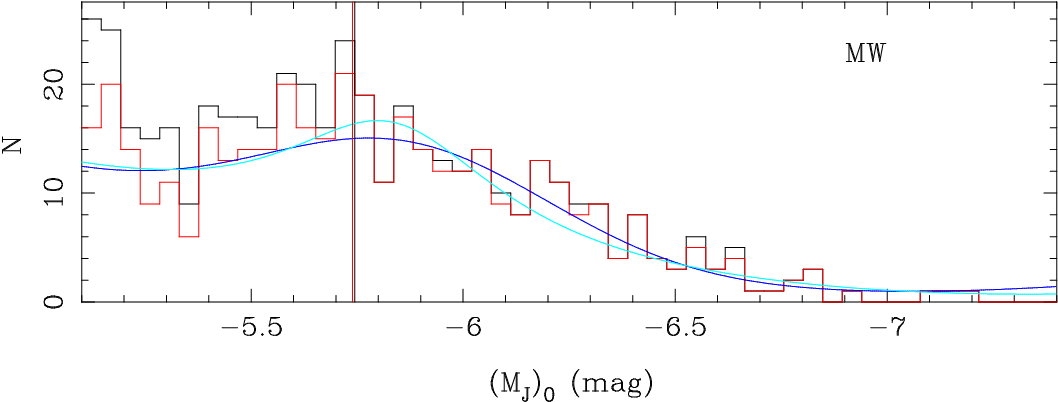}}
    \end{minipage}
        
    \caption{Top three panels: the  \G-2M diagram, the CMD, and the LF for model 101.
      Bottom panel: the LF for model 102.
      }
        \label{Fig:MW101}
    \end{figure}

The results for the MW turn out to be less clear as for the MCs and less straightforward to interpret.tab:f
model 101 in the tables used the same selection box as the standard model, but background terms have been included, and some
terms in the fitting have been fixed to achieve convergence (the width of the Gaussian and the Lorentzian distribution are set to
the average found for LMC and SMC, and the skewness and kurtosis are set to zero).
Specific to the MW models is that absolute magnitudes are calculated from individual distances with errors and that 
an error in the DM $\sigma_{\rm DM} < 0.2$~mag was adopted.
The \G-2M, a  CMD, and the fit to the LF are shown in Fig.~\ref{Fig:MW101}.

Several points are immediately clear.
The number of stars in the selection box is small, smaller even than for the SMC, and the contamination by O-rich stars is close to 50\%.
This high level of contamination is also clear from the CMD and from the LF. The mean, median and the peak of the Gaussian and Lorentzian distribution
differ significantly from that found for the MCs.
The  CMD and \G-2M diagram are also qualitatively different from that in the MCs with fewer bright stars present.
Increasing the lower limit in the $(J-\Ks)$ and $M_{\rm J}$ colour of the selection box reduces the level of contamination (model 102, bottom panel in
Fig.~\ref{Fig:MW101}), but the magnitudes do not really change.

\begin{figure}
    \centering
    \begin{minipage}{0.45\textwidth}
       \resizebox{\hsize}{!}{\includegraphics{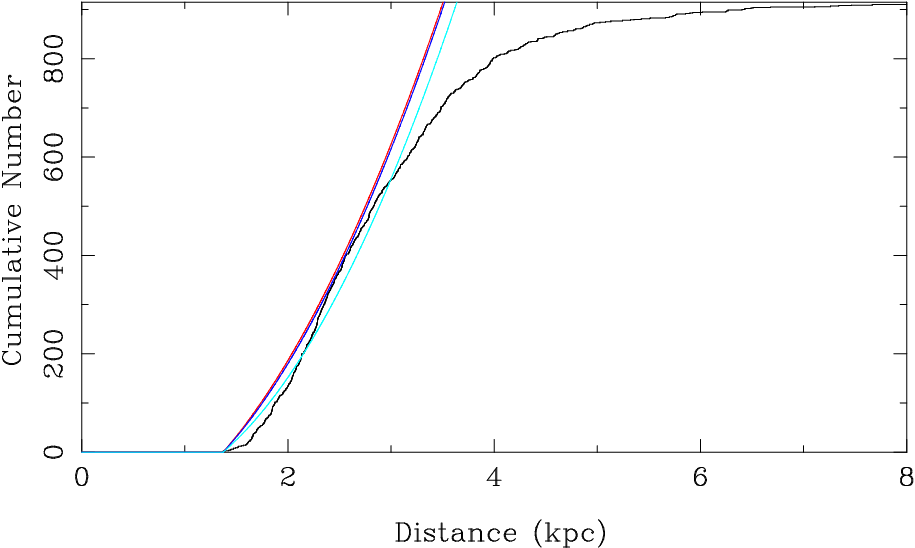}}
    \end{minipage}
        
    \caption{Cumulative number of stars in the selection box.
      Coloured lines indicate model predictions (\citealt{Gr1992}, Eq.~20) for a volume density ($\rho_0$) of
      70~/kpc$^3$ and scale height $H= 200$~pc (red),
      $\rho_0= 28$/kpc$^3$ and $H= 500$~pc (dark blue), and
      $\rho_0= 14$/kpc$^3$ and $H= 1000$~pc (light blue).
      The expected number of stars inside about 1.4~kpc is subtracted  (158, 132, and 95 stars, respectively).
    }
      \label{Fig:DistrSTD}
    \end{figure}

Figure~\ref{Fig:DistrSTD} shows the cumulative number of all stars as a function of distance.
It turns out that the closest star in the selection box is at about 1.4~kpc (and the closest C star is at about 1.7~kpc).
The figure also shows some theoretical models that indicate that the selected sub-sample is incomplete beyond about 2.8~kpc.

That the closest AGB star in the sample is only at 1.4~kpc is surprising.
For example, \citet{Deathstar22} determine distances to 201 AGB stars (of which 188 are within 1.4 kpc) and only one is present in the MW sample.
It turns out, however, that only 5 of the 201 objects have an `AAA' photometric flag in 2MASS (which are located at 4.0, 1.5, 1.1, 1.0, and
0.6 kpc, respectively), the others have quality flags signalling lower quality NIR data, in almost all cases because the sources are very bright.
Three of the five stars do not obey $\sigma_{\pi} / (\pi +0.1) < 0.2$, and one is not listed in the LPV2 catalogue.
As for another example, of the 258 stars in \citet{Whitelock06}, 220 have an 2MASS counterpart (within 6\arcsec) but only 113 have a quality flag of `AAA'.
Eleven of those do not have a parallax listed in the \G\ catalogue, 44 do not obey our parallax selection criteria, and three are not listed in the LPV2 catalogue.
So, in this case, only 55 out of 258 stars are in the MW sample.
It is clear that the MW sample is incomplete for nearby AGB stars, primarily because 2MASS magnitudes are unreliable for bright stars, and secondly because
of the relatively poor parallax determinations, which is a particular problem for AGB stars and red supergiants as convection-related surface dynamics leads to
photocenter shifts that impact the accuracy of the parallax determination (e.g. \citealt{Chiavassa18, Chiavassa22}).

\smallskip
      To remedy this, 1) nearby AGB stars with NIR photometry other than from 2MASS are considered,
      2) for AGB stars in OCs the \G\ parallax of the AGB star is replaced by the more
      accurate parallax of the host cluster, and
      3) a search is made for wide binary systems (WBSs) where the companion to the AGB star has a more accurate parallax.

\subsubsection{Adding MW stars with SAAO photometry}

Regarding the first strategy, AGB stars (of all spectral classes) are added from a variety of papers with photometry on the SAAO system.
The advantage is that it is a significant set of data with homogeneous photometry.
The following data sets have been considered.
From \citet{Whitelock06}   239 C stars and 19 CS stars, peculiar, and uncertain C stars,         
from \citet{Whitelock2000} 193 Mira and semi-regular variables that were observed by Hipparcos,  
from \citet{Whitelock95}   161 late-type stars in the South Galactic Cap, and                    
from \citet{Whitelock94}    61 Miras in the South Galactic Cap.                                  
The combined list has 644 unique entries. The photometry is taken from the most recent work in case of multiple entries.
An additional advantage of this data set is that the reported magnitudes do not result from single-epoch observations, but are the mean magnitudes
derived from a Fourier analysis of the light curves.

This list of sources was treated in the same way as before, that is, correlated with \cite{bailerjones2021} and the various
\G\ catalogues, and the STILISM tool was used to obtain the reddening.
Initially, the SAAO photometry was transformed to the 2MASS system using Eq.~1 in \citet{Koen07} but further investigation led us to derive
transformation equations specific to the present sample, as outlined in Appendix~\ref{App:Trans}.

This procedure results in transformed SAAO data for 475 objects while for 169 objects this was not possible.
Either these objects were not listed in \G\ (mostly because the stars are so red that they are expected to be well fainter than $G= 21$), or
they are listed in \G\ without parallax ({\tt solution\_type}= 3), or
the parallax accuracy was too poor ($R_{\rm plx}$ $<$5), or
they are not listed in the LPV2 catalogue.
Of the 475 objects 147 were already in the MW sample. They are not removed as the photometry is determined independently.
Only 152 of the 475 stars are classified as C-star in the LPV2 catalogue.

Model 103 extends model 102 by including this additional sample of stars.
Figure~\ref{Fig:DistrM103} shows the cumulative number of all stars as a function of distance and indicates that the number of
nearby stars has increased significantly, although it is clear that the sample still is far from volume
complete\footnote{A good deal of the northern hemisphere can not be observed from the SAAO.}.
The total number of C stars inside the selection box that fulfil all criteria is only increased by 38, however.
On average the absolute magnitudes are slightly brighter.
A model where the width of the Gaussian and the Lorentzian distribution are fitted does converge now,
but with large error bars in these parameters and its results are not listed explicitly.

\subsubsection{AGB stars in OCs}

Regarding improved distances for AGB stars in OCs the list of stars and clusters in tables 1 and 2 from \citet{Marigo22} were
considered (excluding the cases listed as doubtful), which results in 51 unique objects. In those tables distances based on
GDR2 data (from \citealt{Cantat-GaudinAA640}) are listed.
In their appendix, \citet{Marigo22} recompute the cluster parallaxes using GDR3 data and considering various PZPOs, but only for a limited
set of clusters. Therefore it was decided to compute OC parallaxes from GEDR3 data for all clusters, see Appendix~\ref{App:ClusterP} for details.
Correlating the position of the AGB stars against the 2MASS catalogue, and retaining only stars with quality flag 'AAA' results
in 22 matches in 18 clusters. All but one were already in the MW sample. One extra object was found as the initial selection was
now relaxed to $R_{\rm plx}$ $>$1, anticipating that the cluster parallax is more accurate than the individual parallax.
This source was added to the MW sample and for the  rest of 21 stars the individual parallaxes were updated with the cluster parallax
from Table~\ref{tab:clusterP}. Only 8 are classified as C-star in the LPV2 catalogue.

These changes are implemented in model 104.
The number of stars that fulfil the criteria is only marginally increased  compared to model 103 (many of the stars are too blue)
and the results have hardly changed.

\subsubsection{Looking for common-propermotion companions}    
\label{subsec:WBS}

One additional option to obtain improved parallaxes for AGB stars is to use the parallax of a physical companion in a WBS.
Initially the catalogue of \cite{ElBadry21} with over a million WBSs was queried, but as a parallax lower limit of 1~mas is
imposed in their work, it was deemed necessary to perform an independent search as most of our objects are located beyond 1~kpc.
To keep a useful and manageable subsample, only the about 20~800 objects with $1.3 < (J-\Ks)_0 < 2.0$~mag were considered.
To test the scripts and codes an equally sized test sample was selected from \cite{ElBadry21} with parallaxes of the primaries and secondaries
close to their limit of 1~mas. 
Details of the procedure are given in Appendix~\ref{App:WBS} and a total of 65 candidate WBSs were found.
For those objects the parallax, parallax error, GoF, RUWE, and distances from \cite{bailerjones2021} were updated with that of the candidate WBS.
These changes are implemented in model 105.
Again, the number of objects has only marginally increased and the results remain unchanged.

\subsubsection{Final remarks and models for the MW}    
\label{subsec:FR}

Of the different procedures to increase the number of C stars with accurate distances inside the selection box, the
inclusion of nearby stars with SAAO photometry proved to be the most efficient. Using AGB stars in OCs or in a WBSs had very little effect.
In fact, it added more O-rich stars in proportion, and the contamination rose from 12\% (model 103) to 19\% (model 105).
In model 106, the lower limit of the selection box is set to $(J-\Ks)_0=$ 1.5~mag, and the contamination of O-rich stars
is reduced to 9.5\%.

So far, no selection on the quality of the astrometric solution was taken into account.
However, many solutions are poor with GoF (or RUWE) parameters outside the recommended range, see e.g. Table~\ref{App:Tab:WBS}.
In model 107 only sources with $-4 <$ GoF $<10$ were kept, which reduced the number of selected sources by 1/3 compared to model 106.

Another potential issue is the adopted reddening. In quite some cases the estimated distance is larger than the largest distance ($d_{\rm max}$)
available in the 3D reddening map.
In all previous MW models, the adopted reddening in such cases has been the reddening at $d_{\rm max}$, which implies an underestimate of the true reddening.
In model 108 only stars are kept where the estimated distance is less than $1.5 \cdot d_{\rm max}$, and this reduces the number of stars in the
final sample by about 55\%.
An alternative is model 109, where the $E(B-V)$ is simply scaled with $d/d_{\rm max}$, which certainly leads to an overestimate of the reddening.

Finally, updated reddening maps of \citet{Lallement22} and \citet{Vergely22} were used\footnote{https://explore-platform.eu} that typically go out to larger
distances than the STILISM maps, that became available only when this work was close to completion.
However, this proved practically impossible to do for all of the  250~000 stars in our sample, and updated reddenings and corresponding values for $d_{\rm max}$ 
were obtained for the about 21~300 stars with $1.3 < (J-\Ks)_0 < 2.5$~mag that could reasonably make it into the selection box.
Model 110 contains the results with the updated reddenings, again only retaining stars where the estimated distance is less than $1.5 \cdot d_{\rm max}$.
In model 111 $E(B-V)$ is again scaled with $d/d_{\rm max}$, with the additional limit that $A_{\rm V} < 1.5$~mag.
Limiting the sample to the most accurate distances ($\sigma_{\rm DM} <0.1$~mag) results in few stars, where the Gaussian and Lorentzian fits
become meaningless (model 112).
Model 113 is a model without background terms (the BIC decreases successively when setting $d$, $d$+$c$, and $d$+$c$+$b$ to zero, respectively)
and with limits $-5.2 < (M_{\rm J})_0 < -6.2$~mag, approximately centred around the fainter magnitude of the MW stars.

In the models presented so far the widths of the Gaussian and Lorentzian distributions have been fixed to the typical value found for the SMC and LMC.
However, \cite{ripoche2020carbon} found a width in the LF of the MW that was much larger than in the MCs.
In model 114 the  widths of the Gaussian and Lorentzian distributions have been fixed to 0.6. The change in the peak magnitudes ($\sim 0.02$~mag) is
much smaller than the error in the peak magnitudes ($\sim 0.09$~mag)

Although there are uncertainties due to the adopted reddening, the quality of {\rm Gaia}'s astrometric solutions, the accuracy of the parallaxes,
models 106 to 114 with a selection of $1.5 < (J-\Ks)_0 < 2.0$~mag and a contamination by O-stars of less than 10\%, are consistent in that
they lead to absolute magnitudes in the MW that are systematically fainter than in the MCs by about 0.2-0.4~mag.
The scatter in the fitted absolute magnitudes due to these different assumptions is of order 0.1~mag however, and much larger than for the MCs.

\subsection{Recommended models for calibration}

The final models are presented that are recommended for calibration. They are driven by the results on the MW, although this result
remains puzzling, see Sect.~\ref{sec:conclusions}.

The models assume $1.5 < (J-\Ks)_0 < 2.0$~mag and a box length in $J$ magnitude of $\Delta (M_{\rm J})_0$= 1.2~mag. No background terms are included.
C stars are selected as those that are C stars according to the \G-2M diagram or according to the classification in the LPV2 catalogue.
For the MW it includes updated parallaxes for AGB stars in OCs and WBS and nearby stars with SAAO photometry transformed to the 2MASS system, 
only retaining solutions with $-4 <$ GoF $< 10$, updated reddenings from \citet{Lallement22} and \citet{Vergely22},
scaling the reddening with the distance relative to $d_{\rm max}$, imposing $A_{\rm V} < 1.5$~mag and $\sigma_{\rm DM} <0.2$~mag.

These are models 25 (LMC), 26 (SMC) and 115 (MW), and the relevant figures are displayed in Fig.~\ref{Fig:AF_M2526115}.
Table~\ref{tab:comparison} compares the results obtained in the present work with values in the literature.

So far in the present work the statistical errors quoted are the errors on the mean.
  To obtain a different estimate of the error bar, Monte Carlo simulations have been
  carried out where the analysis was repeated 1001 times on datasets where the $J$ and $K$-band magnitudes, the $E(B-V)$ reddening, and the distance were varied according to
  Gaussian distributions. The second line in Tables~\ref{Tab:Fit}-\ref{tab:comparison} for models 25, 26, and 115 give the 50 percentile and half the difference between
  the 84 and 16 percentile as error. The error bar calculated in this way can be both larger or smaller than the formal error on the mean depending on the galaxy and the fitted parameter.

For the MCs our final results are largely in agreement with previous estimates in the literature.
The $(M_{\rm J})_0$ is slightly brighter in the LMC than in the SMC, independent which estimator one uses.
The smallest difference between the two is when using the ZP at $J-\Ks = 1.60$~mag, followed by the (weighted) mean magnitude.
One can therefor argue, as is the hypothesis in \citet{madore2020calibration} and \citet{freedman2020application}, that the values
for LMC and SMC can be averaged and that they are consistent with a weighted mean value of $-6.212 \pm 0.021$~mag based on our models,
marginally brighter than the $-6.20$ adopted by Madore, Freedman and collaborators.   

The same is true when considering the peak value of a Gaussian fit as measure.
On the other hand, it is clear that a Gaussian does not provide a good fit to the LF and so one can question the meaning of the fact that the peak values agree.
When considering a Lorentzian function the LF is well fitted, but then one has to conclude that the LF and peak values
are different in SMC and LMC, and we essentially confirm the results in \citet{Parada21,Parada23}.

The results for the MW remain puzzling. We find a fainter value for  $(M_{\rm J})_0$ than in the MCs, and in between the values in 
\citet{ripoche2020carbon} and in \citet{Madore22OC}.
The former is based on Gaia DR2 parallaxes, and a colour cut with a lower bound at $J-\Ks = 1.40$~mag, which should have
a larger fraction of contamination by O-rich stars.
The preferred average quoted in \citet{Madore22OC} is based on the one hand on AGB stars in OCs where a lower bound
of $J-\Ks = 1.2$ mag is used and a sample that is known to contain only few confirmed C stars and in fact known non-C stars,
and on the other hand on a sample of C stars of which the majority has very poor 2MASS photometry.

    \begin{table*}
	\centering
	\tiny 
\setlength{\tabcolsep}{1.9mm}
	\caption{Results on the JAGB method.}
	\label{tab:comparison}
	\begin{tabular}{lllll}
        \hline  
        Study                           & LMC                        & SMC                           &  MW               & Remarks \\
                                        &  (mag)                     & (mag)                         &  (mag)            & \\
        \hline       \hline 
        \citet{madore2020calibration}   & $-6.22 \pm 0.01 \pm 0.03$    & $-6.18 \pm 0.01 \pm 0.05$   &                   & mean \\
        \citet{freedman2020application} & $-6.22 \pm 0.004 \pm 0.026$  & $-6.18 \pm 0.006 \pm 0.048$ &                   & mean  \\
        \citet{ripoche2020carbon}       & $-6.284 \pm 0.004$           & $-6.160 \pm 0.015$          &                   & median, all stars \\
        \citet{ripoche2020carbon}       &                              &                             &  $-5.601 \pm 0.026$  & median, C-stars \\ 
        \citet{Parada21}                & $-6.283 \pm 0.005$           & $-6.160 \pm 0.016$          &                    &  median \\
        \citet{Parada23}                & $-6.256 \pm 0.005$           & $-6.187 \pm 0.014$          &                    &  median \\
        \citet{Zgirski21}               & $-6.212 \pm 0.010 \pm 0.030$ & $-6.201 \pm 0.012 \pm 0.044$ &                   & mean \\
        \citet{lee2021preliminary}      &                              &                              &  $-6.14 \pm 0.05 \pm 0.11$ & median, C-stars \\
        \citet{Madore22OC}              &                              &                              &  $-6.40 \pm 0.11$ & OCs       \\
        \citet{Madore22OC}              &                              &                              &  $-6.19 \pm 0.04$ & Combined with \citet{lee2021preliminary}  \\
 \\        
        This work                       & $-6.2518 \pm 0.0035$         & $-6.1992 \pm 0.0132$         &   $-5.897 \pm 0.023$   &  weighted mean, all stars  \\  %
   (models 25, 26, 115                  & $-6.2609 \pm 0.0039$         & $-6.1863 \pm 0.0147$         &   $-5.853 \pm 0.030$   &  median \\
    from Table~\ref{Tab:Fit})           & $-6.2454 \pm 0.0045$         & $-6.1761 \pm 0.0195$         &   $-5.838 \pm 0.041$   &  peak Gaussian distribution \\
                                        & $-6.3104 \pm 0.0065$         & $-6.1732 \pm 0.0191$         &   $-5.830 \pm 0.086$   &  peak Lorentzian distribution \\
                                        & $-6.2386 \pm 0.0040$         & $-6.1941 \pm 0.0133$         &   $-5.847 \pm 0.022$   &  ZP at $(J-\Ks)_0=1.6$~mag \\
        
        \hline 
    \end{tabular}
\tablefoot{
When two errors are quoted the first is the statistical and the second the systematic error bar. 
\citet{madore2020calibration} take $-6.20 \pm 0.01 \pm 0.04$ as the average of SMC and LMC,  \citet{freedman2020application} quote $-6.20 \pm 0.037$~mag.
For the final models from this work the results from the Monte Carlo simulations are given. 
}
    \end{table*}

\section{Discussion and conclusions}
\label{sec:conclusions}

The main purpose of the present paper is to consider the influence of  contamination by O-rich AGB stars on the
absolute $J$-band magnitude in a colour-magnitude box selected to be (predominantly) C-rich stars.
The blue limit in $(J-\Ks)_0$ colour is the main driver in setting this contamination.
For the MW a limit of 1.5 is required to have a contamination of  $<$8\% (in model 115).
In the MCs the level of contamination is negligible for such a limit.
We determined mean, median, and the peak of Gaussian and Lorentzian profile fits.
Imposing a range in $J$ (or $M_{\rm J}$) of 1.2~mag also reduces the contamination and
ensures that no background terms are required in fitting the Gaussian and Lorentzian profiles.
In practise this means that iterations may be required.
For an initial guess of the DM to an external galaxy the range in $J$ magnitude can be determined
(or no limit is imposed in the first iteration) after which the mean/median/Gaussian/Lorentzian value is determined,
after which a new range can be applied.

The values we find for $(M_{\rm J})_0$ are in agreement with the literature for SMC and LMC.
The two methods proposed in the literature for distance determination seem both plausible.
The LF of SMC and LMC are different, and so one uses a calibration depending on the skewness of the distribution
(a "LMC-like" and "SMC-like" calibration).
On the other hand, for SMC and LMC, the mean magnitudes inside the colour selection box agree to within the
errors and so an average value is formally an accurate mathematical representation.

The result for the MW is puzzling.
A fainter value is found, contrary to theoretical model predictions.
\cite{Eriksson23} calculated synthetic photometry for C-stars based on radiation-hydrodynamics 
simulations of the dust formation for a grid of models with solar composition (with different values for the C/O ratio).
In their Figure~17 they show the $M_{\rm J}$ versus $(J-K)$ diagram and note a `general similarity to Fig.~1 in \citet{Madore22OC}'.
Their figure, however, suggests that in the range  $1.4 < (J-K) < 2$~mag,  $M_{\rm J}$ depends on $(J-K)$.
Using the data in \cite{Eriksson23} for the models with  $1.4 < (J-K) < 2$~mag Table~\ref{Tab:Erik} lists the median $M_{\rm J}$ and
the median-absolute-deviation (MAD) as a function of stellar mass\footnote{This assumes that other parameters namely, effective temperature,
luminosity and carbon excess lead to models that are equally probable from an evolutionary standpoint.}.
As expected, higher mass C-stars have brighter $J$-band magnitude indicating
that the SFH of a galaxy should also influence the average magnitude of C-stars.
The effect of metallicity for a given initial mass appears to be small.
For a 2.5~\msol\ star the maximum luminosity attained during the C-star phase is $L$= 13~000, 10~000, 9400, and 9000~\lsol\
at Z= 0.001, 0.004, 0.008, and solar metallicity, respectively (P. Ventura, private communication, see e.g. \citet{DellAgli15}).
In this case, larger metallicity tends to indeed give lower maximum luminosity, and
in terms of bolometric magnitude this  corresponds to about 0.12~mag change from SMC to solar metallicity.
For lower initial masses the trend with metallicity is less clear and the effect is less.
The straight comparison of the bolometric luminosity ignores the effect of mass-loss and effective temperature on the NIR magnitudes.
Preliminary models using the PARSEC-COLIBRI tracks \citep{Marigo17} including TP-AGB evolution with mass loss \citep{Pastorelli19,Pastorelli20}
and the population synthesis code TRILEGAL \citep{Girardi05} indicate that in the age range 0.63-1~Gyr
for $Z$ in the range 0.006 to 0.014 (corresponding to about 2.2-2.5~\msol\ initial mass) the average $J$-magnitude
during the C-star phase becomes brighter by about 0.4~mag (Pastorelli et al., in prep.).
At solar metallicity this average is about $M_{\rm J} \sim -6.8$~mag for a star of near 2~\msol\ which
is reasonable agreement with the model by \cite{Eriksson23}.

    \begin{table}
	\centering
	\caption{Results of the models by \cite{Eriksson23} for a solar composition.}
	\label{Tab:Erik}
	\begin{tabular}{cccl}
        \hline 
        Mass    & $M_{\rm J}$ & MAD   & N  \\
        (\msol) &  (mag)     & (mag) &    \\
        \hline         \hline 
0.75 & $-6.38$ & 0.21 & 19 \\
1.00 & $-6.60$ & 0.26 & 21 \\
1.50 & $-6.91$ & 0.22 & 10 \\
2.00 & $-7.17$ & 0.09 & $\phantom{1}$6  \\
        \hline 
    \end{tabular}
\tablefoot{
  Stellar mass, median  $M_{\rm J}$ and the MAD in $M_{\rm J}$, and the number of models.
}
    \end{table}

The situation is complex from an empirical point of view. Reddening is an issue, but putting limits on the maximum reddening did
not have a large impact.
We also ran a model with the selective reddening of \cite{wang2019optical} replaced by \citet{Cardelli89}.
This made the absolute magnitudes brighter, but only by $\sim 0.02$~mag.
A larger change ($\sim 0.1$~mag brighter to near $-6.0$~mag) occurred when not using the distance estimate by \cite{bailerjones2021} but
1/parallax instead. This is still fainter than one might expect but possibly reflects the two assumptions in \cite{bailerjones2021}.
The first is that a prior is used in  \cite{bailerjones2021} based on a mock \G\ catalogue that includes all objects.
This prior may not be optimally suited for AGB stars (see \citealt{Deathstar22}).
Also, as the parallaxes of AGB stars are more uncertain than that of non-AGB stars of comparable magnitude and colour, the
prior has a larger influence on the a posteriori distance estimate.
Second, \cite{bailerjones2021} used the PZPO correction of \citet{GEDR3_LindegrenZP}. There is a debate ongoing whether this PZPO
is too small, or in fact, over corrects and is too large, see e.g. Figure~10 in \citet{Molinaro23}, or \citet{Gr23IAUS}.
Ninety percent of the stars in the final MW sample have G magnitudes in the range 7.0-11.5~mag,
and $(Bp-Rp)$ colours that range from 2.4 to 3.9-6.3~mag for the reddest 10\%.
The PZPO is very poorly known in this regime, and e.g. \citet{CRA23} in their study of Cepheids in clusters even exclude entirely
this magnitude range and colours redder than 2.75~mag.

For an unbiased absolute calibration a volume complete sample would be best.
The major obstacle to achieve this is the lack of reliable 2MASS photometry, as the most nearby AGB stars saturate.
Due to the large year-long effort there is a substantial number of AGB stars with SAAO photometry, but the sample is still incomplete.
A dedicated effort to obtain NIR photometry for a few hundred (C- and O-rich) AGB stars in both hemispheres with
good \G\ astrometric solutions, with suitable $(J-K)$ colours and that saturate in 2MASS would be beneficial for an improved calibration
in the future. This would only require small sized telescopes.
A complementary approach is to study the JAGB LF in other galaxies with solar like metallicities and this is work in progress.

\begin{acknowledgements} 

EM would like to thank Martin Groenewegen for making this internship in the Royal Observatory of Belgium possible, allowing me to do this research.
I am grateful for his helpful comments and tips, and for the coffee at noon with his colleagues.
\\
\noindent
The authors thank the referee for a thorough report that has improved the paper.

\noindent
This research was supported by the International Space Science Institute (ISSI) in Bern,
through ISSI International Team project \#490,  SHoT: The Stellar Path to the Ho Tension in the Gaia, TESS, LSST and JWST Era".
In particular the very generous extension of the stay by one day related to the strike of the German railroad in December 2023 is highly appreciated by MG.
\\

\noindent
We thank 
Dr. Katie Hamren and Dr. Steve Goldman for providing their catalogues with spectral classifications and
Dr. Greg Sloan for comments on the paper.

This work presents results from the European Space Agency (ESA) space mission Gaia. Gaia data are being processed by the Gaia Data Processing
and Analysis Consortium (DPAC). Funding for the DPAC is provided by national institutions, in particular the institutions participating in the
Gaia MultiLateral Agreement (MLA). The Gaia mission website is \url{https://www.cosmos.esa.int/gaia}. The Gaia archive website
is \url{https://archives.esac.esa.int/gaia}. 

This research has used data, tools or materials developed as part of the EXPLORE project that has received funding from the
European Union’s Horizon 2020 research and innovation programme under grant agreement No 101004214.

This work makes use of data products from the Two Micron All Sky Survey, which is a joint project of the University of Massachusetts and the
Infrared Processing and Analysis Center/California Institute of Technology, funded by the National Aeronautics and Space Administration and the
National Science Foundation.

This research has made use of the VizieR catalogue access tool, CDS, Strasbourg, France.
    
\end{acknowledgements}

\bibliographystyle{aa.bst}
\bibliography{references.bib}

\begin{appendix}

\section{Radial velocities}
\label{App:RV}

\begin{figure}[h]
    \centering
    \begin{minipage}{0.49\textwidth}
       \resizebox{\hsize}{!}{\includegraphics{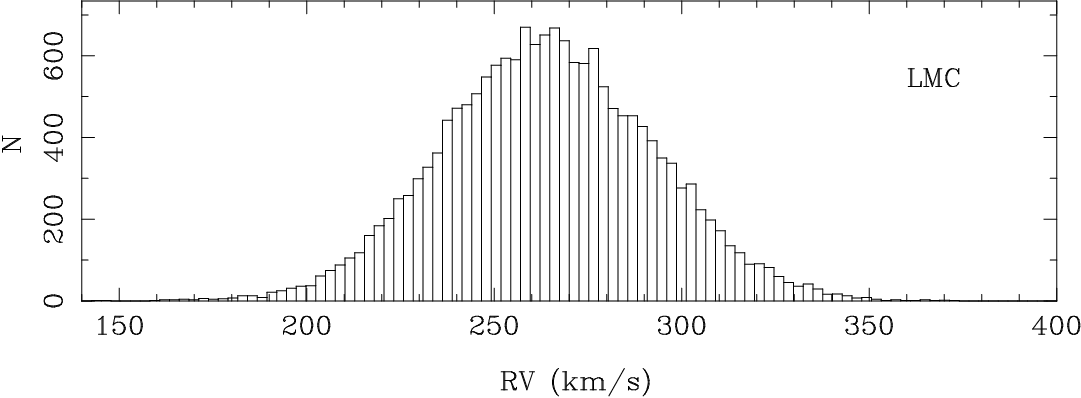}}
    \end{minipage}
    
    \begin{minipage}{0.49\textwidth}
       \resizebox{\hsize}{!}{\includegraphics{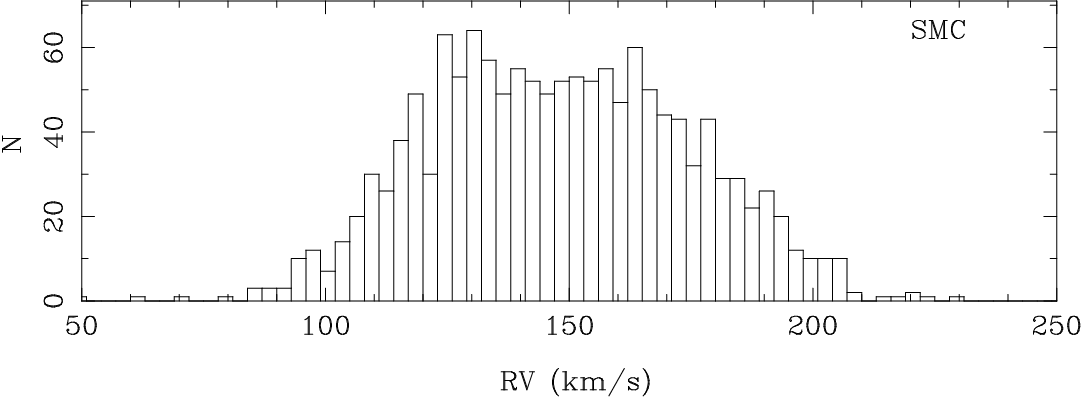}}
    \end{minipage}
        
    \caption{Radial velocity distribution of SMC and LMC stars selected on position, parallax, and proper motions.
    }      
        \label{Fig:RV}
    \end{figure}

Of the 4973 and 39014 stars in the SMC and LMC sample, 1400, respectively, 17731 have an RV listed in the main \G\ catalogue.
Fig.~\ref{Fig:RV} shows the distribution in RVs, which are consistent with the expected values, indicating that the selection on
position, parallax, and proper motion result in reliably selected MC samples.

\section{Notes to Tables~1, 2 and 3}
\label{App:Notes}

\begin{table*}
\caption{\label{Tab:FitNotes} Notes to Table~1.}
\begin{tabular}{l}
\hline
  Column~1: Model number (see details below). \\
  Column~2: Galaxy: LMC, SMC, or MW. \\
  Column~3: Total number of stars in the selection box. \\
  Column~4: Number of C stars in the selection box. \\
  Columns~5 and  6: Slope and offset (at $(J-\Ks)_0= 1.6$~mag) of a linear fit to all stars in the selection box. \\
  Columns~7 and  8: Weighted mean and error in the mean for all stars and the C stars in the selection box.  \\
  Columns~9 and 10: Median value for all stars and the C stars in the selection box. \\
  \\
  
  Models 1 and 2: standard model, that is, the selection box is based on $1.3 < (J-\Ks)_0 < 2.0$~mag and \\

  $-5.0 < (M_{\rm J})_0 < -7.5$~mag. C stars are selected as those that are C stars according to the  \G-2M diagram AND  \\
  according to the classification in the LPV2 catalogue.  No background terms are included in the fitting. \\ 

           Models 3 and 4: $1.2 < (J-\Ks)_0 < 2.0$~mag. \\
           
           Models 5 and 6: $1.4 < (J-\Ks)_0 < 2.0$~mag. \\
           
           Models 7 and 8: $1.5 < (J-\Ks)_0 < 2.0$~mag. \\
           
           Models 9 and 10:  $-5.5 < (M_{\rm J})_0 < -7.0$~mag. \\
           
           Models 11 and 12: $-4.5 < (M_{\rm J})_0 < -8.0$~mag. \\

           Models 13 and 14: Fits include background terms. \\

           Models 15 and 16: Fits include background terms and have $-4.5 < (M_{\rm J})_0 < -8.0$~mag.  \\

           Models 17 and 18: C stars classified as such from the  \G-2M diagram. \\

  Models 19 and 20: C stars classified as such   in the LPV2 catalogue. \\

  Models 21 and 22: C stars selected as those that are C stars according to the  \G-2M diagram OR \\ according to the classification in the LPV2 catalogue.  \\

  Model 23: As model 3 including background terms. \\
  
  Model 24: As model 7 including background terms. \\
  
  Model 101: standard model but including background terms and with $\sigma_{\rm DM} <0.2$~mag. \\

  Model 102: As model 101, with selection box $1.4 < (J-\Ks)_0 < 2.0$~mag and  $-5.1 < (M_{\rm J})_0 < -7.4$~mag. \\

  Model 103: As model 102, including the nearby stars with SAAO photometry transformed to the 2MASS system. \\

  Model 104: As model 103, including updated parallaxes for AGB stars in OCs. \\
 
  Model 105: As model 104, including updated parallaxes for AGB stars in WBSs. \\

  Model 106: As model 105, with $1.5 < (J-\Ks)_0 < 2.0$~mag (and  $-5.1 < (M_{\rm J})_0 < -7.4$~mag). \\

  Model 107: As model 106, only retaining solutions with $-4 <$ GoF $< 10$. \\

  Model 108: As model 107, only retaining sources with a distance less than 1.5 times the maximum distance  \\ in the reddening map. \\

  Model 109: As model 106, scaling the reddening with the distance relative to $d_{\rm max}$. \\
 
  Model 110: As model 108, with updated reddenings from \citet{Lallement22} and \citet{Vergely22}. \\

  Model 111: As model 110, scaling the reddening with the distance relative to $d_{\rm max}$ and with $A_{\rm V} < 1.5$~mag. \\
 
  Model 112: as model 111, and with $\sigma_{\rm DM} <0.1$~mag. \\

  Model 113: as model 111, with $-5.2 < (M_{\rm J})_0 < -6.2$~mag and no background terms. \\ 

  Model 114: as model 113, with width of the Gaussian and Lorenzian profiles fixed to a larger value (of 0.60). \\ 

  Models 25, 26, 115. Final fits. $1.5 < (J-\Ks)_0 < 2.0$~mag. $\Delta (M_{\rm J})_0$= 1.2~mag. No background terms. \\
  C stars selected as those that are C stars according to the  \G-2M diagram OR according to the classification in \\ the LPV2 catalogue
  For the MW it includes updated parallaxes for AGB stars in OCs and WBS and nearby stars \\ with SAAO photometry transformed to the 2MASS system, 
  only retaining solutions with $-4 <$ GoF $< 10$, \\ updated reddenings from \citet{Lallement22} and \citet{Vergely22},
  scaling the reddening with the distance \\ relative to $d_{\rm max}$ and with $A_{\rm V} < 1.5$~mag, and with $\sigma_{\rm DM} <0.2$.  \\
  \hline
\end{tabular}
\end{table*}

\begin{table*}
\caption{\label{Tab:FitGNotes} Notes to Table~2.}
\begin{tabular}{l}
\hline
  Column~1: Model number.  \\
  Columns~2-3: Mean and width of the Gaussian distribution for all stars. \\
  Columns~4-5: Reduced $\chi^2$ and BIC for the Gaussian fit to all stars. \\
  Columns~6-7: Mean and width of the Gaussian distribution for the C stars. \\
  Columns~8-10: Background terms for the Gaussian fit for the C stars. \\
  Columns~11-12: Reduced $\chi^2$ and BIC for the Gaussian fit to the C stars. \\
  An `F' in a column means the parameter was fixed to the value listed. \\
   \\
  Model 13: Background terms for all stars: $b_{\rm all}= 70.6 \pm 4.5$, $c_{\rm all}= +5.0 \pm 1.1$ (mag$^{-1}$), $d_{\rm all}= -40.4 \pm  3.9$ (mag$^{-2}$).  \\

  Model 14: Background terms for all stars: $b_{\rm all}= 15.8 \pm 2.1$, $c_{\rm all}= -0.4 \pm 0.5$, $d_{\rm all}= -9.9 \pm 1.8$.  \\

  Model 15: Background terms for all stars: $b_{\rm all}= 27.3 \pm 2.5$, $c_{\rm all}= +2.5 \pm 0.4$, $d_{\rm all}= -8.1 \pm 1.1$.  \\

  Model 16: Background terms for all stars: $b_{\rm all}=  7.5 \pm 1.4$, $c_{\rm all}= +0.0 \pm 0.3$, $d_{\rm all}= -2.6 \pm 0.7$.  \\

  Model 23: Background terms for all stars: $b_{\rm all}= 110.0 \pm  5.8$,  $c_{\rm all}= 11.2 \pm  1.6$,  $d_{\rm all}= -53.5 \pm  5.0$. \\ 

  Model 24: Background terms for all stars: $b_{\rm all}= 41.3 \pm  3.0$,  $c_{\rm all}=  0.1 \pm 0.7$,  $d_{\rm all}= -25.5 \pm  2.5$. \\ 

  Model 101: Background terms for all stars: $b_{\rm all}=$ 10.1 $\pm$  1.3, $c_{\rm all}=$ 24.1 $\pm$ 1.4, $d_{\rm all}=$ 15.5 $\pm$  1.9.  \\

  Model 102: Background terms for all stars: $b_{\rm all}=$  4.0 $\pm$  1.0, $c_{\rm all}=$  7.7 $\pm$ 1.1, $d_{\rm all}=$  4.8 $\pm$  1.6.  \\

  Model 103: Background terms for all stars: $b_{\rm all}=$  5.9 $\pm$  1.1, $c_{\rm all}=$  1.1 $\pm$ 1.1, $d_{\rm all}=$  3.1 $\pm$  1.6.   \\

  Model 104: Background terms for all stars: $b_{\rm all}=$  5.9 $\pm$  1.2, $c_{\rm all}=$  1.2 $\pm$ 1.0, $d_{\rm all}=$  3.5 $\pm$  1.6.   \\

  Model 105: Background terms for all stars: $b_{\rm all}=$  6.0 $\pm$  1.2, $c_{\rm all}=$  1.2 $\pm$ 1.0, $d_{\rm all}=$  3.8 $\pm$  1.7.  \\

  Model 106: Background terms for all stars: $b_{\rm all}=$  3.2 $\pm$  1.0, $c_{\rm all}=$  1.0 $\pm$ 0.9, $d_{\rm all}=$  3.4 $\pm$  1.5.   \\

  Model 107: Background terms for all stars: $b_{\rm all}=$  6.0 $\pm$  1.4, $c_{\rm all}=$  1.4 $\pm$ 0.8, $d_{\rm all}=$  0.5 $\pm$  1.8.   \\

  Model 108: Background terms for all stars: $b_{\rm all}=$  1.1 $\pm$  0.8, $c_{\rm all}=$  0.8 $\pm$ 0.8, $d_{\rm all}=$  1.3 $\pm$  1.2.  \\

  Model 109: Background terms for all stars: $b_{\rm all}=$  2.5 $\pm$  1.1, $c_{\rm all}=$  1.1 $\pm$ 0.8, $d_{\rm all}=$  2.6 $\pm$  1.5.  \\

  Model 110: Background terms for all stars: $b_{\rm all}=$  0.3 $\pm$  0.7, $c_{\rm all}=$  0.7 $\pm$ 0.7, $d_{\rm all}=$ -0.2 $\pm$  1.2.   \\

  Model 111: Background terms for all stars: $b_{\rm all}=$  0.9 $\pm$  0.7, $c_{\rm all}=$  0.7 $\pm$  0.7, $d_{\rm all}=$ -0.1 $\pm$  1.1.  \\

  Model 112: Background terms for all stars: $b_{\rm all}=$  0.6 $\pm$  1.1,  $c_{\rm all}=$  1.1 $\pm$  0.6,  $d_{\rm all}=$  0.6 $\pm$  1.5.  \\
  \hline
\end{tabular}
\end{table*}

\begin{table*}
\caption{\label{Tab:FitLNotes} Notes to Table~3.}
\begin{tabular}{l}
  \hline
  Column~1: Model number. \\
  Columns~2-5: Mean, width, skewness, and kurtosis of the Lorentzian fit for all stars. \\
  Columns~6-7: Reduced $\chi^2$ and BIC for the  Lorentzian fit to all stars. \\
  Columns~8-11: Mean, width, skewness, and kurtosis of the  Lorentzian fit for the C stars. \\
  Columns~12-13: Reduced $\chi^2$ and BIC for the Lorentzian fit to the C stars. \\
  An `F' in a column means the parameter was fixed to the value listed. \\ \\
  Model 13: Background terms $b_{\rm all}= +0.6 \pm  14.4$, $c_{\rm all}= -7.1 \pm  4.9$ (mag$^{-1}$), $d_{\rm all}= -5.6 \pm 5.6$ (mag$^{-2}$), \\
                             $b_{\rm C}= +6.4 \pm  10.8$,  $c_{\rm C}= -3.8 \pm  2.5$ (mag$^{-1}$), $d_{\rm C}= -8.3 \pm 5.2$ (mag$^{-2}$).  \\

  Model 14: $b_{\rm all}= -1.4 \pm  372$, $c_{\rm all}= +0.3 \pm 136$, $d_{\rm all}= -1.5 \pm 30$,
                             $b_{\rm C}= -5.2 \pm 24.2$,  $c_{\rm C}= +1.8 \pm 18.1$, $d_{\rm C}= +0.7 \pm 3.9$.  \\

  Model 15: $b_{\rm all}= -6.1 \pm  4.7$, $c_{\rm all}= -0.6 \pm 0.7$, $d_{\rm all}= +1.1 \pm 1.5$,
                             $b_{\rm C}= -2.0 \pm 6.2$,  $c_{\rm C}= -0.2 \pm 0.7$, $d_{\rm C}= -1.7 \pm 2.4$.  \\

  Model 16:  $b_{\rm all}= -4.0 \pm 14.7$, $c_{\rm all}= +0.4 \pm 0.6$, $d_{\rm all}= +1.7 \pm 3.2$,
                             $b_{\rm C}= -6.1 \pm 5.0$,  $c_{\rm C}= +0.6 \pm 0.8$, $d_{\rm C}= +1.7 \pm 1.3$.  \\

  Model 23: $b_{\rm all}=$ -10.9 $\pm$ 54.9,  $c_{\rm all}=$ -26 $\pm$ 90,  $d_{\rm all}=$  1.7 $\pm$ 42.1,
  $b_{\rm C}=$ -33.5 $\pm$ 16.6,  $c_{\rm C}=$ -41 $\pm$ 37,    $d_{\rm C}=$ -17.0 $\pm$ 20.1.  \\

  Model 24: $b_{\rm all}=$  4.7 $\pm$ 11.8,  $c_{\rm all}=$ -0.9 $\pm$ 1.2,  $d_{\rm all}=$ -2.4 $\pm$  5.9,
  $b_{\rm C}=$ 22.6 $\pm$  5.0,   $c_{\rm C}=$ -1.3 $\pm$ 0.7,    $d_{\rm C}=$ -15.5 $\pm$  3.3  \\

  Model 101:  $b_{\rm all}=$ 10.0  $\pm$ 1.2, $c_{\rm all}=$ 23.9 $\pm$  1.4, $d_{\rm all}=$ 15.1 $\pm$   1.5 ,
                              $b_{\rm C}=$  3.3  $\pm$  1.1,  $c_{\rm C}=$  6.3   $\pm$ 0.9, $d_{\rm C}=$  4.1  $\pm$   1.5.  \\

  Model 102:  $b_{\rm all}=$ 3.9   $\pm$ 1.0, $c_{\rm all}=$  7.8  $\pm$   1.0,  $d_{\rm all}=$  4.3  $\pm$   1.3,
                              $b_{\rm C}=$ 2.6 $\pm$  1.0,  $c_{\rm C}=$   4.9 $\pm$   0.9, $d_{\rm C}=$  2.4  $\pm$  1.2.  \\

  Model 103:  $b_{\rm all}=$  6.0 $\pm$  1.1,  $c_{\rm all}=$  8.0 $\pm$  1.0,  $d_{\rm all}=$  2.2 $\pm$  1.3,
                              $b_{\rm C}=$  3.1 $\pm$  1.0,    $c_{\rm C}=$  5.0 $\pm$  0.9,    $d_{\rm C}=$  2.1 $\pm$  1.2.  \\

  Model 104:  $b_{\rm all}=$  6.1 $\pm$  1.1,  $c_{\rm all}=$  8.4 $\pm$  1.0,  $d_{\rm all}=$  2.5 $\pm$  1.3,
                              $b_{\rm C}=$  3.2 $\pm$  1.0,    $c_{\rm C}=$  5.1 $\pm$  0.9,    $d_{\rm C}=$  2.2 $\pm$  1.2.  \\

  Model 105:  $b_{\rm all}=$  6.3 $\pm$  1.2,  $c_{\rm all}=$  8.8 $\pm$  1.0,  $d_{\rm all}=$  2.7 $\pm$  1.3,
                              $b_{\rm C}=$  3.2 $\pm$  1.0,    $c_{\rm C}=$  5.2 $\pm$  0.9,    $d_{\rm C}=$  2.2 $\pm$  1.2.  \\

  Model 106:  $b_{\rm all}=$  3.3 $\pm$  1.0,  $c_{\rm all}=$  5.4 $\pm$  0.8,  $d_{\rm all}=$  2.5 $\pm$  1.2,
                              $b_{\rm C}=$  2.5 $\pm$  0.9,    $c_{\rm C}=$  3.9 $\pm$  0.8,    $d_{\rm C}=$  1.8 $\pm$  1.1.  \\

  Model 107:  $b_{\rm all}=$  6.0 $\pm$  1.1,  $c_{\rm all}=$  6.1 $\pm$  0.8,  $d_{\rm all}=$  0.6 $\pm$  1.3,
                              $b_{\rm C}=$  6.0 $\pm$  1.1,    $c_{\rm C}=$  6.1 $\pm$  0.7,   $d_{\rm C}=$  0.6 $\pm$  1.2.   \\

  Model 108:  $b_{\rm all}=$  0.7 $\pm$  0.8,  $c_{\rm all}=$  1.4 $\pm$  0.7,  $d_{\rm all}=$  1.6 $\pm$  1.0,
                              $b_{\rm C}=$  0.2 $\pm$  0.7,    $c_{\rm C}=$ -0.2 $\pm$  0.5,   $d_{\rm C}=$  0.3 $\pm$  0.8.  \\

  Model 109:  $b_{\rm all}=$  2.2 $\pm$  1.0,  $c_{\rm all}=$  3.8 $\pm$  0.7,  $d_{\rm all}=$  2.4 $\pm$  1.2,
                              $b_{\rm C}=$  1.6 $\pm$  1.0,    $c_{\rm C}=$  2.9 $\pm$  0.6,   $d_{\rm C}=$  1.9 $\pm$  1.2.   \\

  Model 110:  $b_{\rm all}=$  0.1 $\pm$  0.7,  $c_{\rm all}=$ -0.7 $\pm$  0.6,  $d_{\rm all}=$  0.0 $\pm$  0.9,
                              $b_{\rm C}=$  0.1 $\pm$  0.7,    $c_{\rm C}=$ -0.6 $\pm$  0.7,   $d_{\rm C}=$  0.5 $\pm$  1.0.   \\

  Model 111:  $b_{\rm all}=$  0.8 $\pm$  0.7,  $c_{\rm all}=$ -0.1 $\pm$  0.6,  $d_{\rm all}=$ -0.1 $\pm$  0.9,
                              $b_{\rm C}=$  0.5 $\pm$  0.9,    $c_{\rm C}=$ -0.0 $\pm$  0.6,    $d_{\rm C}=$  0.7 $\pm$  1.2.  \\

  Model 112:  $b_{\rm all}=$  1.0 $\pm$  1.0,  $c_{\rm all}=$ -0.3 $\pm$  0.5,  $d_{\rm all}=$  0.1 $\pm$  1.2,
                              $b_{\rm C}=$  1.0 $\pm$  1.0,   $c_{\rm C}=$ -0.3 $\pm$  0.6,    $d_{\rm C}=$  0.1 $\pm$  1.4.   \\
\hline
\end{tabular}
\end{table*}

\FloatBarrier   
\clearpage
 
\section{Transformation from SAAO to 2MASS photometry}
\label{App:Trans}

Initially, the SAAO photometry (dereddenned) was transformed to the 2MASS system using Eq.~1 in \citet{Koen07}.
For the 147 stars that were also in the original MW sample it was possible to compare 
the 2MASS photometry to the transformed SAAO photometry. Taking $J$ as example the median difference was 0.005 mag but
the largest differences were more than a magnitude in some cases.
A closer inspection revealed that this was primarily so for the reddest sources. They are outside the validity range of the transformation
formulae that are $-0.043 < (J-H) < 0.992$, $-0.087 < (J-K) < 1.390$, and $-0.044 < (H-K) < 0.503$ \citep{Koen07}.
The differences are also larger for sources where the 
{\tt trimmed\_range\_mag\_g\_fov} (IQR5) is large, suggesting the effect of variability.
Restricting the comparison to $(J-K) < 2.0$ and IQR5 $<1$~mag (non-Mira variables), the median difference of $J$ is $-0.015$ and for 60 of the 63 sources
the absolute difference is 0.21~mag or less. In $K$ these numbers are +0.037 and 0.17~mag, respectively.

In a next step the original SAAO photometry was directly compared to the 2MASS photometry (for sources with quality flag 'AAA') for 160 sources.
This sample is slightly larger as the condition that the star was in the LPV2 catalogue was not imposed.
The condition on the parallax error was imposed in order to have a distance estimate from \cite{bailerjones2021} so that the reddening could be estimated.
As in the application of the JAGB method a limit on $(J-K)$ is used the same limit is imposed in deriving the transformation formula.
To limit the influence of variability a limit on the absolute difference between the magnitudes is imposed following Figs.~4, 5, and 6 in \cite{Koen07}.
Figure~\ref{Fig:Transformation} shows the results which are typically based on 55-60 stars.

In $J$ the derived slope is formally not significant but agrees with \cite{Koen07} and the adopted formula is
\begin{equation}
  \begin{aligned}
  J_{\rm 2MASS} - J_{\rm SAAO}  =   \\
 (-0.055 \pm 0.040) + (-0.039 \pm 0.030) \; (J-K)_{\rm SAAO},
  \end{aligned}
\end{equation}
with an rms of 0.043~mag.

In the $H$-band the slope is not significant
\begin{equation}
  \begin{split}
    H_{\rm 2MASS} - H_{\rm SAAO} =   \\
                                 (+0.031 \pm 0.035) + (-0.025 \pm 0.026)\; (J-K)_{\rm SAAO}, 
   \end{split}
\end{equation}
and a constant offset
\begin{equation}
  H_{\rm 2MASS} - H_{\rm SAAO} = +0.001~{\rm mag} \;\; {\rm rms= 0.044}
\end{equation}
was adopted, as in \cite{Koen07}.

In the $K$-band the derived slope is significant and the transformation formula was
\begin{equation}
  \begin{split}
  K_{\rm 2MASS} - K_{\rm SAAO} =       \\
                              (-0.243 \pm 0.045) + (+0.122 \pm 0.032) \; (J-K)_{\rm SAAO}, 
  \end{split}
\end{equation}
with an rms of 0.060~mag.

After using these formula the comparison of the 2MASS to the transformed SAAO photometry restricted to $(J-K) < 2.0$ and IQR5 $<1$~mag
(non-Mira variables) gives a median difference in $J$ of  $-0.001$~mag and for 60 of the 63 sources the absolute
difference is 0.19~mag or less. In $K$ these numbers are +0.007 and 0.22~mag, respectively.

    \begin{figure}[th]
    \centering
    \begin{minipage}{0.45\textwidth}
       \resizebox{\hsize}{!}{\includegraphics{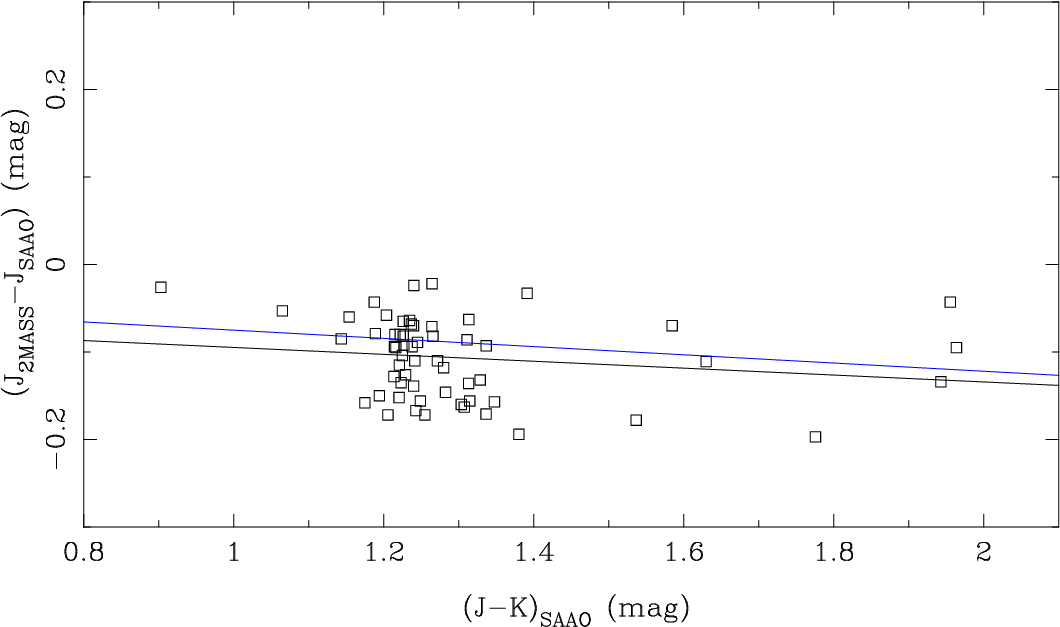}}
    \end{minipage}
    \begin{minipage}{0.45\textwidth}
       \resizebox{\hsize}{!}{\includegraphics{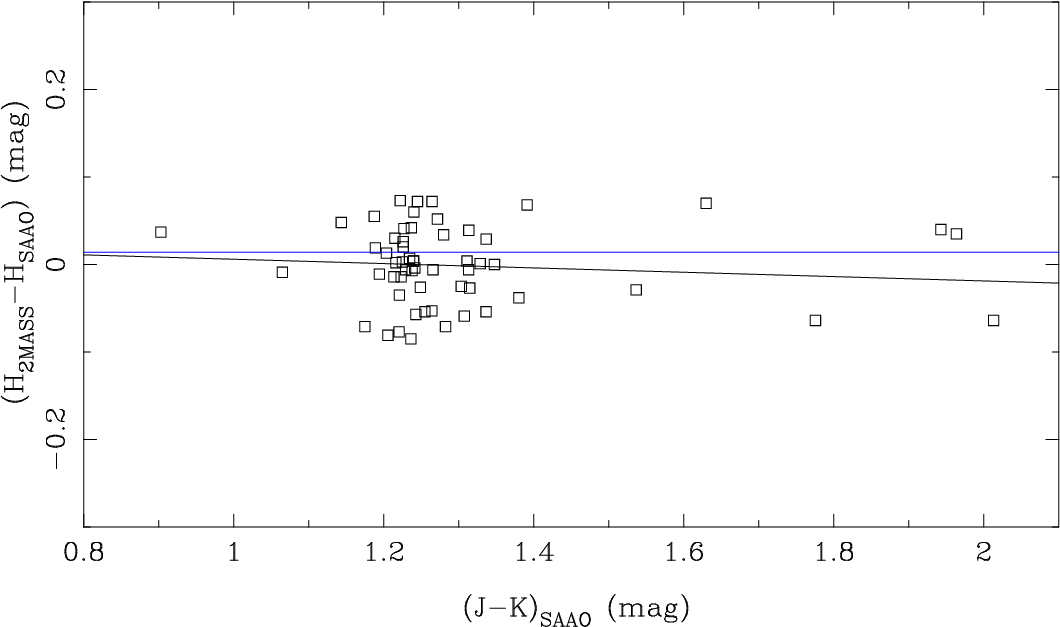}}
    \end{minipage}
    \begin{minipage}{0.45\textwidth}
       \resizebox{\hsize}{!}{\includegraphics{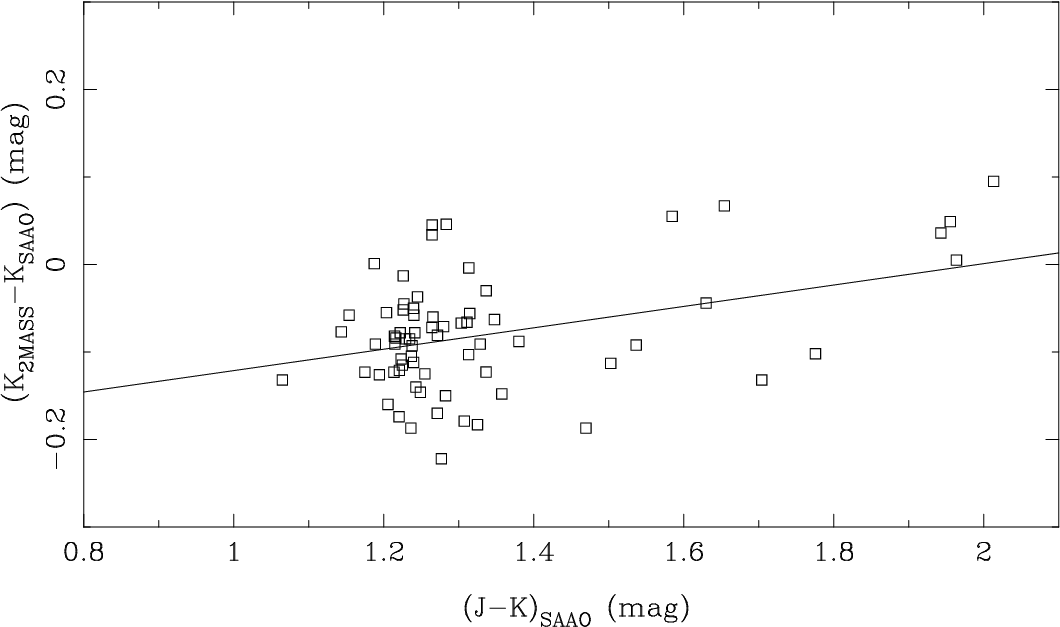}}
    \end{minipage}

    \caption{Difference between 2MASS and SAAO photometry in $JHK$ against SAAO $(J-K)$ photometry.
      The black lines represent the formal fit (see text), and the blue line the result from \citet{Koen07}.
      In the $K$-band the latter is not plotted as the transformation formula in \citet{Koen07} depends on $(H-K)$ and $(J-H)$.
    }
      
        \label{Fig:Transformation}
    \end{figure}

\FloatBarrier
\clearpage
    
\section{Cluster parallaxes}
\label{App:ClusterP}

Cluster parallaxes are a way to improve upon the parallax of individual AGB stars that are in open clusters.
\citet{Marigo22} re-investigated the population of AGB stars in Galactic OCs using {\it Gaia} data.
They presented results mainly based on GDR2 data and an update using GDR3 data was only performed for a subset of clusters.
An independent analysis of the cluster parallaxes is presented here, largely following \citet{Marigo22}.
Full details on the analysis and for a much larger set of clusters will be presented elsewhere (Groenewegen 2024, in prep., hereafter Gr24).

For the clusters analysed here (and in \citealt{Marigo22}) cluster members and cluster membership probabilities are taken from
\citet{CantatGaudin2020}. 
A dedicated script and Fortran code is used to do the analysis.
The list of members is read and the GDR3 main catalogue queried on position (the DR2 position) using a search radius of 0.15\arcsec,
and various quantities are retrieved, like the {\tt source\_id}, the parallax and error, the {\tt astrometric\_gof\_al} (GOF) and {\tt RUWE} quality
  flags, the $G$, $Bp$ and $Rp$ magnitudes, the non-single star (NSS) flag, the galactic end ecliptic coordinates, and the
  {\tt astrometric\_params\_solved}, {\tt nu\_eff\_used\_in\_astrometry}, and {\tt pseudocolour} flags.

\citet{Marigo22} presented results using no PZPOs and using the prescriptions in \citet{GEDR3_LindegrenZP} (hereafter L21) and \cite{Gr21} (hereafter G21).
As the distances in the MW sample are by default based on \cite{bailerjones2021} that uses the L21 correction, only that correction is considered here.
In a first pass the median value and standard deviation (calculated as 1.4826 $\times$ MAD) of the GOF and parallax are calculated.
In a second pass the final sample is selected. Stars with $\mid \pi - \overline{\pi}\mid/ \sigma_{\pi}$ $>$4 were excluded, as
well as stars with a GOF less than $-3.5$, or larger than the median value + 3 times the standard deviation (with a minimum of +3.5),
stars with a non-zero NSS flag, and stars with a membership probability $<$0.5.
Stars were also restricted to the range $0.15 < Bp-Rp < 3.0$ to stay within the validity range of the L21 correction.

For this sample the weighted mean parallax (including the L21 correction) and error in the mean is calculated, as well as the median parallax
and the standard deviation.
The error in the mean is typically very small, but one has to take into account the angular correlation in the parallax (e.g. \citealt{Vasiliev21})
and this sets the floor to the accuracy that can be achieved.
To estimate this effect the distance between all members was calculated and the spatial correlation calculated according to \citet{Vasiliev21}, and
the error was then taken as the median value over all pairs.

Finally the finite size of the cluster was taken into account as the position of any given member is unknown.
To estimate this effect the distance between any member and the centre of the cluster (itself taken as the median in right ascension and declination)
was calculated, and the 68\% percentile was taken as the typical size. This angular size on the sky was converted to a line-of-sight change in parallax
assuming a sphere. This effect is negligible apart from a few large, nearby clusters.

The results are compiled in Table~\ref{tab:clusterP} where the cluster parallax and its final error are listed as well as the
different components to this error bar, the size of the cluster and the number of selected members.
In almost all cases the error bar is set by the limit due to the spatial correlation.
For comparison, the values in \citet{Marigo22} are listed in the format of the original paper, that is, quoting the 68\% confidence intervals.
\citet{Marigo22} appear to quote the standard deviation and in column~8 their and our values are compared directly and the agreement is excellent,
typically to 0.01~mas or better, except in a few clusters with few members, and where the results may depend on details in the selection of the final sample.

 \begin{table*}
	\centering
	\caption{Cluster parallaxes.}
\setlength{\tabcolsep}{1.7mm}
	\label{tab:clusterP}
	\begin{tabular}{lrrrrrrcc}
        \hline 
    Cluster    &  $\pi$            & std.dev. & spat.corr. & depth   & size      & N & $\pi$                       & Remark \\
               &  (\muas)          & (\muas)  & (\muas)    & (\muas) & (\arcmin) &   &  (mas)   & \\
               &                   &          &            &         &           &   &   Marigo et al. (This work)   &        \\
    \hline
    \hline 
     Berkeley 9 &  586.8 $\pm$ 12.5 &  58.99 & 10.95 &  0.90 &  5.3 &  101 & 0.52-0.65 (0.53-0.65)  \\ 
    Berkeley 14 &  234.4 $\pm$ 12.6 &  66.45 & 10.95 &  0.35 &  5.1 &  118 & 0.15-0.30 (0.17-0.30)  \\ 
    Berkeley 29 &   83.2 $\pm$ 19.0 &  71.23 & 10.95 &  0.25 & 10.2 &   21 &                        \\ 
    Berkeley 34 &  160.2 $\pm$ 16.0 &  78.68 & 10.95 &  0.11 &  2.4 &   45 & 0.12-0.27 (0.08-0.24)  \\ 
    Berkeley 53 &  301.9 $\pm$ 12.0 &  84.14 & 10.94 &  2.11 & 24.0 &  358 & 0.22-0.37 (0.22-0.39)  & $\ast$  \\ 
    Berkeley 54 &  159.4 $\pm$ 12.9 &  68.16 & 10.95 &  0.24 &  5.1 &  102 &                        & $\ast$  \\ 
    Berkeley 72 &  203.5 $\pm$ 13.8 &  89.50 & 10.94 &  0.81 & 13.6 &  113 & 0.11-0.25 (0.11-0.29)  \\ 
    Berkeley 91 &  256.9 $\pm$ 17.4 &  81.21 & 10.95 &  0.26 &  3.5 &   36 &                       \\ 
          BH 55 &  250.1 $\pm$ 12.7 &  63.42 & 10.95 &  0.38 &  5.2 &   95 & 0.19-0.30 (0.19-0.31)   \\ 
          BH 67 &  156.9 $\pm$ 13.9 &  61.06 & 10.95 &  0.12 &  2.5 &   51 & 0.07-0.20 (0.10-0.22)  & $\ast$  \\ 
          BH 92 &  423.8 $\pm$ 12.1 &  34.81 & 10.95 &  0.28 &  2.3 &   47 &                       \\ 
          BH 99 & 2269.0 $\pm$ 23.6 &  45.52 & 10.91 & 20.80 & 31.5 &  301 &                       \\ 
   Collinder 74 &  403.2 $\pm$ 11.9 &  50.81 & 10.95 &  0.83 &  7.0 &  119 &  & $\ast$ \\ 
     Czernik 37 &  411.5 $\pm$ 11.7 &  61.50 & 10.95 &  0.81 &  6.8 &  245 &                       \\ 
         Dias 2 &  248.9 $\pm$ 15.5 &  81.38 & 10.95 &  0.56 &  7.7 &   55 & 0.18-0.32 (0.17-0.33) & $\ast$   \\ 
       FSR  154 &  264.0 $\pm$ 11.9 &  48.90 & 10.95 &  0.43 &  5.6 &  114 &                      & $\ast$   \\ 
       FSR  172 &  332.5 $\pm$ 12.0 &  38.89 & 10.95 &  0.28 &  2.9 &   63 & 0.30-0.37 (0.29-0.37)  \\ 
       FSR 1521 &  281.1 $\pm$ 13.0 &  64.92 & 10.95 &  0.34 &  4.2 &   88 &                       & $\ast$  \\ 
       FSR 1530 &  281.6 $\pm$ 19.7 &  94.39 & 10.95 &  0.29 &  3.6 &   33 &   & $\ast$  \\ 
    Gulliver 16 &  233.0 $\pm$ 12.3 &  46.39 & 10.95 &  0.48 &  7.1 &   69 &                       \\ 
     Haffner 14 &  274.5 $\pm$ 11.6 &  37.80 & 10.95 &  1.36 & 17.0 &  114 & 0.25-0.31 (0.24-0.31)  & $\ast$  \\ 
        IC 1311 &  158.0 $\pm$ 11.2 &  49.60 & 10.95 &  0.24 &  5.2 &  504 &                       \\ 
Juchert Saloran 1 & 209.0 $\pm$ 13.2 &  62.27 & 10.95 &  0.45 & 7.4 &   73 & 0.16-0.29 (0.15-0.27) \\ 
        King 11 &  339.0 $\pm$ 11.6 &  59.12 & 10.95 &  0.43 &  4.4 &  262 & 0.27-0.40 (0.28-0.40) & $\ast$  \\ 
     Melotte 66 &  219.1 $\pm$ 11.3 &  55.53 & 10.95 &  0.49 &  7.7 &  404 &                       & $\ast$  \\ 
        NGC 559 &  358.1 $\pm$ 11.2 &  47.32 & 10.95 &  0.63 &  6.1 &  528 &   &  $\ast$ \\ 
        NGC 663 &  371.5 $\pm$ 11.2 &  33.70 & 10.94 &  1.98 & 18.3 &  750 & 0.35-0.40 (0.34-0.41)  \\ 
        NGC 743 &  942.7 $\pm$ 11.7 &  26.27 & 10.95 &  2.18 &  7.9 &   60 &                       \\ 
       NGC 1798 &  240.2 $\pm$ 11.6 &  59.28 & 10.95 &  0.44 &  6.3 &  241 & 0.19-0.28 (0.18-0.30) \\ 
       NGC 2345 &  392.2 $\pm$ 11.2 &  41.70 & 10.95 &  1.25 & 11.0 &  513 &                       \\ 
       NGC 2506 &  315.6 $\pm$ 11.1 &  52.36 & 10.95 &  0.71 &  7.8 & 1447 &                       & $\ast$  \\ 
       NGC 2516 & 2455.7 $\pm$ 25.8 &  33.84 & 10.92 & 23.37 & 32.7 &  400 &                       \\ 
       NGC 2533 &  381.7 $\pm$ 11.3 &  29.90 & 10.95 &  0.79 &  7.1 &  119 & 0.36-0.41 (0.35-0.41)  \\ 
       NGC 2660 &  367.2 $\pm$ 11.1 &  38.63 & 10.95 &  0.34 &  3.2 &  426 & 0.32-0.41 (0.33-0.41)  \\ 
       NGC 5662 & 1327.6 $\pm$ 13.8 &  37.29 & 10.93 &  8.12 & 21.0 &  249 &                       \\ 
       NGC 6242 &  805.9 $\pm$ 11.4 &  50.65 & 10.95 &  2.24 &  9.5 &  481 &                       \\ 
       NGC 6649 &  507.4 $\pm$ 11.2 &  57.97 & 10.95 &  0.77 &  5.2 &  594 &                       \\ 
       NGC 7654 &  636.1 $\pm$ 11.2 &  31.67 & 10.95 &  1.96 & 10.6 & 1008 &                       \\ 
       NGC 7789 &  481.3 $\pm$ 30.4 &  63.32 & 10.95 &  1.53 & 10.9 &    5 & 0.47-0.54 (0.42-0.54)  \\ 
       Pismis 3 &  473.2 $\pm$ 11.3 &  49.92 & 10.95 &  1.36 &  9.9 &  356 & 0.42-0.53 (0.42-0.52)  & $\ast$  \\ 
    Ruprecht 37 &  216.6 $\pm$ 12.8 &  61.55 & 10.95 &  0.19 &  3.0 &   88 & 0.16-0.27 (0.16-0.28) & $\ast$   \\ 
    Ruprecht 42 &  189.4 $\pm$ 12.6 &  60.30 & 10.95 &  0.36 &  6.6 &   94 &                       \\ 
    Ruprecht 83 &  287.4 $\pm$ 11.6 &  41.81 & 10.95 &  0.35 &  4.2 &  126 &                       \\ 
    Ruprecht 91 &  973.4 $\pm$ 12.1 &  29.55 & 10.94 &  4.72 & 16.7 &  160 &                       \\ 
   Ruprecht 107 &  283.8 $\pm$ 11.6 &  32.14 & 10.95 &  0.34 &  4.1 &   71 &                       \\ 
   Ruprecht 112 &  382.8 $\pm$ 11.1 &  36.47 & 10.95 &  1.11 &  9.9 &  455 &                       & $\ast$  \\ 
         SAI 47 &  228.4 $\pm$ 17.4 &  59.16 & 10.95 &  0.19 &  2.8 &   19 &                       \\ 
    Teutsch 106 &  166.5 $\pm$ 12.1 &  46.06 & 10.95 &  0.20 &  4.1 &   83 &                       \\ 
     Tombaugh 1 &  426.4 $\pm$ 11.5 &  31.01 & 10.95 &  0.95 &  7.6 &   85 & 0.39-0.45 (0.40-0.46)  \\ 
     Tombaugh 2 &  140.3 $\pm$ 11.7 &  66.82 & 10.95 &  0.14 &  3.4 &  266 &                       & $\ast$  \\ 
     Trumpler 5 &  336.5 $\pm$ 11.6 &  86.14 & 10.94 &  1.83 & 18.7 &  713 & 0.26-0.41 (0.25-0.42) & $\ast$  \\ 
        \hline 
    \end{tabular}
\tablefoot{
  Column~1: Cluster name.
  Column~2: Cluster parallax and adopted error.
            The error is calculated as the standard deviation in Col.~3 divided by the
            square root of the number of selected members in Col.~7, added in quadrature to the numbers
            in Cols.~4 and 5.
  Column~3: Standard deviation in the cluster parallax.
  Column~4: Floor in the cluster parallax error due to the spatial covariance.
  Column~5: Depth effect due to the size of the cluster.
  Column~6: Size of the cluster (defined as containing 68\% of the selected members).
  Column~7: Total number of selected stars.
  Column~8: Cluster parallax according to \citet{Marigo22} including the \citet{GEDR3_LindegrenZP} correction
            and quoted as the 68\% confidence interval (from their table 10).
            Between parenthesis our result is quoted based on the median and the standard deviation in Column~3 and using the same format.
            Column~9: An asterisks indicates that the AGB star(s) in the cluster is (are) in the MW sample and that the parallax
            has been updated with that of the cluster.
}
    \end{table*}
    
      
\clearpage

\section{Wide-binary systems}
\label{App:WBS}

In this section the search for candidate WBSs is outlined, which largely follows the procedure in \citet{ElBadry21} (hereafter ElB21).
To test the procedure described below a test sample of similar size was selected from the ElB21 catalogue with parallaxes close to 1~mas (as the
AGB stars are typically at larger distances). Of the two stars the one with the larger parallax error was taken as the target star.

We first describe the procedure followed for the test sample, and then indicate the changes made for the AGB sample (which are stricter).
Following ElB21, all objects within 1 pc (or 206.3 kAU) were selected using ($\pi + \sigma_{\pi}$) of the target star
to convert physical distance to angular distance.
All objects within this angular distance of the target object were selected from the \G\ main catalogue.
Following ElB21, a limit $R_{\rm plx}$ $\ge 5$ is imposed. 
One can define a $\chi^2$ statistic
\begin{equation}
\chi^2_{\pi} = ({\pi}_{\rm t} - {\pi}_{\rm WBS})^2 /( ({\sigma}_{{\pi}_{\rm t}})^2 + ({\sigma}_{{\pi}_{\rm WBS}})^2 ) 
\end{equation}
for the parallax, where the subscript {\rm t} refers to the target and {\rm WBS} to the WBS candidate,
and similarly  $\chi^2$ statistics for the proper motion (PM) in RA and the PM in Declination.
Following ElB21, an initial limit $\chi^2_{\pi} \le 36$ is imposed.
ElB21 does not impose limits on $\chi^2_{\rm PM RA}$ and $\chi^2_{\rm PM DE}$ but this is done here.
In the case of the test sample they are very generous and have no impact on the results ($\chi^2_{\rm PM RA} < 335$ and $\chi^2_{\rm PM DE} < 335$).
The output of this first step are the parameters of the candidate WBSs for each target. 
Using these selection criteria counterparts are found for 99.8\% of the targets in the test sample.
At this point there can still be multiple candidate WBSs

In a second step, additional criteria are employed, and a single WBS candidate is determined.
Requiring that all candidate binaries have proper motion differences within 3$\sigma$ of the maximum velocity difference
expected for a system of total mass 5~\msol\ with circular orbits leads to the condition \citep{ElBadry18,ElBadry21}
\begin{equation}
  \label{eq:deltamu}
  \Delta \mu \le \Delta {\mu}_{\rm orbit} + 3 \cdot {\sigma}_{\mu},
\end{equation}
where the quantities  $\Delta \mu$, $\Delta {\mu}_{\rm orbit}$, and ${\sigma}_{\mu}$
are calculated following Eqs.~4, 5, and 6 in \cite{ElBadry18}.
In addition, ElB21 impose
\begin{equation}
  \label{eq:par}
\mid {\pi}_{\rm t} - {\pi}_{\rm WBS} \mid \; \le f \cdot (\sqrt{  ({\sigma}_{{\pi}_{\rm t}})^2 + ({\sigma}_{{\pi}_{\rm WBS}})^2 }),
\end{equation}
where $f$= 6 for separations less than 4\arcsec\, and $f$= 3 otherwise.

If an object obeys Eqs.~\ref{eq:deltamu}, Eqs.~\ref{eq:par} and the physical separation
(= angular separation in arcsec/ (${\pi}_{\rm WBS}+{\sigma}_{{\pi}_{\rm WBS}}$)) is less than 206 kAU the object is considered as a
valid WBS candidate.

If there is more than one candidate an additional selection step is required. 
\cite{ElBadry21} does not discuss this situation in detail, and the following procedure was devised in order
to retrieve the correct counterpart as listed in ElB21 in the overall majority of cases.
A total $\chi^2_{\rm total}$ is calculated based on the $\chi^2$ in parallax and the two PM components.
\begin{equation}
\chi^2_{\rm total}  = \chi^2_{\pi} + \chi^2_{\rm PM RA} + \chi^2_{\rm PM DE}. 
\end{equation}
A selection on $\chi^2_{\rm total} < 55$ is imposed.
However, the retrieval of the correct counterpart was increased by adding more weight to the $\chi^2$ in parallax.
In the case of multiple candidate counterparts the one listed by ElB21 was almost always the one closest
to the target, even if it did not have the lowest $\chi^2$ in our analysis.
In the end, the candidate WBS is the one with the smallest value of $\chi^2$,
\begin{equation}
\chi^2 = \chi^2_{\rm total} + 0.25 \cdot \chi^2_{\pi} +  P
\end{equation}
where $P$ is a penalty function which favours counterparts closer to the target,
$P= 12 \cdot (d/300\arcsec)^2$, where $d$ is the distance between the target and the WBS candidate in seconds of arc.

With these procedure, of the 21600 objects in the test sample, 14400 have a single WBS candidate (in all cases the
one listed by ElB21), and in the 7200 cases where there are multiple candidates the one listed by ElB21
is picked in 99.8\% of the cases. The remaining 15 cases were checked in detail, and the WBS component found by our procedure
is equally, or more, likely than the one listed in ElB21. As briefly mentioned by them, the target star can be in a multiple system and the
procedure may ultimately find another WBS candidate that is part of the same hierarchical system.

For the AGB sample, slightly stricter criteria have been adopted in step 1.
Objects with {\tt solution\_type}= 3 were excluded (i.e. objects without parallax in the \G\ catalogue. This is not relevant for the test sample).
The physical distance limit was (somewhat arbitrarily) reduced to 65 kAU, to lower the probability of a change alignment.
With this limit the likely common-proper motion component around R Scl was retrieved (at 159\arcsec\ on the sky, see below).
%
Stricter limits were used on the PMs, $\chi^2_{\rm PM RA} < 165$ and $\chi^2_{\rm PM DE} < 165$.
It was imposed that the errors in the parallax and the two PM components are smaller in the WBS candidate than in the target star as
the hypothesis is that these errors are larger in the target star simply because of its very AGB nature.
Applying the criteria in step 1 and step 2, 65 candidate WBS were found, and they are reported in Table~\ref{App:Tab:WBS}

Possibly the most interesting specific result is the discovery of candidate companions to R Scl (2.703 $\pm$ 0.017~mas) and R Hya (7.79 $\pm$ 0.20~mas).
An independent distance to R Scl was derived by \cite{Maercker18} of 361 $\pm$ 44~pc based on the phase-lag between the variations
of the star and of the dust-scattering in the resolved dust shell. This distance corresponds to 2.77 $\pm$ 0.34~mas and is in good agreement
with the parallax derived for the WBS here.
R Hya has a parallax determined from VLBI measurements of $7.93 \pm 0.18$~mas \citep{VERA20}, again a result in good agreement
with the parallax derived for the WBS here.
Formally, the GDR3 parallaxes of R Scl (2.54 $\pm$ 0.08~mas) and R Hya (6.74 $\pm$ 0.46~mas) are also in agreement with
the independent estimates at the 0.6 and 2.4$\sigma$ level, respectively (combining the parallax error of \G\ and of the
independent estimate in $\sigma$), but the parallax determinations of their companions are more precise and are
in agreement with the independent estimate at the 0.2 and 0.5$\sigma$ level, respectively.
The astrometric solution of the companion to R Hya is poor indicating that the companion may itself be in a multiple system.

\longtab[1]{
  \begin{landscape}
  \scriptsize
\setlength{\tabcolsep}{1.4mm}
\begin{longtable}{lcrrrrllrcrrrr}
  \caption{\label{App:Tab:WBS} WBS candidates. } \\
  \hline  \hline
        \multicolumn{7}{c}{AGB star} & \multicolumn{7}{c}{WBS candidate} \\
        SID & $\pi$ & PMRA     & PMDE      & GoF & RUWE & Name & SID & $\theta$ & $\pi$ & PMRA & PMDE & GoF & RUWE \\
            & (mas) & (mas/yr) & (mas/yr)  &     &      &      &     & (\arcsec) &  (mas) & (mas/yr) & (mas/yr)  & &   \\
        \hline         \hline 
\hline
\endfirsthead
\caption{continued.}\\
\hline\hline
        \multicolumn{7}{c}{AGB star} & \multicolumn{7}{c}{WBS candidate} \\
        SID & $\pi$ & PMRA     & PMDE      & GoF & RUWE & Name & SID & $\theta$ & $\pi$ & PMRA & PMDE & GoF & RUWE \\
            & (mas) & (mas/yr) & (mas/yr)  &     &      &      &     & (\arcsec) &  (mas) & (mas/yr) & (mas/yr)  & &   \\
        \hline         \hline 
\endhead
\hline
\endfoot

4156710718195701120 &  0.579 $\pm$ 0.131 & -2.07 $\pm$  0.13 & -1.64 $\pm$  0.11 &    7.56 &  1.385 & & 4156710791220375168 &   9.1 &  0.396 $\pm$ 0.044 & -2.06 $\pm$  0.04 & -1.51 $\pm$  0.04 &   -0.66 &  0.967 \\
2061777232014283136 &  0.215 $\pm$ 0.051 & -2.02 $\pm$  0.06 & -4.57 $\pm$  0.06 &    2.27 &  1.084 & & 2061777236328348032 &   5.8 &  0.375 $\pm$ 0.051 & -2.17 $\pm$  0.06 & -4.35 $\pm$  0.06 &    0.29 &  1.010 \\
5526894724021302272 &  0.232 $\pm$ 0.039 & -3.06 $\pm$  0.04 & +3.82 $\pm$  0.05 &    1.37 &  1.048 & & 5526894728323115904 &  14.9 &  0.377 $\pm$ 0.036 & -3.08 $\pm$  0.04 & +3.93 $\pm$  0.05 &    0.10 &  1.003 \\
5307412755502488704 &  0.414 $\pm$ 0.065 & -6.17 $\pm$  0.08 & +4.89 $\pm$  0.08 &    2.64 &  1.105 & & 5307412759818232576 &  26.4 &  0.537 $\pm$ 0.056 & -6.13 $\pm$  0.07 & +5.16 $\pm$  0.07 &    2.29 &  1.090 \\
5720176674770929536 &  0.398 $\pm$ 0.061 & -2.73 $\pm$  0.07 & +3.69 $\pm$  0.06 &   -0.44 &  0.981 & V554 Pup & 5720176674770929408 &  23.7 &  0.443 $\pm$ 0.032 & -2.66 $\pm$  0.03 & +3.56 $\pm$  0.03 &   -0.17 &  0.992 \\
5523967897082962304 &  0.544 $\pm$ 0.033 & -3.70 $\pm$  0.04 & +1.40 $\pm$  0.04 &    0.32 &  1.010 & & 5523967965802436480 &  17.1 &  0.555 $\pm$ 0.032 & -3.67 $\pm$  0.04 & +1.27 $\pm$  0.04 &   25.55 &  2.002 \\
2061033927794150144 &  0.191 $\pm$ 0.046 & -3.98 $\pm$  0.05 & -6.47 $\pm$  0.06 &   -1.67 &  0.938 & & 2061033829018303872 &  10.6 &  0.247 $\pm$ 0.034 & -3.84 $\pm$  0.03 & -6.31 $\pm$  0.04 &    2.73 &  1.100 \\
1972050829719149952 &  0.473 $\pm$ 0.106 & -3.43 $\pm$  0.10 & +0.03 $\pm$  0.10 &    8.69 &  1.345 & V1243 Cyg & 1972050834022984064 &  31.5 &  0.688 $\pm$ 0.089 & -3.60 $\pm$  0.08 & +0.09 $\pm$  0.09 &    0.39 &  1.014 \\
5869385724125867008 &  0.371 $\pm$ 0.069 & -5.58 $\pm$  0.05 & -0.24 $\pm$  0.06 &   -5.93 &  0.824 & & 5869385728425084928 &  24.3 &  0.382 $\pm$ 0.042 & -5.50 $\pm$  0.03 & -0.16 $\pm$  0.04 &   -0.36 &  0.989 \\
5351589625192577792 &  0.249 $\pm$ 0.029 & -6.83 $\pm$  0.04 & +3.31 $\pm$  0.03 &   -2.48 &  0.908 & & 5351589625192580736 &  15.0 &  0.266 $\pm$ 0.025 & -6.85 $\pm$  0.03 & +3.31 $\pm$  0.03 &   -0.04 &  0.998 \\
4251703162723263104 &  0.411 $\pm$ 0.070 & +0.53 $\pm$  0.09 & -0.59 $\pm$  0.07 &   -0.41 &  0.979 & & 4251703162723233792 &  21.8 &  0.492 $\pm$ 0.055 & +0.51 $\pm$  0.07 & -0.51 $\pm$  0.06 &   -0.75 &  0.963 \\
5866284727695849984 &  0.442 $\pm$ 0.094 & -8.29 $\pm$  0.06 & -2.67 $\pm$  0.07 &    2.83 &  1.068 & & 5866284727695857664 &  27.9 &  0.394 $\pm$ 0.067 & -8.33 $\pm$  0.04 & -2.84 $\pm$  0.05 &    1.80 &  1.042 \\
6054857981262594688 &  0.319 $\pm$ 0.076 & -5.51 $\pm$  0.07 & -0.27 $\pm$  0.08 &    1.35 &  1.048 & & 6054857976929568640 &  13.2 &  0.557 $\pm$ 0.064 & -5.32 $\pm$  0.06 & -0.04 $\pm$  0.06 &    0.24 &  1.008 \\
2199405958625523200 &  0.135 $\pm$ 0.038 & -2.51 $\pm$  0.05 & -2.10 $\pm$  0.04 &    2.75 &  1.101 & & 2199405958625523712 &   7.7 &  0.200 $\pm$ 0.028 & -2.60 $\pm$  0.03 & -2.16 $\pm$  0.03 &   -1.12 &  0.958 \\
4042375046648286208 &  0.406 $\pm$ 0.073 & -4.40 $\pm$  0.08 & -2.93 $\pm$  0.05 &  -10.63 &  0.633 & & 4042375046823739904 &  20.7 &  0.439 $\pm$ 0.040 & -4.58 $\pm$  0.04 & -2.90 $\pm$  0.03 &   -0.02 &  0.998 \\
2058300443118232832 &  0.358 $\pm$ 0.084 & -0.48 $\pm$  0.08 & -3.69 $\pm$  0.09 &    9.95 &  1.331 & & 2058300438810196224 &   6.0 &  0.554 $\pm$ 0.034 & -0.50 $\pm$  0.03 & -3.89 $\pm$  0.04 &   -1.29 &  0.958 \\
5335984978263991296 &  0.274 $\pm$ 0.053 & -7.53 $\pm$  0.06 & +2.15 $\pm$  0.05 &    1.30 &  1.044 & & 5335983500751376256 &  13.5 &  0.309 $\pm$ 0.044 & -7.72 $\pm$  0.05 & +2.07 $\pm$  0.04 &    2.56 &  1.089 \\
5940923425302686848 &  0.401 $\pm$ 0.074 & -1.40 $\pm$  0.08 & -2.41 $\pm$  0.06 &    6.13 &  1.211 & & 5940923420958391296 &  12.4 &  0.606 $\pm$ 0.038 & -1.27 $\pm$  0.04 & -2.31 $\pm$  0.03 &   -0.40 &  0.986 \\
1863771955405265920 &  0.258 $\pm$ 0.042 & -1.35 $\pm$  0.04 & -3.10 $\pm$  0.04 &    4.13 &  1.139 & & 1863771959715452032 &   9.2 &  0.294 $\pm$ 0.030 & -1.31 $\pm$  0.03 & -3.04 $\pm$  0.03 &    0.53 &  1.017 \\
4042804405935260416 &  0.327 $\pm$ 0.073 & -0.42 $\pm$  0.08 & -4.62 $\pm$  0.06 &   -4.33 &  0.811 & & 4042804410270019456 &  12.2 &  0.223 $\pm$ 0.042 & -0.50 $\pm$  0.05 & -4.63 $\pm$  0.04 &   -1.66 &  0.922 \\
5851194957078593792 &  0.305 $\pm$ 0.054 & -6.38 $\pm$  0.04 & -3.32 $\pm$  0.05 &   -1.64 &  0.952 & & 5851194961442724224 &  21.1 &  0.434 $\pm$ 0.029 & -6.37 $\pm$  0.02 & -3.31 $\pm$  0.03 &    5.55 &  1.170 \\
4515937457837006080 &  0.365 $\pm$ 0.074 & -0.53 $\pm$  0.06 & -4.92 $\pm$  0.07 &   -0.76 &  0.981 & & 4515937457837008000 &  23.2 &  0.657 $\pm$ 0.068 & -0.67 $\pm$  0.06 & -4.91 $\pm$  0.06 &   -0.76 &  0.980 \\
5337723301786834688 &  0.300 $\pm$ 0.051 & -4.29 $\pm$  0.05 & +1.66 $\pm$  0.05 &    0.23 &  1.007 & & 5337723306129120000 &  14.8 &  0.315 $\pm$ 0.018 & -4.19 $\pm$  0.02 & +1.66 $\pm$  0.02 &    7.43 &  1.276 \\
5308236843797941248 &  0.222 $\pm$ 0.034 & -5.34 $\pm$  0.04 & +2.96 $\pm$  0.04 &   -3.71 &  0.869 & & 5308236843797942144 &   9.6 &  0.202 $\pm$ 0.034 & -5.37 $\pm$  0.04 & +2.94 $\pm$  0.04 &   -0.56 &  0.978 \\
4095204832339129984 &  0.338 $\pm$ 0.086 & -3.82 $\pm$  0.09 & -4.75 $\pm$  0.07 &   -4.60 &  0.786 & & 4095204832339122944 &  17.7 &  0.289 $\pm$ 0.033 & -3.98 $\pm$  0.03 & -4.68 $\pm$  0.03 &   -1.52 &  0.928 \\
5317707452818888320 &  0.432 $\pm$ 0.068 & -3.18 $\pm$  0.09 & +3.69 $\pm$  0.09 &   13.10 &  1.542 & & 5317707452820149376 &  19.1 &  0.336 $\pm$ 0.062 & -3.08 $\pm$  0.08 & +3.71 $\pm$  0.09 &    2.67 &  1.105 \\
202632231998270208 &  0.306 $\pm$ 0.065 & -0.30 $\pm$  0.08 & -0.45 $\pm$  0.06 &    6.14 &  1.196 & & 202632983615613312 &  14.8 &  0.284 $\pm$ 0.050 & -0.20 $\pm$  0.06 & -0.68 $\pm$  0.05 &   -0.28 &  0.991 \\
5341952268352601344 &  0.316 $\pm$ 0.048 & -8.16 $\pm$  0.05 & +1.20 $\pm$  0.05 &   -2.08 &  0.928 & & 5341952062194178176 &  16.7 &  0.399 $\pm$ 0.028 & -8.16 $\pm$  0.03 & +1.20 $\pm$  0.03 &    0.69 &  1.024 \\
2546193059886033408 &  0.908 $\pm$ 0.057 & +7.56 $\pm$  0.08 & -3.33 $\pm$  0.04 &    3.58 &  1.139 & & 2546193059886033024 &  23.9 &  0.994 $\pm$ 0.025 & +7.58 $\pm$  0.03 & -3.36 $\pm$  0.02 &    2.24 &  1.086 \\
5016138145186249088 &  2.544 $\pm$ 0.079 & -9.21 $\pm$  0.06 & -30.81 $\pm$  0.04 &   39.61 &  2.548 & R Scl & 5016150308533630080 & 159.0 &  2.703 $\pm$ 0.017 & -9.19 $\pm$  0.01 & -31.10 $\pm$  0.01 &    0.80 &  1.025 \\
6195030801635544704 &  6.736 $\pm$ 0.464 & -54.21 $\pm$  0.47 & +11.79 $\pm$  0.30 &   29.38 &  2.913 & R Hya & 6195030801634430336 &  21.8 &  7.788 $\pm$ 0.196 & -55.68 $\pm$  0.21 & +13.28 $\pm$  0.14 &  143.51 & 12.066 \\
5939379504806105600 &  0.446 $\pm$ 0.105 & -1.17 $\pm$  0.13 & -2.43 $\pm$  0.10 &    2.70 &  1.111 & & 5939379401726879872 &  26.7 &  0.561 $\pm$ 0.021 & -1.25 $\pm$  0.02 & -2.58 $\pm$  0.02 &   -0.58 &  0.975 \\
6057419980808629504 &  0.226 $\pm$ 0.055 & -7.33 $\pm$  0.06 & +0.51 $\pm$  0.07 &   -2.71 &  0.901 & & 6057419980808627968 &   4.9 &  0.191 $\pm$ 0.036 & -7.35 $\pm$  0.04 & +0.41 $\pm$  0.05 &    0.69 &  1.026 \\
4308248335643989248 &  0.432 $\pm$ 0.081 & -2.60 $\pm$  0.08 & -4.56 $\pm$  0.07 &    6.85 &  1.344 & & 4308248335643991040 &  10.4 &  0.378 $\pm$ 0.049 & -2.88 $\pm$  0.05 & -4.57 $\pm$  0.05 &   -0.24 &  0.987 \\
4093993307977298560 &  0.406 $\pm$ 0.098 & -0.79 $\pm$  0.13 & -1.78 $\pm$  0.09 &   -2.18 &  0.900 & & 4093993303636552448 &   8.9 &  0.712 $\pm$ 0.083 & -0.59 $\pm$  0.11 & -1.65 $\pm$  0.08 &    2.30 &  1.109 \\
5359650213650619264 &  0.262 $\pm$ 0.063 & -6.09 $\pm$  0.07 & +3.15 $\pm$  0.07 &    4.43 &  1.146 & & 5359650213650620288 &  10.6 &  0.311 $\pm$ 0.021 & -5.91 $\pm$  0.02 & +3.07 $\pm$  0.02 &    3.86 &  1.131 \\
5352094777754135680 &  0.346 $\pm$ 0.068 & -4.90 $\pm$  0.08 & +3.22 $\pm$  0.08 &    4.91 &  1.186 & & 5352094782073697408 &   6.7 &  0.296 $\pm$ 0.057 & -4.97 $\pm$  0.06 & +3.46 $\pm$  0.06 &    2.76 &  1.105 \\
5350293713307017472 &  0.340 $\pm$ 0.047 & -7.08 $\pm$  0.05 & +2.77 $\pm$  0.05 &   -2.07 &  0.926 & & 5350293713280660352 &   2.6 &  0.455 $\pm$ 0.041 & -6.87 $\pm$  0.04 & +2.72 $\pm$  0.05 &    8.94 &  1.355 \\
4119106321018338944 &  0.368 $\pm$ 0.089 & +0.25 $\pm$  0.10 & -2.16 $\pm$  0.07 &    0.09 &  1.003 & & 4119106321018346624 &  18.7 &  0.271 $\pm$ 0.028 & +0.32 $\pm$  0.03 & -2.38 $\pm$  0.02 &   -1.05 &  0.953 \\
5309054365035430784 &  0.200 $\pm$ 0.025 & -4.14 $\pm$  0.03 & +3.45 $\pm$  0.03 &    3.86 &  1.139 & & 5309054365035432192 &   9.7 &  0.192 $\pm$ 0.020 & -4.10 $\pm$  0.03 & +3.50 $\pm$  0.02 &   -1.14 &  0.959 \\
6055060871179774464 &  0.293 $\pm$ 0.068 & -8.33 $\pm$  0.07 & -0.92 $\pm$  0.07 &    2.67 &  1.089 & & 6055060875530685184 &  22.8 &  0.361 $\pm$ 0.046 & -8.14 $\pm$  0.04 & -0.80 $\pm$  0.05 &   -1.02 &  0.963 \\
5977054781796737664 &  0.461 $\pm$ 0.083 & -2.85 $\pm$  0.11 & -4.21 $\pm$  0.07 &   -5.34 &  0.857 & & 5977054781796726784 &  14.1 &  0.632 $\pm$ 0.043 & -2.64 $\pm$  0.06 & -4.20 $\pm$  0.04 &    0.06 &  1.001 \\
5862013846618739840 &  0.224 $\pm$ 0.055 & -6.03 $\pm$  0.05 & -2.00 $\pm$  0.06 &   -2.42 &  0.919 & & 5862013846618732160 &  10.3 &  0.263 $\pm$ 0.021 & -6.13 $\pm$  0.02 & -2.06 $\pm$  0.02 &    0.28 &  1.009 \\
2058769075592632192 &  0.289 $\pm$ 0.069 & -2.73 $\pm$  0.07 & -4.90 $\pm$  0.07 &    5.45 &  1.181 & & 2058769006873141248 &  22.1 &  0.355 $\pm$ 0.011 & -2.68 $\pm$  0.01 & -4.79 $\pm$  0.01 &   -0.71 &  0.976 \\
5255469558465315456 &  0.257 $\pm$ 0.052 & -6.41 $\pm$  0.06 & +2.43 $\pm$  0.05 &   -1.01 &  0.960 & & 5255469768946334208 &  18.8 &  0.351 $\pm$ 0.015 & -6.34 $\pm$  0.02 & +2.52 $\pm$  0.01 &   -1.71 &  0.933 \\
4096445558165111040 &  0.473 $\pm$ 0.093 & -0.06 $\pm$  0.11 & -4.17 $\pm$  0.08 &    4.78 &  1.220 & & 4096445523805362816 &  35.7 &  0.525 $\pm$ 0.080 & -0.05 $\pm$  0.09 & -4.10 $\pm$  0.07 &    2.12 &  1.103 \\
5254728114016438016 &  0.252 $\pm$ 0.062 & -5.42 $\pm$  0.07 & +2.65 $\pm$  0.07 &    1.14 &  1.041 & & 5254728114016439808 &  15.9 &  0.330 $\pm$ 0.043 & -5.24 $\pm$  0.05 & +2.64 $\pm$  0.05 &    0.35 &  1.012 \\
3344069512424833408 &  0.324 $\pm$ 0.042 & +0.44 $\pm$  0.04 & -1.46 $\pm$  0.03 &    0.39 &  1.016 & & 3344069508125016576 &  23.0 &  0.322 $\pm$ 0.035 & +0.46 $\pm$  0.04 & -1.40 $\pm$  0.03 &   -0.78 &  0.965 \\
4054567840322057856 &  0.705 $\pm$ 0.102 & -1.76 $\pm$  0.11 & -3.40 $\pm$  0.08 &    3.25 &  1.142 & & 4054567775916101248 &  34.1 &  0.536 $\pm$ 0.054 & -2.12 $\pm$  0.06 & -3.55 $\pm$  0.04 &    0.11 &  1.004 \\
5836472599463344768 &  0.353 $\pm$ 0.082 & -3.29 $\pm$  0.08 & -4.46 $\pm$  0.07 &   -0.79 &  0.974 & & 5836472672586591488 &  22.4 &  0.437 $\pm$ 0.058 & -3.52 $\pm$  0.06 & -4.41 $\pm$  0.05 &    0.49 &  1.015 \\
5940199877886921728 &  0.293 $\pm$ 0.078 & -2.43 $\pm$  0.10 & -3.50 $\pm$  0.09 &   -0.48 &  0.982 & & 5940200629505998208 &  19.6 &  0.441 $\pm$ 0.023 & -2.19 $\pm$  0.03 & -3.62 $\pm$  0.03 &    2.26 &  1.086 \\
5876067318844428800 &  0.203 $\pm$ 0.057 & -3.11 $\pm$  0.06 & -2.85 $\pm$  0.06 &   -2.56 &  0.916 & & 5876067529384509696 &   7.6 &  0.321 $\pm$ 0.053 & -2.99 $\pm$  0.06 & -2.78 $\pm$  0.06 &    1.68 &  1.056 \\
5864937340604715904 &  0.254 $\pm$ 0.056 & -7.42 $\pm$  0.05 & -1.67 $\pm$  0.06 &   -6.73 &  0.778 & & 5864937379379958784 &  18.2 &  0.280 $\pm$ 0.043 & -7.27 $\pm$  0.04 & -1.78 $\pm$  0.04 &   18.81 &  1.712 \\
5884180516435054080 &  0.407 $\pm$ 0.099 & -3.61 $\pm$  0.10 & -4.48 $\pm$  0.09 &   -2.41 &  0.933 & & 5884180585154534272 &  12.9 &  0.267 $\pm$ 0.043 & -3.38 $\pm$  0.04 & -4.32 $\pm$  0.04 &    1.98 &  1.058 \\
5960032284573314944 &  0.522 $\pm$ 0.111 & -0.08 $\pm$  0.12 & -5.02 $\pm$  0.09 &    3.10 &  1.146 & & 5960032383296721408 &  18.5 &  0.502 $\pm$ 0.062 & -0.28 $\pm$  0.07 & -4.94 $\pm$  0.06 &   -1.66 &  0.924 \\
5836353135067293440 &  0.381 $\pm$ 0.084 & -2.01 $\pm$  0.08 & -3.17 $\pm$  0.07 &    0.54 &  1.017 & & 5836329667362543232 &  18.9 &  0.374 $\pm$ 0.061 & -2.14 $\pm$  0.06 & -2.98 $\pm$  0.05 &   -3.53 &  0.884 \\
5972172228614200832 &  0.382 $\pm$ 0.078 & -1.24 $\pm$  0.10 & -3.37 $\pm$  0.07 &    3.50 &  1.113 & & 5972172228614202368 &  12.1 &  0.201 $\pm$ 0.025 & -1.01 $\pm$  0.03 & -3.37 $\pm$  0.02 &    4.51 &  1.149 \\
5241834897073094912 &  0.332 $\pm$ 0.079 & -6.04 $\pm$  0.09 & +3.41 $\pm$  0.09 &   -3.22 &  0.882 & & 5241829021557424896 &  22.5 &  0.298 $\pm$ 0.056 & -5.90 $\pm$  0.06 & +3.29 $\pm$  0.06 &   -1.45 &  0.945 \\
5835467748915111808 &  0.307 $\pm$ 0.058 & -3.80 $\pm$  0.07 & -4.57 $\pm$  0.05 &   -3.38 &  0.898 & & 5835465519922201984 &  21.6 &  0.467 $\pm$ 0.023 & -3.90 $\pm$  0.03 & -4.60 $\pm$  0.02 &   -1.97 &  0.939 \\
2033144785333019776 &  0.217 $\pm$ 0.045 & -2.44 $\pm$  0.04 & -4.35 $\pm$  0.05 &   -1.44 &  0.953 & & 2033144785333014400 &  16.9 &  0.270 $\pm$ 0.032 & -2.35 $\pm$  0.03 & -4.40 $\pm$  0.04 &   -1.43 &  0.953 \\
4153647758021649024 &  0.644 $\pm$ 0.100 & -0.95 $\pm$  0.13 & -3.51 $\pm$  0.10 &    1.10 &  1.054 & & 4153635903911879936 &  46.4 &  0.697 $\pm$ 0.072 & -1.05 $\pm$  0.08 & -3.29 $\pm$  0.06 &    0.91 &  1.043 \\
4064997502764163072 &  0.355 $\pm$ 0.084 & -2.64 $\pm$  0.11 & -3.97 $\pm$  0.08 &   -5.15 &  0.735 & & 4064997502764137472 &  14.4 &  0.470 $\pm$ 0.066 & -2.40 $\pm$  0.09 & -4.07 $\pm$  0.07 &    3.04 &  1.170 \\
5876417895568050560 &  0.482 $\pm$ 0.106 & -5.62 $\pm$  0.11 & -2.92 $\pm$  0.13 &   -0.37 &  0.987 & & 5876417891265274624 &  32.0 &  0.794 $\pm$ 0.037 & -5.71 $\pm$  0.04 & -2.53 $\pm$  0.04 &   -1.10 &  0.962 \\
5982009219865174144 &  0.391 $\pm$ 0.066 & -2.70 $\pm$  0.07 & -3.06 $\pm$  0.06 &    5.20 &  1.193 & & 5982009219865174912 &  16.3 &  0.494 $\pm$ 0.054 & -2.65 $\pm$  0.06 & -3.06 $\pm$  0.05 &    0.38 &  1.013 \\
4043397802970235136 &  0.364 $\pm$ 0.082 & -1.96 $\pm$  0.09 & -2.84 $\pm$  0.06 &   -8.33 &  0.689 & & 4043397798622539776 &  16.9 &  0.403 $\pm$ 0.022 & -1.91 $\pm$  0.03 & -2.75 $\pm$  0.02 &   -4.74 &  0.813 \\

        \hline         
\end{longtable}
\tablefoot{
  Columns~1-7 refer to the target star from the LPV2 catalogue, Columns~8-14 to the potential companion, and list
  \G\ source Id, parallax, proper motion, GoF and RUWE parameters.
  Column~7 lists a common name for the target star, Column~9 gives the distance of the companion to the target star.
}
  \end{landscape}
}

\newpage
 \FloatBarrier

\section{Additional figures and tables}
\label{App:AF}

    \begin{figure}
    \centering
    \begin{minipage}{0.45\textwidth}
       \resizebox{\hsize}{!}{\includegraphics{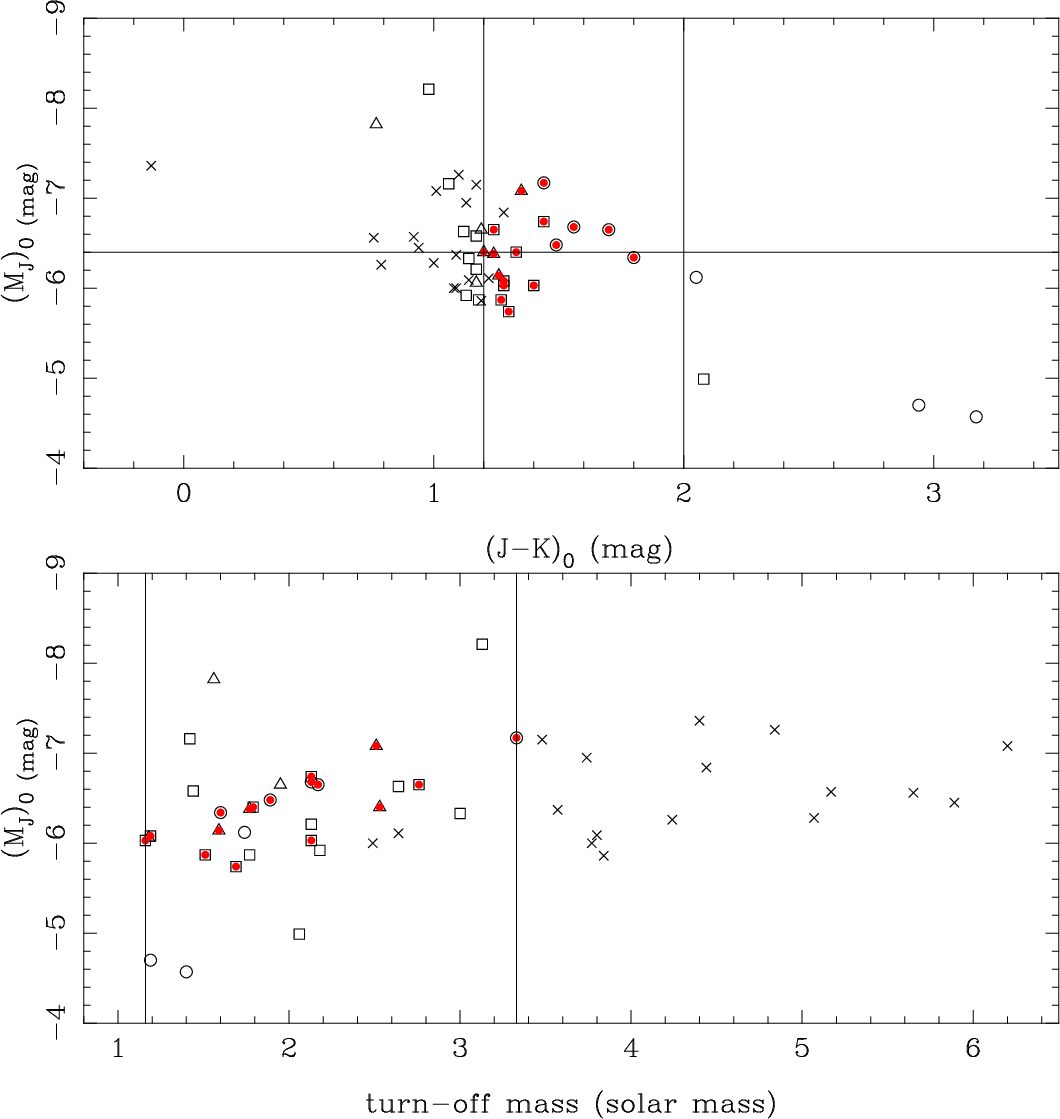}}
    \end{minipage}
        
    \caption{Absolute $J$ magnitude plotted against $(J-K)_0$ colour and OC turn-off mass,
      following Fig.~1 in \citet{Madore22OC}.
      Stars classified as "TP-AGB" in \citet{Marigo22} are plotted as
      C (open circles), M, S, or MS (open triangles), and stars of unknown spectral type (open squares), while stars
      classified as "E-AGB" are plotted as crosses.
      Red filled-in symbols indicate stars selected for the calculation of the absolute magnitude,
      following \citet{Madore22OC}. Lines indicate colours of 1.2 and 2.0~mag and masses of 1.16 and 3.3~\msol, see \citet{Madore22OC}.
    }
      \label{Fig:Marigo}
    \end{figure}

    \begin{figure}
    \centering
    \begin{minipage}{0.45\textwidth}
       \resizebox{\hsize}{!}{\includegraphics{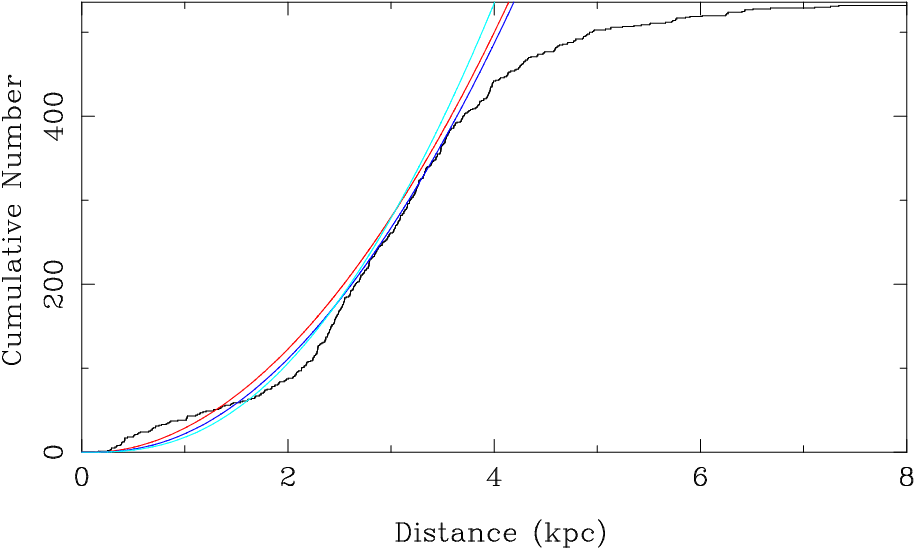}}
    \end{minipage}
        
    \caption{Cumulative number of stars in the selection box for Model 103 (cf. Fig.~\ref{Fig:DistrSTD}).
      Models are for 
      $\rho_0= 25$/kpc$^3$, $H=  200$~pc (red),
      $\rho_0= 10$/kpc$^3$, $H=  500$~pc (dark blue), and
      $\rho_0=  6$/kpc$^3$, $H= 1000$~pc (light blue).
      The number of stars within 1.4~kpc is not subtracted.
    }
      \label{Fig:DistrM103}
    \end{figure}

\begin{figure*}
    \centering
    \begin{minipage}{0.32\textwidth}
       \resizebox{\hsize}{!}{\includegraphics{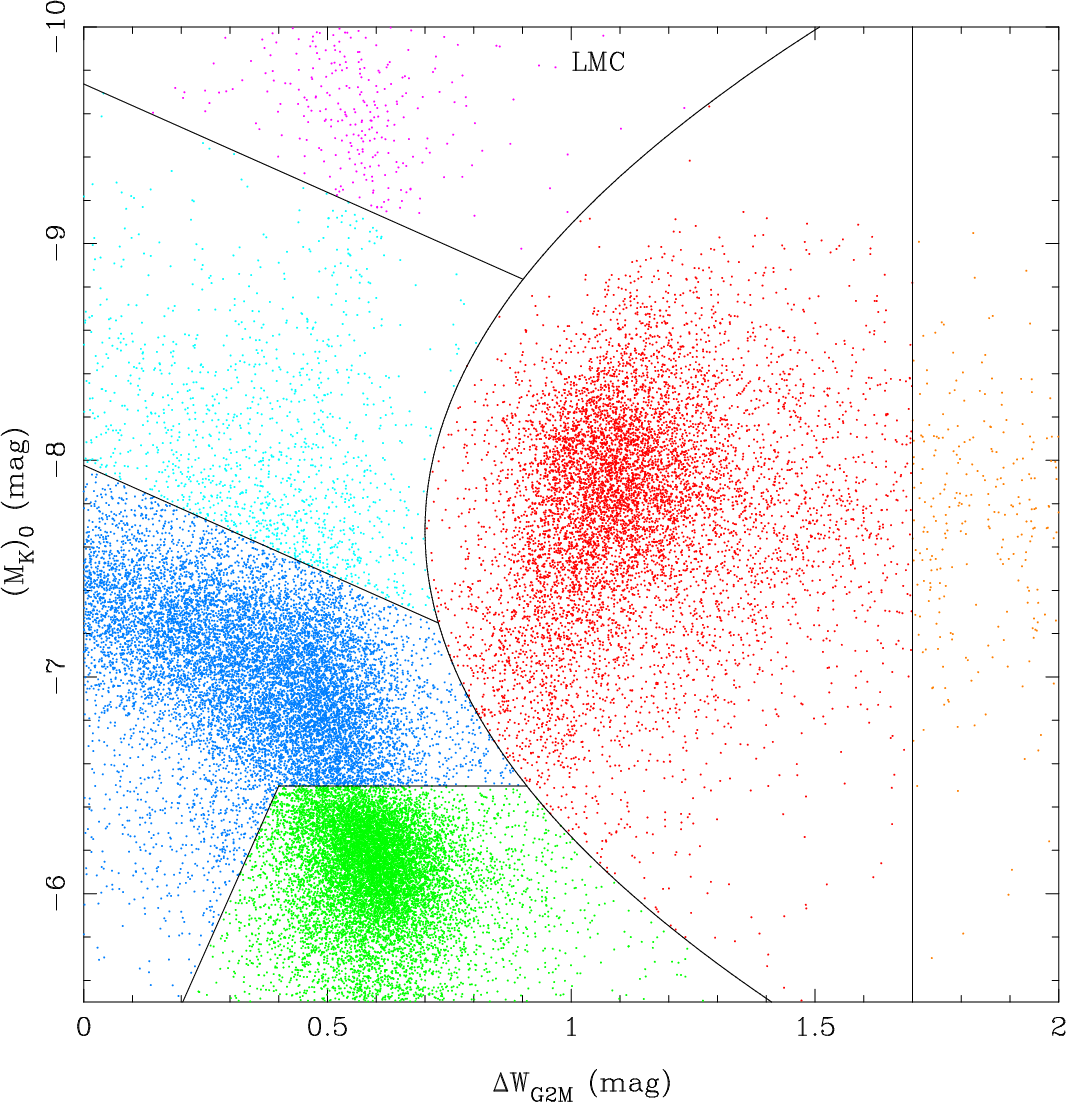}}
    \end{minipage}
    \begin{minipage}{0.32\textwidth}
       \resizebox{\hsize}{!}{\includegraphics{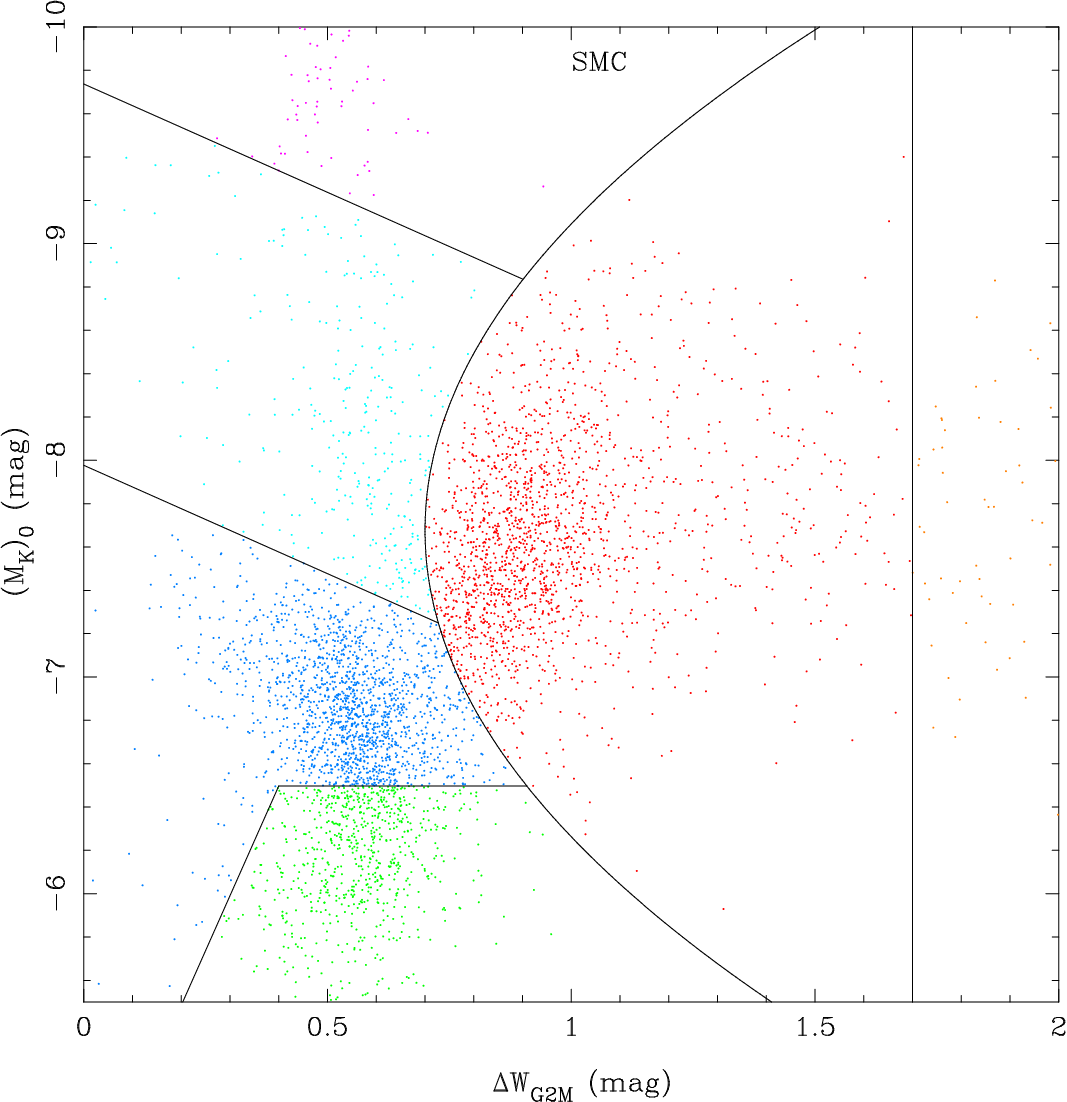}}
    \end{minipage}
    \begin{minipage}{0.32\textwidth}
       \resizebox{\hsize}{!}{\includegraphics{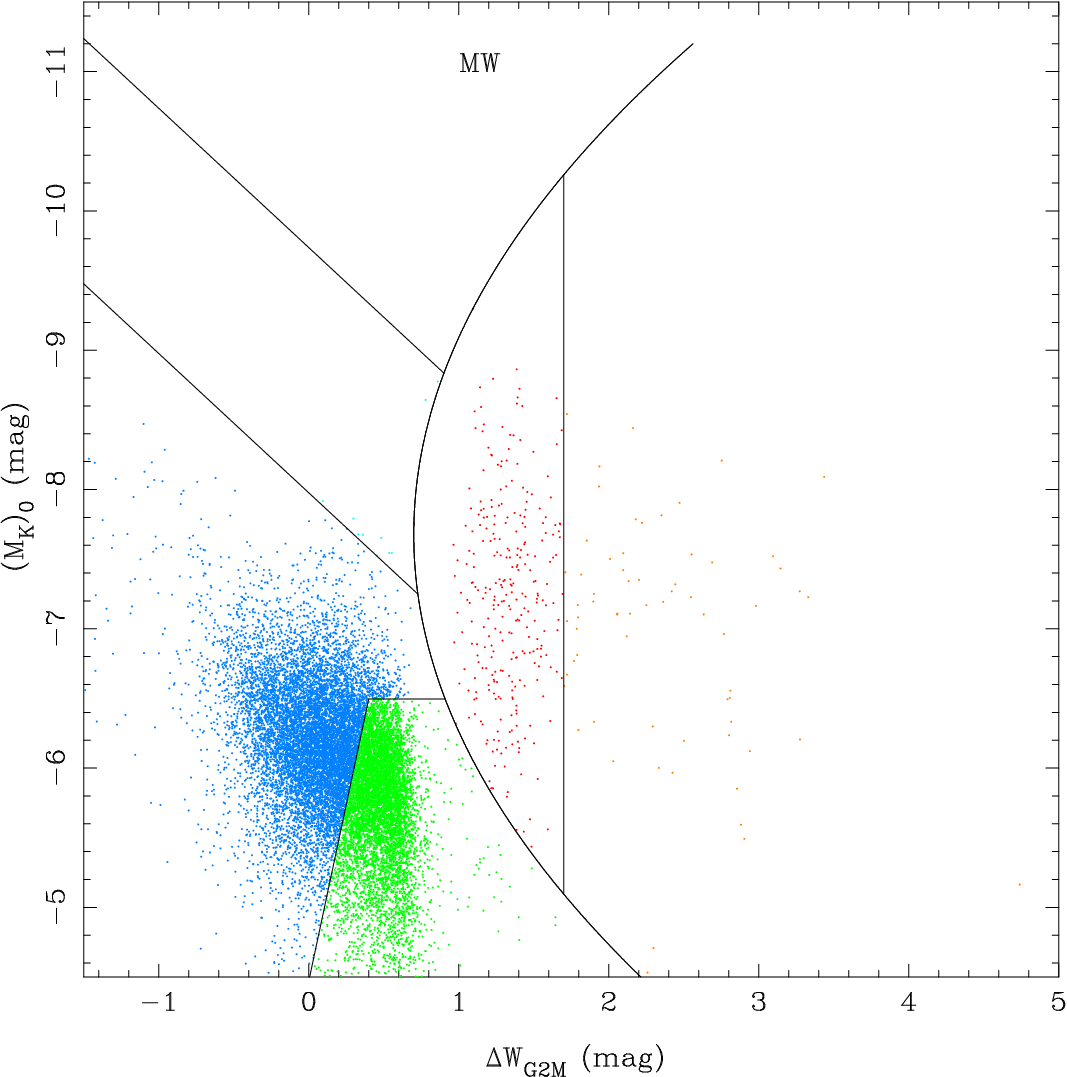}}
    \end{minipage}
    
    \begin{minipage}{0.32\textwidth}
       \resizebox{\hsize}{!}{\includegraphics{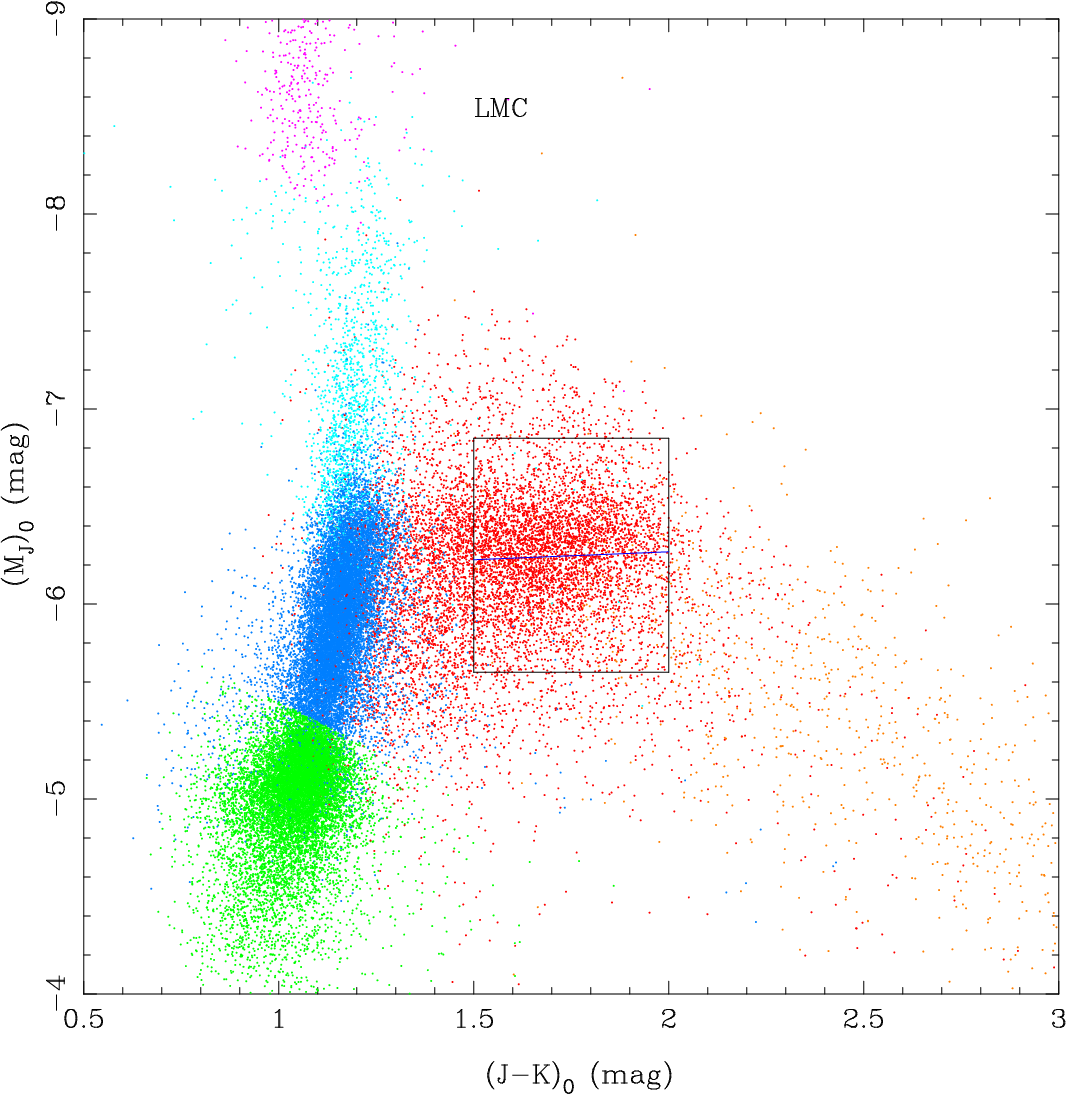}}
    \end{minipage}
    \begin{minipage}{0.32\textwidth}
       \resizebox{\hsize}{!}{\includegraphics{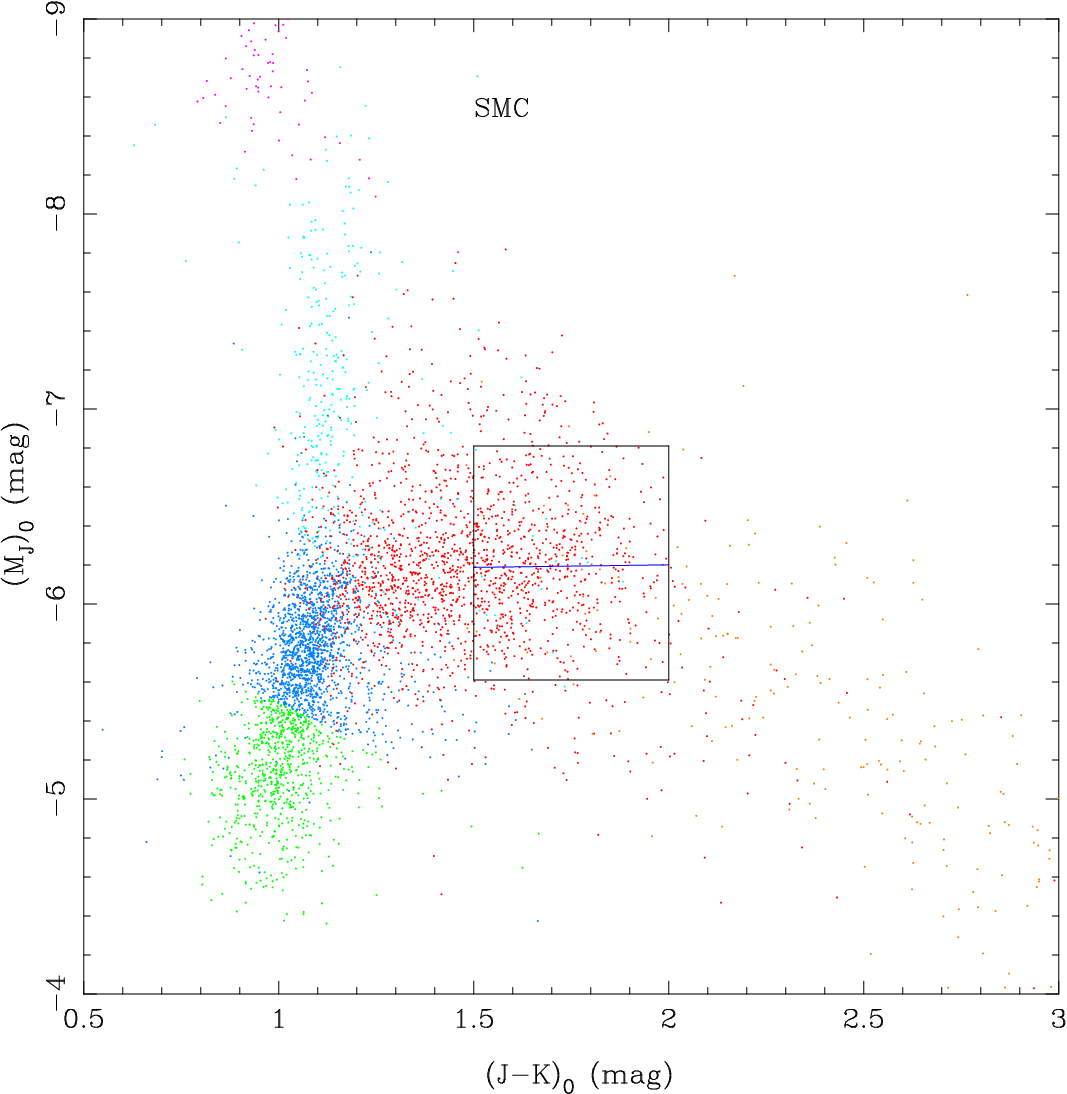}}
    \end{minipage}
    \begin{minipage}{0.32\textwidth}
       \resizebox{\hsize}{!}{\includegraphics{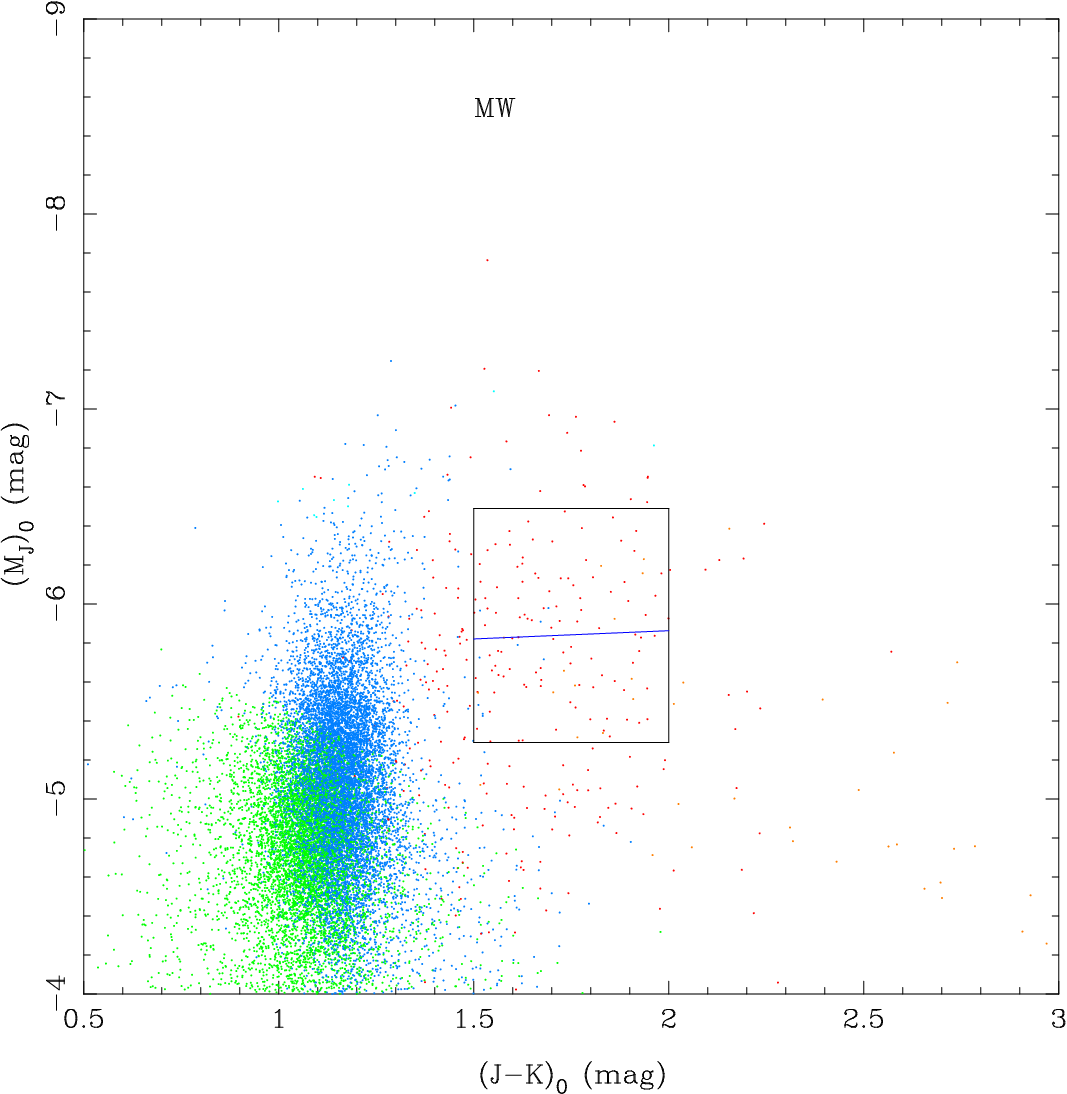}}
    \end{minipage}

    \vspace{5mm}
    
    \begin{minipage}{0.32\textwidth}
       \resizebox{\hsize}{!}{\includegraphics{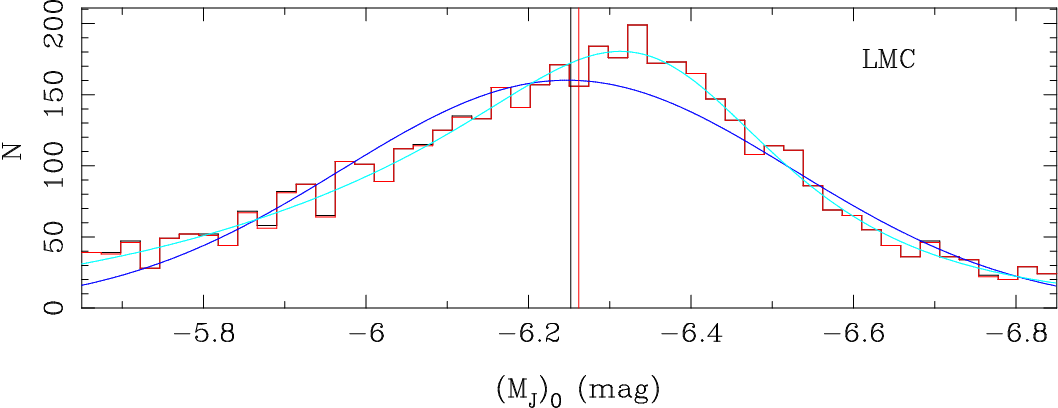}}
    \end{minipage}
    \begin{minipage}{0.32\textwidth}
       \resizebox{\hsize}{!}{\includegraphics{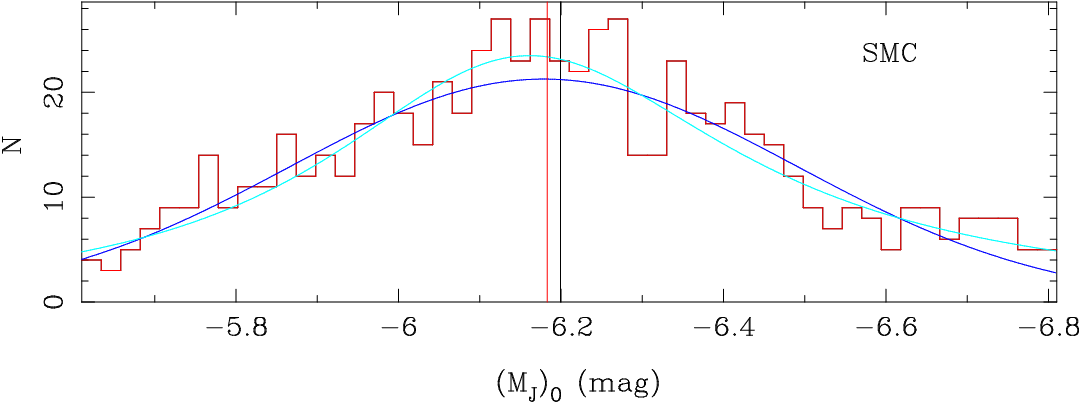}}
     \end{minipage}
    \begin{minipage}{0.32\textwidth}
       \resizebox{\hsize}{!}{\includegraphics{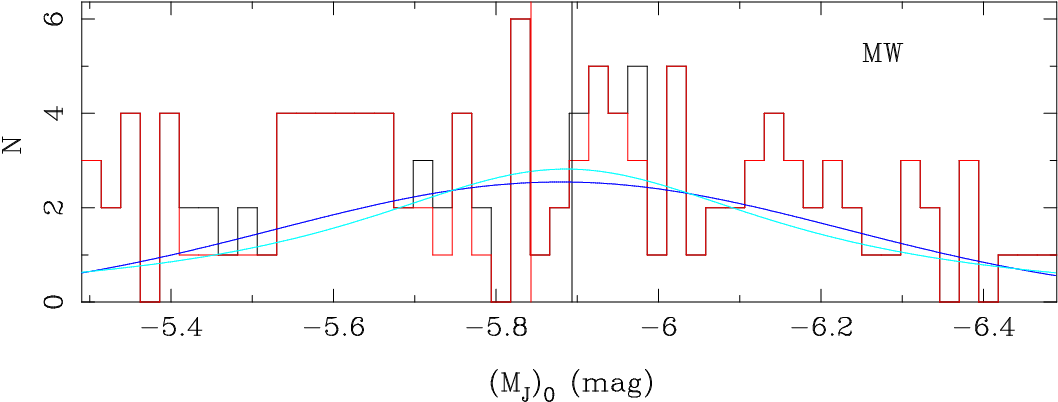}}
     \end{minipage}

    \caption{CMDs, \G-2M diagrams and fits to the LF for models 25 (LMC), 26 (SMC), and 115 (MW)
    }
      
        \label{Fig:AF_M2526115}
    \end{figure*}

   \begin{table*}
	\centering
	\tiny 
	\caption{Criteria to distinguish the different regions in the \G-2M diagrams, rescaled to absolute magnitude of the LMC (DML= 18.477).}
	\label{App:Tab:Boun}
	\begin{tabular}{ll }
        \hline 
        Group & Criterium \\
        \hline         \hline 
        O rich & $W_{RP} - W_{K} \leq 0.7 + 0.15 \cdot (\Ks - 10.8 + {\rm DML})^2 $ \\
        C rich & $W_{RP} - W_{K} >    0.7 + 0.15 \cdot (\Ks - 10.8 + {\rm DML})^2 $ \\
        \hline 
        Faint AGB and RGB & O rich \\
        & $\Ks \geq 11.98 - {\rm DML}$ \\
        & $\Ks \geq 13.50 - 5.067 \cdot (W_{RP} - W_{K} - 0.1) - {\rm DML}$ \\
        Low-mass AGB & O rich \\
        & $\Ks \geq 10.50 + (W_{RP} - W_{K}) - {\rm DML}$ \\
        & [$\Ks < 11.98 - {\rm DML}$ or $\Ks < 13.50 - 5.067 \cdot (W_{RP} - W_{K} - 0.1) - {\rm DML}$ ] \\
        Intermediate-mass AGB & O rich \\
        & $\Ks \geq 8.74 + 2 \cdot (W_{RP} - W_{K}) - {\rm DML}$ \\
        & $\Ks \leq 10.50 + (W_{RP} - W_{K}) - {\rm DML}$ \\
        Massive AGB and RSG & O rich \\
        & $\Ks \leq  8.74 + 2 \cdot (W_{RP} - W_{K}) - {\rm DML}$ \\
        \hline 
        Non-extreme C star & C rich \\
        & $W_{RP} - W_{K} \leq 1.7$ \\
        Extreme C star     & C rich \\
        & $W_{RP} - W_{K}  >   1.7$ \\
        \hline 
        
    \end{tabular}
    \end{table*}
    
    \FloatBarrier

\end{appendix}

\end{document}